\documentclass[amsmath,amssymb,floatfix,twocolumn,citeautoscript,showpacs,notitlepage,superscriptaddress]{revtex4-1}

\usepackage[english]{babel}
\usepackage[hidelinks]{hyperref}
\addto\extrasenglish{%

}
 \addto\captionsenglish{}
\usepackage{amsmath,amssymb,amsthm}
\usepackage{bbm}
\usepackage{mathtools}
\usepackage[dvipsnames]{xcolor}
\usepackage{graphicx}
\usepackage{bm}
\usepackage{multirow}
\usepackage{tikz} 
\usepackage{txfonts}
\usepackage{ulem}
\usepackage{lipsum}
\usepackage{blindtext}
\usepackage{subfiles}
\usepackage{tikz-feynman}
\usepackage{amssymb}
\usepackage{amsmath}
\usepackage{eurosym}

\let\origunderline\underline
\renewcommand{\underline}[1]{\origunderline{\smash{#1}}}

\newcommand{\ul}[1]{\underline{#1}}
\newcommand{\bs}[1]{{\boldsymbol{#1}}} 
\newcommand{\mc}[1]{\mathcal{#1}}

\renewcommand{\d}{\mathrm{d}}
\renewcommand{\i}{\mathrm{i}}
\newcommand{\e}{\mathrm{e}}

\newcommand{\br}{\bs{r}}
\newcommand{\bk}{\bs{k}}
\newcommand{\bq}{\bs{q}}

\newcommand{\mf}{\overline}
\newcommand{\nF}{n_\mathrm{F}}
\newcommand{\bra}[1]{\left\langle #1\right|}
\newcommand{\ket}[1]{\left| #1\right\rangle}
\renewcommand{\mathbb}{\mathbbm}

\newcommand{\pdd}[2]{\frac{\partial #1}{\partial #2}}
\newcommand{\pddx}[3]{\left.\pdd{#1}{#2}\right|_{#3}}
\newcommand{\pddd}[3]{\frac{\partial^2 #1}{\partial #2\partial #3}}
\newcommand{\pdddx}[4]{\left.\pddd{#1}{#2}{#3}\right|_{#4}}

\newcommand{\app}[1]{Appendix~\ref{#1}}

\newcommand{\eq}[1]{Eq.~\eqref{#1}}

\renewcommand{\dag}{\dagger}
\newcommand{\nodag}{{\vphantom\dag}}

\newcommand{\cc}{c^{\nodag}}
\newcommand{\cd}{c^\dag}
\newcommand{\ff}{f^{\nodag}}
\newcommand{\fd}{f^\dag}

\newcommand{\ttau}{\underline\tau}

\newcommand{\pp}{\underline p}
\newcommand{\zz}{\underline z}

\DeclareMathOperator{\tr}{tr}
\DeclareMathOperator{\Tr}{Tr}

\newcommand{\ed}{e^\dag}
\newcommand{\dd}{d^\dag}
\newcommand{\pd}{p^\dag}
\newcommand{\pn}{p^{\nodag}}
\newcommand{\ppuu}{\sum_{\mu=0}^3\pd_{\mu} \pn_{\mu}}

\let\Re\relax
\let\Im\relax
\DeclareMathOperator{\Re}{Re}
\DeclareMathOperator{\Im}{Im}

\begin{document}

\title{Slave-Boson analysis of the 2D Hubbard model}

\newcommand{\affeth}{\affiliation{
Institut f\"ur Theoretische Physik, ETH Z\"urich,
8093 Z\"urich, Switzerland}}
\newcommand{\affuw}{\affiliation{
Institute for Theoretical Physics, University of W\"urzburg,
97074 W\"urzburg, Germany}}
\newcommand{\affpr}{\affiliation{
Princeton Center for Theoretical Science, Princeton University, 
Princeton, NJ 08544, USA}}
\newcommand{\affuzh}{\affiliation{
Department of Physics, University of Z\"urich, 
8057 Z\"urich, Switzerland}}
\newcommand{\affka}{\affiliation{
Institute for Theory of Condensed Matter and Institute for Nanotechnology,
Karlsruhe Institute of Technology,
76021 Karlsruhe, Germany}}

\author{David~Riegler}
\email{david.riegler@physik.uni-wuerzburg.de}
\address{
Institute for Theoretical Physics and Astrophysics, University of W\"urzburg, Am Hubland, D-97074 W\"urzburg, Germany
}

\author{Michael~Klett}
\address{
Institute for Theoretical Physics and Astrophysics, University of W\"urzburg, Am Hubland, D-97074 W\"urzburg, Germany
}

\author{Titus~Neupert}
\address{
Department of Physics, University of Zurich, Winterthurerstrasse 190, 8057 Zurich, Switzerland
}

\author{Ronny~Thomale}
\email{rthomale@physik.uni-wuerzburg.de}
\address{
Institute for Theoretical Physics and Astrophysics, University of W\"urzburg, Am Hubland, D-97074 W\"urzburg, Germany
}

\author{Peter W\"olfle}
\email{peter.woelfle@kit.edu}
\address{
Institute for Theory of Condensed Matter and Institute for Nanotechnology, Karlsruhe Institute of Technology, 76021 Karlsruhe, Germany
}

\date{\today}

\begin{abstract}
We present a comprehensive study of the 2D one-band Hubbard model applying the spin rotation invariant slave-boson method.
We utilize a spiral magnetic mean field and fluctuations around a paramagnetic mean field to determine the magnetic phase diagram and find the two approaches to be in good agreement. 
Apart from the commensurate phases characterized by ordering wave vectors $\bs Q=(\pi,\pi)$, $(0,\pi)$, and $(0,0)$ we find incommensurate phases where the ordering wave vectors $\bs Q=(Q,Q)$, and $(Q,\pi)$ vary continuously with filling, interaction strength or temperature. The mean field quantities magnetization and effective mass are found to change discontinuously at the phase boundaries separating the $(Q,Q)$ and $(Q,\pi)$ phases, indicating a first order transition. The band structure and Fermi surface is shown in selected cases. The dynamic spin and charge susceptibilities as well as the structure factors are calculated and discussed, including the emergence of collective modes of the zero sound and Mott insulator type. The dynamical conductivity is calculated in dependence of doping, interaction strength and temperature. Finally, a temperature-interaction strength phase diagram is established.
\end{abstract}

\maketitle


\section{Introduction}

Strongly correlated electrons on a lattice has proven to be
one of the most interesting and challenging topics of contemporary physics,
following the discovery of heavy electron systems and of the high $T_c$
superconductors. Such systems show a plethora of interesting properties such
as metal-insulator transitions, the emergence of long range order such as
magnetic, charge or nematic order, and possible non-Fermi liquid behavior of
the quasiparticles, in particular near a quantum phase transition, not to
mention the up to now still not fully understood electronic pairing
mechanism of the high $T_c$ superconductors. Most electronic systems in the
metallic phase form a Landau Fermi liquid \cite{landau1957,baym1991}, a state
adiabatically connected to the weakly coupled limit, which at low energies
in slowly varying external fields is characterized by a only a
few parameters $-$ effective mass and Landau interaction functions. The
phenomenological Fermi liquid theory appears to work in extreme strong
coupling situations as represented, e.g., by the heavy quasiparticle system
found in heavy fermion compounds. These successes of \ Fermi \ liquid theory
not withstanding, a microscopic theory of the renormalizations expressed by
the Fermi liquid functions is still largely missing, despite decades of
research efforts.

The archetypical model of a correlated Fermi system is given by the Hubbard
model \cite{hubbard1963,hubbard1964} $-$ a one-band model of electrons on a
lattice subject to on-site interaction $U$. For large $U$\ the model
captures the competition between kinetic energy favoring mobility and local
interaction forcing localization, giving rise to a Mott metal-insulator
transition \cite{mott1949}. In order to treat strong correlations,
non-perturbative methods are required. A first successful approach, capable
of treating the Mott-Hubbard transition, is the so-called Gutzwiller
approximation \cite{gutzwiller1963,brinkman1970,vollhardt1984}. Initially
formulated as a variational problem for the approximate calculation of the
energy expectation value of a correlated wave function, it has later been
rederived in a slave-boson mean field approximation \cite{kotliar_new_1986}.
A further widely used approach starts from the limit of infinite
coordination number \cite{metzner1989,georges1996}, the Dynamical Mean Field
Theory (DMFT). DMFT successfully captures local correlations, and has been
widely applied, with remarkable success. The treatment of longer range
correlations such as present with incommensurate magnetic order or superconductivity is more difficult within the DMFT framework. Among the
many other methods that have been proposed to deal with strongly correlated
systems we like to mention a diagrammatic approximation scheme proposed by B%
\"{u}nemann et al.  \cite{B_nemann_2012} for evaluating Gutzwiller
projected states. The method has been applied, e.g., to study
superconductivity within the 2D Hubbard model~\cite{PhysRevB.88.115127}.

The difficulty in treating models of strongly correlated electrons on the
lattice is that the dynamics of an electron depends on the occupation of the
site it is residing on, which can either be empty ($|0\rangle $), singly occupied ($\ket{\uparrow}$, $\ket{\downarrow }$) or doubly occupied ($|2\rangle $).
For the Hubbard model with large repulsive on-site interaction $U$, the doubly occupied states will be pushed far up
in energy, and will not contribute to the low energy physics. This leads
effectively to a projection of Hilbert space onto a subspace without doubly
occupied sites. It turns out to be difficult to effect this projection
within conventional many-body theory. A powerful technique for describing
the projection in Hilbert space is the method of auxiliary particles \cite%
{fresard2012}: One assigns an auxiliary field or particle to each of the
four states $|0\rangle $, $\ket{\uparrow}$, $\ket{\downarrow }$, $%
|2\rangle $ at a given lattice site (considering one strongly correlated
orbital per site). The fermionic nature of the electrons requires that two of
the auxiliary particles are fermions, e.g., the ones representing $\ket{\uparrow}$,
$\ket{\downarrow } $, and the remaining two are bosons. There are
various ways of defining auxiliary particles for a given problem. This
freedom may be used to choose the one which is best adapted to the physical
properties of the system. A more complex representation of electron
operators in terms of auxiliary particle operators, incorporating the result
of the Gutzwiller approximation \cite{gutzwiller1963} on the slave-boson
mean field level, has been developed by Kotliar and Ruckenstein (KRSB) \cite%
{kotliar_new_1986}. Further extensions to multi-band Hubbard models have
been introduced as well \cite{fresard1997,lechermann2007}. A generalization
of the KRSB method to manifestly spin rotation invariant form \cite%
{woelfle_spin_rotation_1989,woelfle_spin_1992} (SRIKR) has been developed,
allowing one to address non-collinear spin configurations and transverse
spin fluctuations. In particular, the method has been used to describe
antiferromagnetic \cite{lilly1990}, spiral \cite%
{fresarddz1991,igoshev2013,igoshev2015,fresardw1992,moellerd1993}, and
striped \cite%
{seibold1998,lorenzana2002,lorenzana2003,lorenzana2005,raczkowski2006,raczkowski2007,fleck2001}
phases. Furthermore, the competition between spiral and striped phases has
been studied \cite{raczkowskif2006}. In the limit of large $U>60t$, it has
been found that the spiral order continuously evolves toward ferromagnetic
order.

In this paper we present a detailed derivation of the SRIKR slave-boson
formalism (\autoref{sec2} and appendices \ref{Chapter:App:Operatorlevel}-\ref{Chapter:App:MaME}). In \autoref{sec:Results:ZeroT} we apply it to calculate
slave-boson mean field solutions and two-particle response functions for the
one-band Hubbard model on a two-dimensional square lattice at zero temperature. The magnetic phase diagram in the interaction-density-plane within the manifold of spiral magnetic states is obtained from
the mean field analysis. We discuss the energy spectrum and the mass
enhancement of quasiparticles at the Fermi level, as well as the Fermi
surfaces. The static spin susceptibility is parametrized in terms of a
Landau interaction function. The dynamic spin susceptibility is calculated
and parametrized in terms of a Landau damping function. At the phase
transition, the spin susceptibility at the ordering wave vector is found to
diverge as $\chi (\bs Q,0)\propto (n-n_{c})^{-\alpha }$ where $n_{c}$
is the critical doping. We determine the phase boundaries to the
paramagnetic phase (i) from the mean field equations and (ii) from the
divergence of the paramagnetic spin susceptibility at finite wave vector $%
\bs Q$. The two methods provide consistent results, where in the case of second order transitions, method (ii) is
more efficient, whereas first order transitions can only be found with method (i). The ordering wave vector $\bs Q$ is found to vary continuously over large
parts of the phase diagram, but suffers occasional jumps signaling first
order phase transitions. The charge response function is employed to
calculate the dynamical conductivity.

In \autoref{sec:Results_finiteT} we present results at finite temperature. We determine the
magnetic phase diagram in the temperature $T-$ doping $n$ plane at fixed
interaction $U$. We determine the phase boundaries separating the
magnetically ordered phases from the paramagnetic phase and also separating\
different ordered states. A continuous change of the ordering wave vector as
the temperature and doping are varied is determined. The static spin
susceptibility at fixed $U$ and $n$ is found to diverge at the transition as 
$\chi (\bs Q,0)\propto (T-T_{c})^{-1}$, where $T_{c}$ is the critical
temperature.

We compare our results with available benchmark results, in particular a
recent study of the two-dimensional Hubbard model using Density Matrix
Embedded Theory (DMET) \cite{PhysRevB.93.035126}, finding remarkably good agreement.


\section{Model and method}\label{sec2} 
This section defines the Hamiltonian and summarizes the most important aspects of spin rotation invariant slave-boson formalism, while a detailed derivation of the method can be found in the appendices \ref{Chapter:App:Operatorlevel}-\ref{Chapter:App:MaME}.
We investigate the one band Hubbard model in two spatial dimensions (2D), 
\begin{align}\label{Model:Hamilton}
H =&-\sum_{i,j,\sigma }t_{i,j}c_{i,\sigma }^{\dagger}c^\nodag_{j,\sigma }
-\mu_0\sum_{i,\sigma}c_{i,\sigma }^{\dagger}c^\nodag_{i,\sigma }
+U\sum_{i}c_{i,\uparrow }^\dagger c^\nodag_{i,\uparrow }c^\dagger_{i,\downarrow }c_{i,\downarrow }^\nodag,
\end{align}
where the operator $c_{i,\sigma }^{\dagger }$ creates a fermion on site $i$ with spin $\sigma =\{\uparrow ,\downarrow \}$. We allow hopping terms between nearest neighboring sites denoted by $t$ and next nearest neighboring sites by $t'$.
Further we employ an on-site Hubbard interaction $U$. All energy scales are given in units of $t$.

\subsection{Slave-boson representation}\label{sec2.A} 
We apply the SRIKR slave-boson representation, where the original fermionic operator $c^{(\dagger)}_{i,\sigma}$
is expressed as a combination of the pseudofermion operator $f_{i,\sigma }$ and bosonic operators $e_{i},d_{i},p_{i,\mu }$ with $\mu \in \{0,1,2,3\}$, labeling empty, doubly and singly
occupied states, respectively, as 
\begin{align}
c^\dagger_{i,\sigma} &\equiv \sum_{\sigma'} z^\dag_{i,\sigma\sigma'} \fd_{i,\sigma'},
\end{align}
where
\begin{subequations}
\begin{align}
\zz = (e^\dag\underline{L} M\underline{R} \ \pp+\tilde\pp^\dag \underline{L}M\underline{R} d), 
\end{align}
with
\begin{align}
\underline{L} &= \left((1-d^\dag d)\ttau_0-2 \pp^\dag \pp\right)^{-1/2}, \\
M &= \left(1 + d^\dag d +e^\dag e +\sum\limits_{\mu} p^\dag_\mu p^\nodag_\mu \right)^{1/2}, \\
\underline{R} &= \left((1-e^\dag e)\ttau_0-2 \tilde{\pp}^\dag \tilde{\pp}\right)^{-1/2}.
\end{align} 
\end{subequations}
The underbar denotes $2\times 2$ spin matrices, specifically
\begin{equation}
 \underline{p}=\frac{1}{2}\sum_{\mu=0}^3 \pn_\mu \ttau^\mu,
\end{equation}
where $\underline{\tau }^{\mu }$ are the Pauli matrices including the unit matrix $\underline{\tau }^{0}$, and $\underline{\tilde{p}}$ is the time reversed operator of $\underline{p}$. 
This form ensures spin rotation invariance (compare \app{Chapter:App:Operatorlevel}) and the correct non-interacting limit within the mean field approximation (compare \app{Chapter:App:MF}).

The slave-boson representation is complemented by the local constraints
\begin{subequations}
\begin{align} 
1&= \ed_i e_i^\nodag+\dd_i d_i^\nodag + \sum_{\mu=0}^3p^\dagger_{i,\mu} p^\nodag_{i,\mu}\ ,\label{Eq:MM:Constraint1}\\ 
\sum_{\sigma}\fd_{i,\sigma} f^\nodag_{i,\sigma}&= \sum_{\mu=0}^3p^\dagger_{\mu,i} p^\nodag_{\mu,i}+2\dd_i d^\nodag_i \ , \\
\sum_{\sigma \sigma'}^{\vphantom{3}} \bs \tau_{\sigma \sigma'} \fd_{i,\sigma'}f^\nodag_{i,\sigma}&= \pd_{0,i} \bs p_i^\nodag + \bs p_i^\dagger p_{0,i}^\nodag-\i \bs p_i^\dagger \times \bs p_i^\nodag\;, 
\end{align}
\end{subequations}
where $\bs \tau$ is the three-vector of Pauli matrices. These constraints project onto the physical subspace and are
enforced on each lattice site $i$, using the Lagrange multipliers $\alpha_i$, $\beta_{0,i}$ and $\bs \beta_i$ within the path integral formulation (compare \app{Chapter:App:Pathintegral}).
In \app{Chapter:App:Atomic} we show that the SRIKR slave-boson representation exactly recovers the atomic limit by integrating out all fields.

\subsection{Mean field approximation}\label{sec2.B}
In the slave-boson representation, the interaction term of the Hamiltonian becomes quadratic, at the cost of a non quadratic hopping contribution in bosonic operators. Therefore we employ a (para-) magnetic mean field. In this approximation, the space and time dependent slave-boson
	fields are replaced by their static expectation value $\psi \rightarrow \langle \psi \rangle$ with $\partial_\tau \langle \psi \rangle=0$.
A suitable mean field ansatz incorporating a spin spiral with ordering vector $\bq$  is given by \cite{PhysRevB.48.10320},
\begin{align}
\begin{split}
e_i&\rightarrow \langle e \rangle \in \mathbb{R}_0^+,\\
p_{0,i}&\rightarrow \langle p_0\rangle \in \mathbb{R}_0^+,\\
d_i&\rightarrow \langle d\rangle \in \mathbb{R}_0^+,\ \partial_\tau \langle d \rangle = 0, \\
i\beta_{0,i}&\rightarrow \langle \beta_0\rangle \in \mathbb{R},\\
i\alpha_{i}&\rightarrow \langle\alpha\rangle \in  \mathbb{R}, \\
\bs p_i&\rightarrow\langle p\rangle
\begin{pmatrix}
\cos(\phi_i)\\
\sin(\phi_i)\\
0
\end{pmatrix},\ \langle p \rangle \in \mathbb{R}_0^+, \\
i\beta_i&\rightarrow\langle \beta \rangle
\begin{pmatrix}
\cos(\phi_i)\\
\sin(\phi_i)\\
0
\end{pmatrix},\ \langle \beta \rangle \in \mathbb{R},\\
\phi_i&\equiv \bs q \bs r_i.
\end{split}
\end{align}
To keep the notation short we drop the brackets $\langle \rangle$ in the following. Within this mean field ansatz the Lagrangian is given by
\begin{subequations}
\begin{equation}\label{key}
\begin{gathered}
\mathcal{L}_{\bq}=\sum_\bk \bs f^\dagger_\bk \left(\underline{H}_\bk[\bq,\psi] + \partial_\tau \right) \bs f^\nodag_\bk +N \big[ U d^2 -2\beta p_0 p\\-\beta_{0}(p_{0}^2+p^2+2d^2) +\alpha(e^2+p_{0}^2+d^2-1+p^2)\big],
\end{gathered}
\end{equation}
with 
\begin{gather}
\bs f_\bk=
\begin{pmatrix}
f_{\uparrow,\bk} \\
f_{\downarrow,\bk-\bq} 
\end{pmatrix},
\end{gather}
\end{subequations}
where $\underline{H}_\bk[\bq,\psi]$ is the non interacting Hamiltonian of the pseudo fermions in dependence of the slave-boson mean fields $\psi$ and $N$ is the total number of lattice sites. 
Due to the form of the constraints, the Free energy per lattice site $F_\bq/N$ only depends on two independent bosonic fields, which we choose to be $p_0$, $p$ without loss of generality, and the two Lagrange multiplier fields $\beta$ and $\mu_\text{eff}=\mu_0-\beta_0$.
After integrating out the pseudofermionic degrees of freedom, it is given by (compare \app{Chapter:App:MaME})
\begin{equation}\label{key}
\begin{gathered}
\frac{F_\bq}{N}=-\frac TN\sum_{\bk,\pm} \ln\left[1+e^{-\epsilon_{\bk,\pm}/T}\right]\\-\frac{U}{2}(p_0^2+p^2-n)+\mu_{\text{eff}}n-2\beta p_0 p,
\end{gathered}
\end{equation}
where $\epsilon_{\bk,\pm}$ are the eigenvalues of the matrix $\underline{H}_\bk[\bq,\psi]$, $n$ is the electron filling per lattice site and $T$ is the temperature.\\
The saddle point solution for the ground state is determined by minimizing the free energy with respect to $p_0$, $p$ and ordering vector $\bq$ while maximizing with respect to $\beta$ and $\mu_\text{eff}$.
The according mean field values of the bosonic fields are denoted by $\bar{\psi}$ and can be characterized by $\bar{p}=0$ describing a paramagnet (PM) and $\bar{p}\neq 0$ yielding magnetic order. 

\subsection{Fluctuations around the paramagnetic saddle point} \label{sec2.C}
In order to calculate response functions, we consider Gaussian fluctuations around the paramagnetic saddle point ~\cite{Li1991,woelfle_spin_1997}, i.e., we expand the action to the second order in bosonic fields $\psi_\mu$ around the mean field solution
\begin{subequations}
\begin{equation}
\delta\mc S^{(2)}=\sum_{q,n}\delta \psi_\mu(-\bs q,-i\omega_n) \mathcal{M}_{\mu\nu}(q) \delta\psi_\nu(\bs q,i\omega_n),
\end{equation}
with the fluctuation matrix
\begin{equation}
 \mathcal{M}_{\mu\nu}(\bs q,\omega_n)
\equiv\frac{\delta^2 \mathcal{S}(\psi)}{\delta \psi_\mu(-\bs q,-i\omega_n)\delta \psi_\nu(\bs q,i\omega_n)}\ ,
\end{equation}
\end{subequations}
where $q=(\bs q, \omega_n)^T$ and $\omega_n=2\pi Tn$ ($ n\in \mathbb{Z} $) is a bosonic Matsubara frequency. 
The phases of the $e, p_0$ and $\bs p$ fields can be removed by a gauge 
transformation (compare \app{Chapter:App:Gauge}), such that only the $d$-field remains complex valued in position space $d\equiv d_1+id_2$. 

Since the fluctuations are calculated by means of functional
derivatives, they violate the constraints which are exactly enforced only at the saddle point. Such violations are actually necessary in order to resolve correlations
and evaluate whether the system will relax back to the paramagnetic mean field solution or whether it features an instability.
Since the Lagrange multipliers are part of the effective field theory, one needs to consider the fluctuation of $\beta_0$ and $\bs \beta$ as well. However, fluctuations in $\alpha$ would yield bosonic occupations per lattice site unequal to one, which can be associated with a violation of the Pauli principle. This needs to be avoided by replacing an arbitrary slave-boson field (we choose $p_0$ w.l.o.g) via \eq{Eq:MM:Constraint1}, i.e. fluctuating on the subspace where the $\alpha$ constraint is exactly fulfilled (compare \app{Chapter:App:Fluc}).
We apply the following convention for the ten bosonic fields: 
$\psi_{1}=e$, $\psi_{2}=d_1$, $\psi_{3}=d_2$, $\psi_{4}=\beta_0$, $\psi_{5,6,7}=p_{1,2,3}$ and $\psi_{8,9,10}=\beta_{1,2,3}$. 

Dynamical response functions can be calculated using the path integral (compare \app{Chapter:App:Correlation})
\begin{align}
\begin{split}
 \langle \delta \psi_\mu^*(q) \delta\psi^\nodag_\nu(q)\rangle &=
 \frac{1}{Z^{(2)}}\int D[\psi^*,\psi] \delta \psi^*_\mu(q) \delta\psi^\nodag_\nu(q) e^{-\delta \mathcal{S}^{(2)}}\\
 &=\mathcal{M}^{-1}_{\mu\nu}(q) \\
\text{with} \quad Z^{(2)} &=\int D[\psi^*,\psi]  e^{-\delta \mathcal{S}^{(2)}}.
\end{split}
\end{align}
To evaluate these quantities, we apply a Wick rotation $i\omega_n\rightarrow \omega+i\eta$, where $\eta\rightarrow 0^+$ regularizes diverging terms and needs to be kept finite for most numerical calculations.
\subsubsection{Spin susceptibility}
The spin susceptibility is defined by
\begin{subequations}
\begin{equation}
 \chi^{\alpha\beta}_{\text{s}}(q)=  \langle  \delta S^\alpha_{-q} \delta S^\beta_q\rangle.
\end{equation}
where $\bs S$ is the spin density, which can be written in terms of slave-bosons (compare \app{Chapter:App:Correlation})
\begin{equation}\label{Eq:Model:Spinoperator}
 \bs S =  \check{\bs p} p_0  \quad  \text{with}\ \check{\bs p}=(p_1,-p_2,p_3)^T. 
\end{equation}
In the one band Hubbard model, the fluctuation matrix is block diagonal in ($\beta_\alpha, p_\alpha)$ and $(e,d_1,d_2,\beta_0)$ since spin and charge sector are decoupled. Hence, the spin susceptibility takes the
simple diagonal form
\begin{equation}
 \chi_{\text{s}}^{\alpha\beta}(q)= \bar{p}_0^2\frac{ \mathcal{M}_{10,10}(q)}{\mathcal{M}_{7,7}(q)\mathcal{M}_{10,10}(q)-\mathcal{M}_{7,10}(q)\mathcal{M}_{10,7}(q)}\delta^{\alpha\beta}.
\end{equation}
\end{subequations}

\subsubsection{Bare susceptibility and Charge susceptibility}
\begin{subequations}
The bare susceptibility can be determined analogously and is defined by
\begin{equation}
\begin{gathered}
 \chi_0(q)\equiv \frac{1}{Z^{(0)}}\int D[f^*,f] n_{- q}n_ q  e^{-\mathcal{S}^{(0)}}=-2\mathcal{M}_{4,4}(q)\\
\text{with} \quad Z^{(0)} =\int D[f^*,f]  e^{-\mathcal{S}^{(0)}}
\end{gathered}
\end{equation}
where $n_q=\sum_\bk f^*_{\bk + \bq}f^\nodag_\bk$ is the pseudofermion density and $\mathcal{S}^{(0)}$ is the mean field action.

The charge susceptibility is defined by
	\begin{equation}
		\chi_c(q)= \langle \delta n_{-q} \delta n_q\rangle
	\end{equation}
where $n$ is the charge density, which can be written in terms of slave-bosons
\begin{equation}
 n = 1+d^2-e^2. 
\end{equation}	
Hence, we find
	\begin{equation}
		\chi_c(q)=2\bar{d}^{2}_{1}\mathcal{M}^{-1}_{2,2}(q)+2\bar{e}^2
		\mathcal{M}^{-1}_{1,1}(q)-2\bar{d}_{1}\bar{e}\left(\mathcal{M}^{-1}_{1,2}(q)+\mathcal{M}^{-1}_{2,1}(q)\right).
	\end{equation}
\end{subequations}

\subsubsection{Structure factors}
To describe thermal fluctuations at the frequency $\omega$, we define the structure factor, which is given by the real quantity~\cite{auerbach2012interacting}
\begin{equation}
 \label{Model:Spinstructurefactor}
S_{\alpha}(\bq)= -\int\limits_{-\infty }^{\infty}\frac{d\omega }{\pi }\frac{\mathrm{Im}\ \chi _{\alpha}(\bs q,\omega+i\eta)}{1-e^{-\omega /T}},
\end{equation}
where $\alpha=s$ is called spin structure factor and $\alpha=c$ charge structure factor.

\subsection{Dynamical conductivity}\label{section:conductivity}
With our results for the charge susceptibility $\chi_c$, we can calculate the dynamical conductivity as
\begin{subequations}
 \begin{equation}
	\sigma(\omega+i\eta)=e^2\lim_{\bq\rightarrow 0}\frac{-i\omega+\eta}{\bq^2}\chi_c(\bq,\omega+i\eta),\label{Model:Conductivity}
\end{equation}
where we have performed the analytical continuation $i\omega_n \rightarrow \omega+i\eta$.
The convergence parameter $\eta$ can be identified with an inverse scattering time $\tau=1/\eta$ of a Drude conductivity
\begin{equation}
\sigma_D(\omega,\tau)= \frac{\sigma_0}{1+\omega^2\tau^2}+i\frac{\sigma_0\omega\tau}{1+\omega^2\tau^2}.\label{Model:Drude}
\end{equation}
\end{subequations}
where $\sigma_0 \propto \tau$. By data fitting, we can determine the DC-conductivity $\sigma_0$ and assign sensible results for $\eta \neq 0$ whereas $\sigma_0 \rightarrow \infty$ for $\eta\rightarrow 0$.



\section{Results at zero temperature}\label{sec:Results:ZeroT}
This section discusses the slave-boson mean field and fluctuation results applied to the 2D Hubbard model at zero temperature.
\subsection{Mean field approximation (MFA)}

\begin{figure}[t]
	\includegraphics[width=0.48\textwidth]{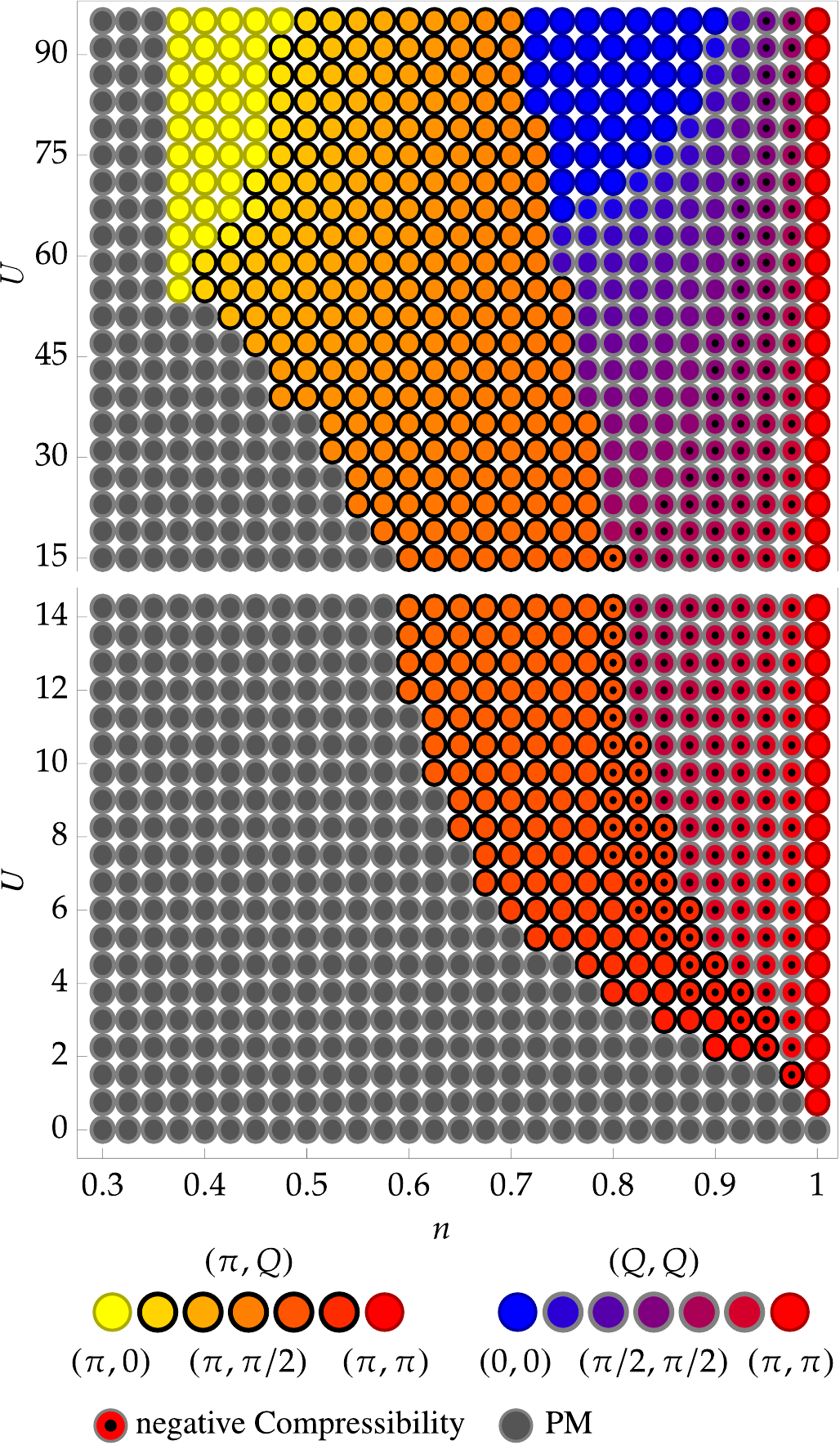}
	\caption{Magnetic mean field phase diagram for the 2D Hubbard model
		as function of the interaction $U$ and the filling $n$ without next nearest
		neighbor hopping ($t^{\prime }=0$) at zero temperature ($T=0$). We only show 
		$n\leq 1$ since the phase diagram for $n>1$ is mirrored due to particle hole
		symmetry ($n\leftrightarrow 2-n$). It features three distinct phases, namely
		the PM (gray), a ($\protect\pi ,Q$) phase denoted by black circles
		filled with coloring from red to yellow and a ($Q,Q$) phase denoted by gray
		circles filled with coloring from red to blue. The ordering vector within
		one phase regime changes continuously with $U$ and $n$ visualized by the
		color scheme as indicated in the plot legend. The phase diagram features
		three commensurate magnetic orderings which are special cases of the above
		described phases, namely the antiferromagnet (red circle), the ferromagnet
		(blue circle) and stripe magnetism (yellow circle). }
	\label{Fig:Phasediagram_t2=0}
\end{figure}

\begin{figure*}[t]
	\includegraphics[width=\textwidth]{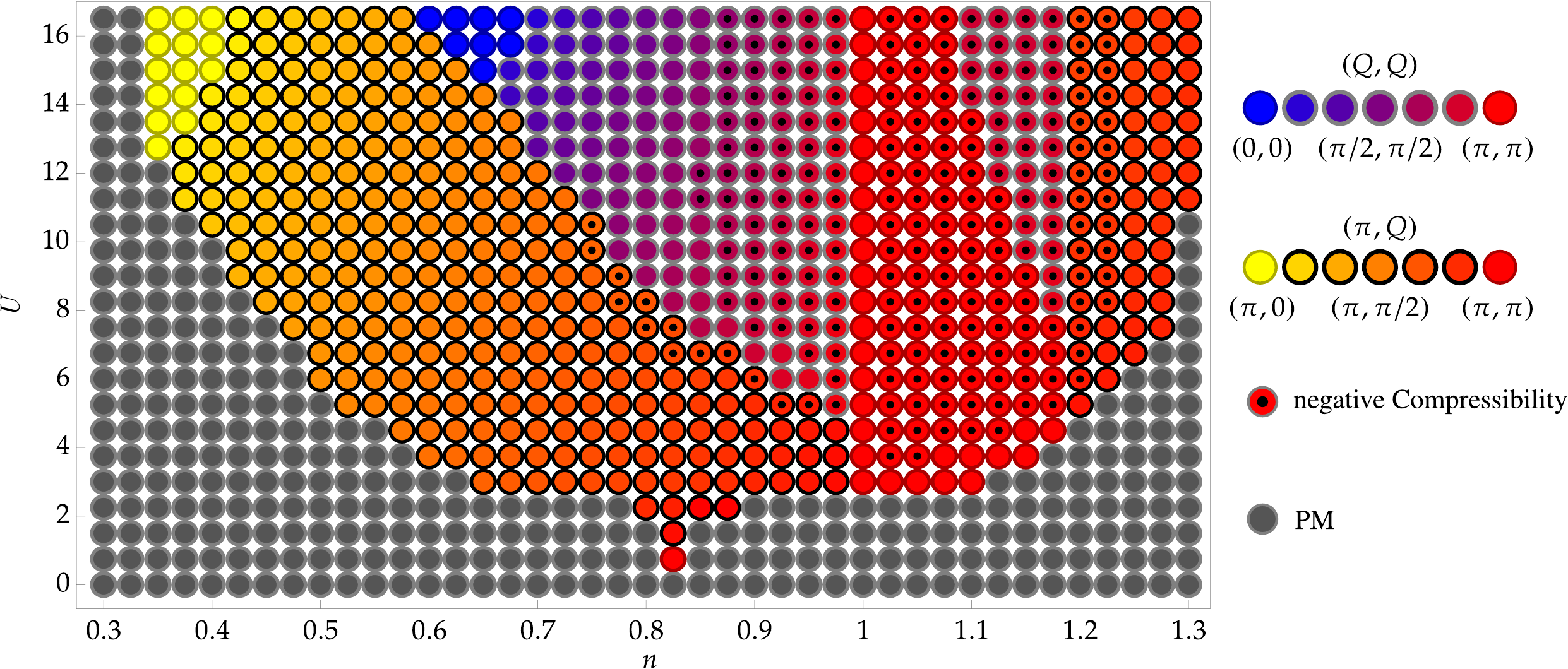}
	\caption{Magnetic mean field phase diagram for the 2D Hubbard model
		as function of the interaction $U$ and the filling $n$ with next nearest
		neighbor hopping ($t^{\prime }=-0.2$) at zero temperature ($T=0$). It features three distinct phases, namely
		the paramagnet (gray), a ($\protect\pi ,Q$) phase denoted by black circles
		filled with coloring from red to yellow and a ($Q,Q$) phase denoted by gray
		circles filled with coloring from red to blue. The ordering vector within
		one phase regime changes continuously with $U$ and $n$ visualized by the
		color scheme as indicated in the plot legend. The phase diagram features
		three commensurate magnetic orderings which are special cases of the above
		described phases, namely the antiferromagnet (red circle), the ferromagnet
		(blue circle) and stripe magnetism (yellow circle).}
	\label{Fig:Phasediagram_t2=-0.2}
\end{figure*}

We have solved the slave-boson mean field equations presented in \autoref%
{sec2.B} and described in more detail in \app{Chapter:App:MaME}, for the
paramagnetic and spiral magnetic phases. We found the minimum of \ the
resulting free energy, determining the thermodynamically stable phase. In
this way we established a phase diagram, presented in the next subsection. An
important property of the mean field solution is the renormalization of the
fermionic excitation spectrum, given by the factors $z_{0}$ for the
paramagnet and $\mathcal{Z}_{+,-}$ for the spiral magnet. Results on the $z-$factors
and on several other mean field parameters are presented in the subsequent
part. We also show examples of the electronic band structure and the Fermi
surface. In addition to magnetic order, charge density wave order may
appear. In this work we do not address the case of several types of order
being present simultaneously. We do, however, identify signals for the
probable appearance of charge order in the presence of magnetic order by
studying the compressibility, which is found to turn negative in a certain
portion of the magnetically ordered phases.

\subsubsection{\protect\bigskip Magnetic mean field phase diagram}

The phase diagram shown in \autoref{Fig:Phasediagram_t2=0} has been
calculated by means of the magnetic mean field theory defined in \autoref{sec2.B},
putting $t^{\prime }=0$. It complements phase diagrams shown in the
literature \cite{fresardw1992,igoshev2013} which have only been calculated
for smaller $U$. At the phase boundary to the paramagnet, the order
parameters $p$ and $\beta $ vanish continuously, i.e., the paramagnetic
solution is recovered via a continuous phase transition.\newline
At half filling, the ordering is antiferromagnetic for every finite
interaction $U>0$. 
Away from half filling, within the ordered phase regime, the ordering vector 
$\bs Q$ evolves continuously as function of the filling $n$ and interaction $%
U$. The transitions observed between ($\pi ,Q$) and ($Q,Q$) phases are of
first order since the order parameter $p$ is found to be discontinuous at
the phase boundaries. Furthermore, the phase diagram
features three different commensurate magnetic phases, namely the
antiferromagnet [$\bs Q=(\pi ,\pi )$], ferromagnet [$\bs Q=(0,0)$] and stripe
magnetism [$\bs Q=(0,\pi )$]. The ferromagnet features a vanishing double
occupancy $d=0$ and $p=p_{0}=\sqrt{n/2}$ which yields the maximum possible
magnetization per lattice site $m=pp_{0}=n/2$ (in units of the Bohr
magneton) for a given filling. The contribution of fluctuations is expected
to lead to $d\neq 0$ and to $m<n/2$ . Every non-ferromagnetic state has a
finite double occupancy $d\neq 0$ even on the mean field level.

\autoref{Fig:Phasediagram_t2=-0.2} shows the magnetic phase diagram for $t^{\prime }=-0.2$, in the extended density regime $0.3<n<1.3$ and is in very good agreement with a previous Slave-boson study \cite{igoshev2013}. At half
filling, the tendency towards the antiferromagnet is reduced, because finite 
$t^{\prime }$ prevents the perfect nesting of the Fermi surface, yielding a
paramagnetic regime at weak interaction. On the other hand, at $U\gtrsim 2$,
the tendency towards magnetic order is generally increased for larger $|n-1|$, due to the increased hopping range. The next nearest neighbor hopping $t^{\prime
}=-0.2$ moves the van Hove singularity, giving rise to an enhanced tendency
for magnetic order at $n\approx 0.825$.

\subsubsection{Fermionic $z-$factors}

The factor $z_{0}$ renormalizes the fermionic band structure in the
paramagnetic phase as

\begin{subequations}
\begin{equation}
\epsilon _{\bk}=z_{0}^{2}\xi _{\bk}-\mu _{\text{eff}}
\end{equation}
where for nearest neighbor hopping $\xi _{\bk}=-2\left[ \cos (k_{1})+\cos
(k_{2})\right] $. The factor $z_{0}$ governs the bandwidth $W=8z_{0}^{2}$
and the effective mass at the Fermi level, e.g. along the x-axis $m^{\ast
}(k_{1F})=[2z_{0}^{2}\cos (k_{1F})]^{-1}=[2z_{0}^{2}+\mu _{\text{eff}}]^{-1}$, where $k_{1F}$ is the Fermi wavenumber.

In the magnetically ordered phase the fermionic dispersion is given by

\begin{equation}
\epsilon _{\bk,\pm }=\frac{1}{4}\left[\zeta _{+}\xi _{\bk,+}\pm 
\sqrt{(\zeta _{+}^{2}-\zeta _{-}^{2})\xi _{\bk,-}^{2}+(\zeta _{-}\xi
_{\bk,+}+4\beta )^{2}}\right]-\mu _{\text{eff}}
\end{equation}%
where $\zeta _{\pm }=z_{+}^{2}\pm z_{-}^{2}$, and $\xi _{\bk,\pm }=\xi _{%
\bk}\pm \xi _{\bs k+\bs Q}.$

\begin{figure}[t]
\includegraphics[width=0.48\textwidth]{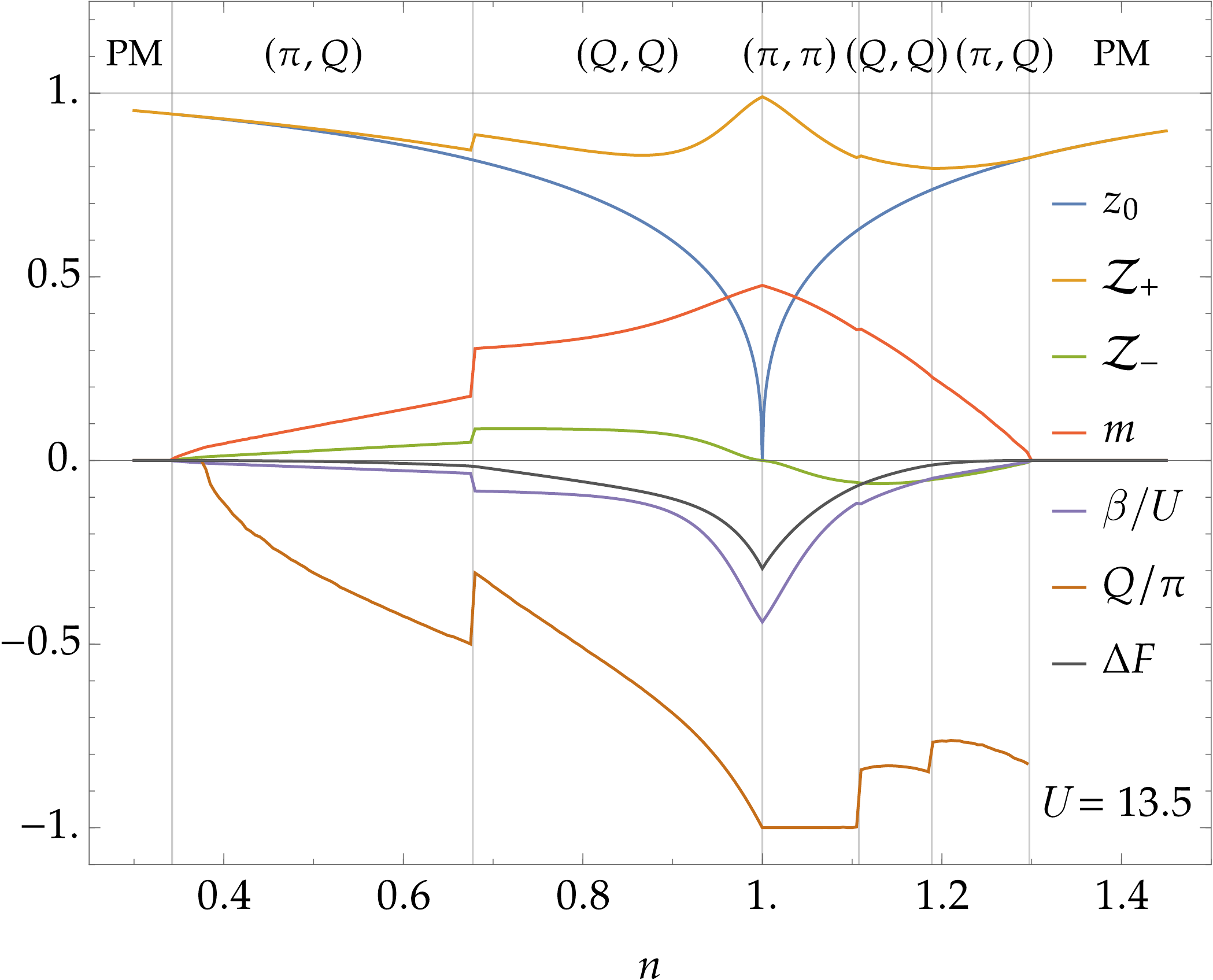}
\caption{Magnetic mean field parameters $z_{0}$, $\mathcal{Z}_{\pm}=(z_{+} \pm
z_{-})/2 $, magnetization $m=p p_{0}$, $\ \protect\beta/U$, ordering vector component $Q$ and the relative difference between the corresponding magnetic and
non-magnetic free energy $\Delta F$, versus filling $n$ at $U=13.5$
and at $t^{\prime }=-0.2$. The vertical grid lines indicate a phase transition and the respective
phases are denoted in the upper part of the plot.\label{Fig:zfactorsU13.5}}
\end{figure}

\begin{figure}[t]
\includegraphics[width=0.48\textwidth]{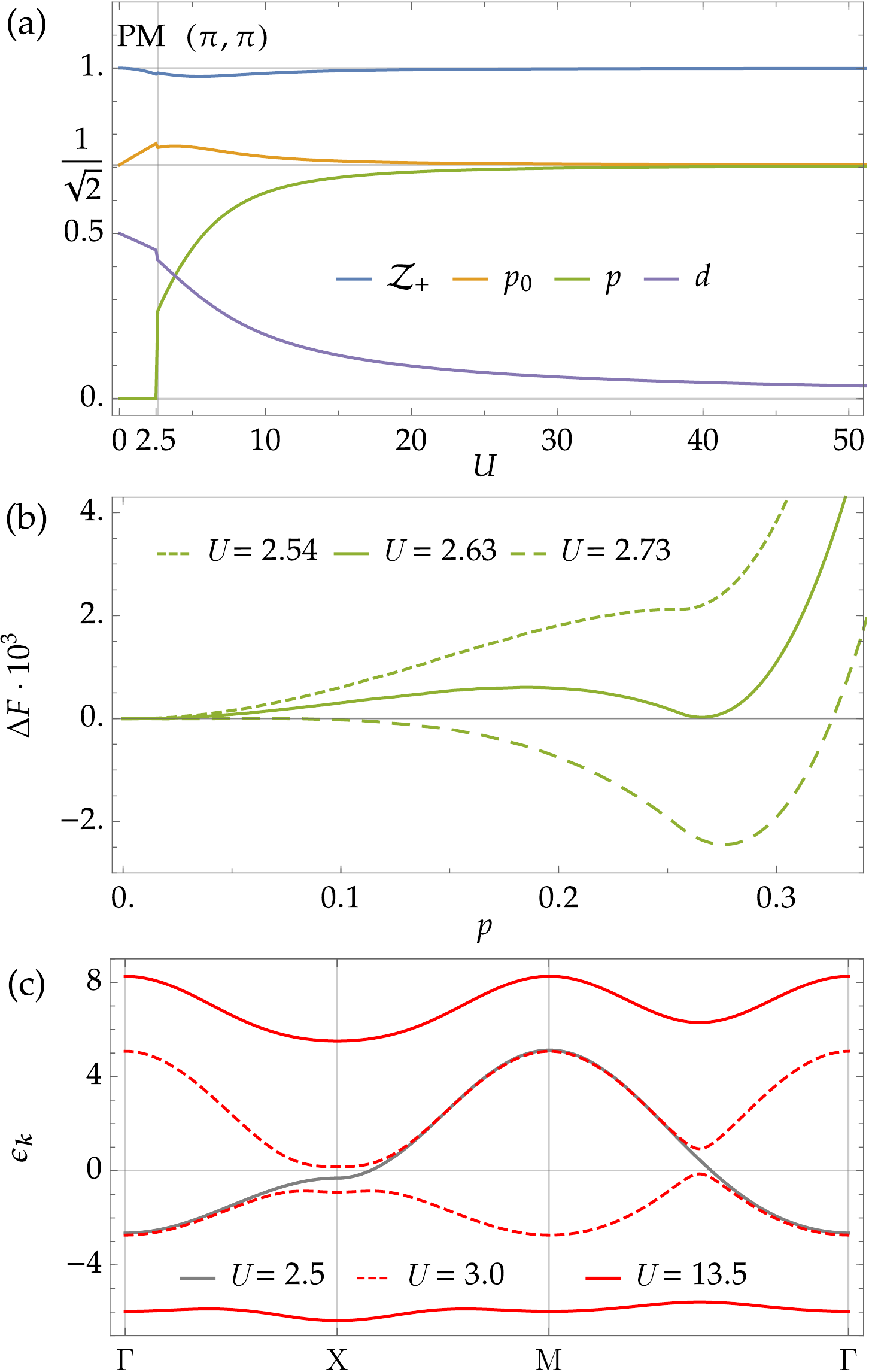}
\caption{(a) Magnetic mean field parameters $\mathcal{Z}_{+},p_{0},p,d$ at half filling $n=1$ versus $U$
at $t^{\prime }=-0.2$. The vertical grid lines indicate phase transitions and the respective
phases are denoted in the upper part of the plot.
(b) Free energy as a function of order parameter $p$ at $n=1$ for
various $U$ at $t^{\prime }=-0.2$. The coexistence of two different minima of equal energy at $U=2.63$ indicates a phase transition of first order.
(c) Band structure at $n=1$ for various $U$ at $t^{\prime }=-0.2$ on the high symmetry path $\Gamma$--X--M--$\Gamma$ of the non-magnetic Brillouin zone. We show only one spin degenerate band for the paramagnetic case (gray), whereas the magnetic spectrum (red) splits due to translation- and spin rotation symmetry breaking.}
\label{Fig:MF_n1}
\end{figure}

\autoref{Fig:zfactorsU13.5} shows the quasiparticle renormalization
factors $z_{0}$ for the paramagnet and $\mathcal{Z}_{\pm }=(z_{+}\pm z_{-})/2$ for magnetic state as a
function of $n$ for fixed $U=13.5$ for $t^{\prime }=-0.2$. Also shown are
the quantities $m=pp_{0}, \ \beta /U,$ 
and the condensation energy $\Delta F=F_{\text{mag}}-F_{\text{para}}$ of the magnetic
phases.\ The various ordered phases are indicated and their respective
incommensurate wave vector components $Q$ are also shown as functions of density $%
n $.\ The interaction is chosen to be greater than the critical value $U_{c}$
separating the metallic and the Mott insulating phase in a hypothetical
paramagnetic phase at half-filling (we determine the critical value of $U$
as $U_{c}=128/\pi ^{2}\approx 12.97$ for $t^{\prime }=0$ and $U_{c}\approx
13.1$ for $t^{\prime }=-0.2$). Therefore $z_{0}$ is found to vanish for $%
n\rightarrow 1$, $z_0^2\propto |n-1|$, causing the effective mass to
diverge, $m^{\ast }\propto 1/|n-1|$. By contrast, in the magnetically
ordered phase $z_{\pm }$ stays finite, but $z_{+}-z_{-}\rightarrow 0$, as $%
n\rightarrow 1$. Consequently, the two dispersions take the
limiting values $\epsilon _{\bk,\pm }=\frac{1}{4}\left(\zeta _{+}\xi _{\bk,+}\pm 
\sqrt{\zeta _{+}^{2}\xi _{\bk,-}^{2}+4\beta ^{2}}\right)-\mu _{\text{eff}}$
as $n\rightarrow 1$\ , indicating a band insulator with
excitation gap $2|\beta| $. However, as shown next, in the limit of large $U$
one finds $2|\beta| \rightarrow U$, which is the signature of a Mott insulator.

To supplement the discussion of the energy bands at half-filling, we show in %
\autoref{Fig:MF_n1}a the parameters $\mathcal{Z}_{+},p_{0},p,d$ as a function of $%
U $. In the limit of large $U$ these quantities approach the values $%
Z_{+}\rightarrow 1$, $p_{0}\rightarrow p\rightarrow 1/\sqrt{2}$, $%
d\rightarrow 0$. At small $U$ the behavior in the neighborhood of the
magnetic transition indicates a first order transition at $U\approx 2.63$.
This is clearly seen in the behavior of the free energy as a function of the
magnetic order parameter $p$ shown in \autoref{Fig:MF_n1}b. Analogously close to half-filling, 
the transition from the $(\pi ,\pi )-$state to the adjacent $(Q,Q)-$state is
also first order, since the ordering wave vector is found to jump from $%
Q=\pi $ to $Q=0.844\pi $ at $n=1.105$ (see \autoref{Fig:zfactorsU13.5}).

\subsubsection{Electronic band structure}

\begin{figure}[t]
\includegraphics[width=0.48\textwidth]{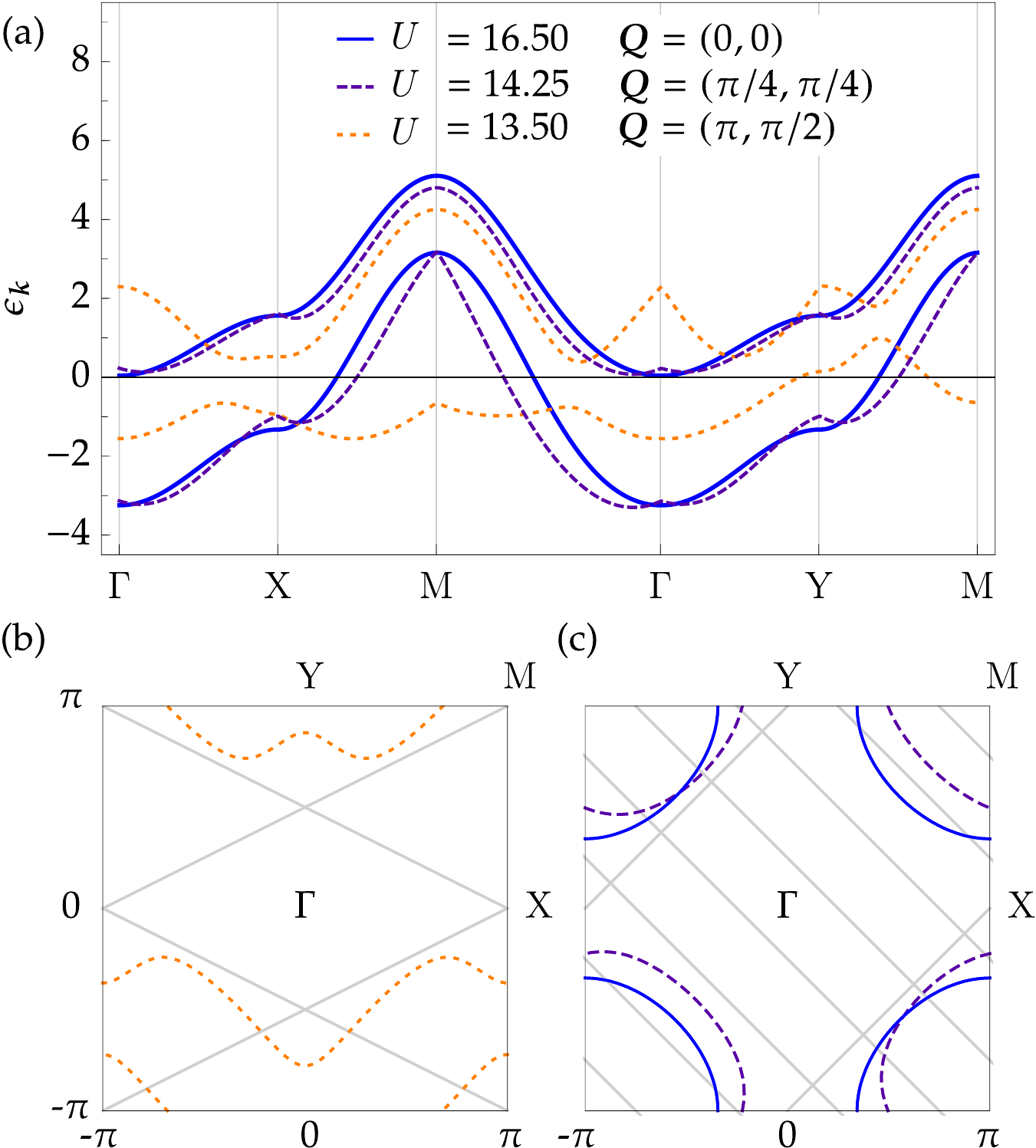}
\caption{(a) Band structure at filling $n=0.675$ for the magnetic ordering vectors $(\protect%
\pi/2,\protect\pi)$ for $U=13.50$, $(\protect\pi/4,\protect\pi/4)$ for $U=14.25$
and $(0,0)$ for $U=16.50$ with $t^{\prime }=-0.2$ on the high symmetry path $\Gamma$--X--M--$\Gamma$--Y--M [X $=(\pi,0)$, Y $=(0,\pi)$, M $=(\pi,\pi)$]
of the non magnetic Brillouin zone. Notice, that the ordering vectors $(\protect\pi/2,\protect\pi)$
and $(\protect\pi/4,\protect\pi/4)$ break the $C_{4v}$ symmetry of the non magnetic phase. 
(b) Fermi surface to the corresponding band structure shown in (a) for $U=13.50$. The gray lines indicate the backfolded Brillouin zone of the magnetic unit cell for the ordering vector $(\protect\pi,\protect\pi/2)$.
(c) Fermi surface to the corresponding band structures shown in (a) for $U=14.25$ and $U=16.50$. The gray lines indicate the backfolded Brillouin zone of the magnetic unit cell for the ordering vector $(\protect\pi/4,\protect\pi/4)$.}
\label{Fig:BS_FS}
\end{figure}

The dependence of the electronic dispersion on interaction strength and
filling is demonstrated in \autoref{Fig:MF_n1}c and \autoref{Fig:BS_FS}%
. In \autoref{Fig:MF_n1}c we consider the case of half-filling, taking $%
t^{\prime }=-0.2$. For $U=2.25$, $3.00$, and $13.5$ one observes the splitting of the
bands by the onset of magnetic order and the smooth transition of the spectrum as $U$ moves
beyond $U=U_{c}$.

To demonstrate the character of the electronic bands in various magnetically
ordered phases in \autoref{Fig:BS_FS}a, we fix the density at $n=0.675$, take $t^{\prime }=-0.2$ and plot $\epsilon _{k,\pm }$ along $\Gamma$--X--M--$\Gamma$--Y--M . We choose the interaction such that three different
orderings are realized: at $U=13.50$ we have $\bs Q=(\pi /2,\pi )$, at $U=14.25$
we have $\bs Q=(\pi /4,\pi /4)$ and at $U=16.50$ the ferromagnetic phase is
reached, with $\bs Q=(0,0)$.
The corresponding Fermi surfaces are shown in \autoref{Fig:BS_FS}b ($U=13.5$) and 
\autoref{Fig:BS_FS}c ($U=14.25$ and $U=16$).

\subsubsection{Compressibility}

\begin{figure}[t]
\includegraphics[width=0.48\textwidth]{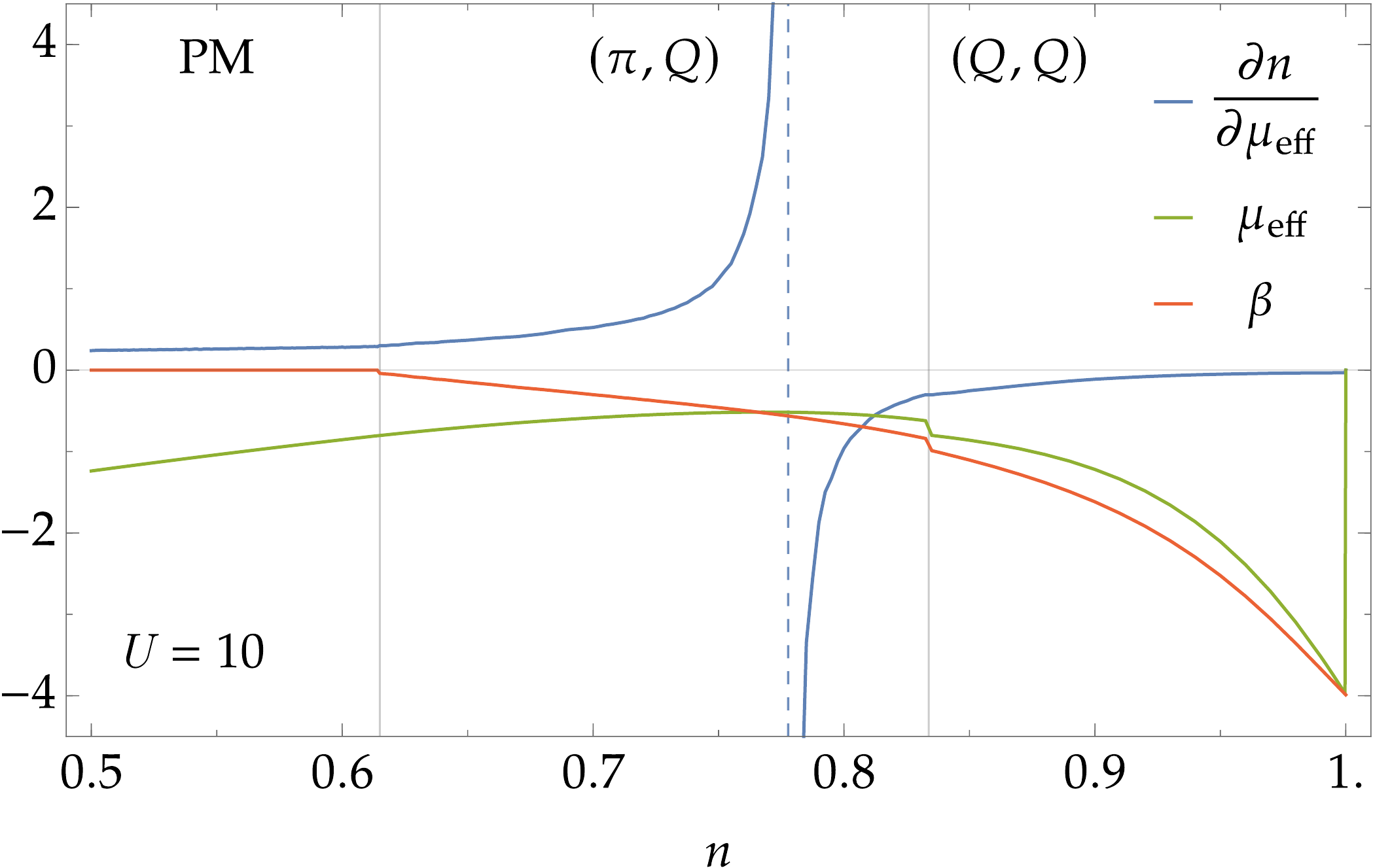}
\caption{$\partial n/\partial \mu _{\text{eff}}$, chemical potential $\mu _{\text{eff}}$ and $\protect\beta$ versus filling $n$
at $U=10$ and $t^{\prime }=0$. The compressibility $\kappa _{T}\propto \partial n/\partial \mu _{\text{eff}}$ diverges at the maximum of $\mu _{\text{eff}}$. The vertical gray grid lines indicate phase transitions where the respective
phases are denoted in the upper part of the plot and the vertical blue grid line indicates the divergence of the compressibility.}
\label{Fig:Compressibility}
\end{figure}

The mean field results allow the calculation of the isothermal
compressibility $\kappa _{T}$, or equivalently $\partial n/\partial \mu
_{\text{eff}}=n^2\kappa _{T}$, as obtained from
\begin{equation}
\frac{\partial n}{\partial \mu _{\text{eff}}}=\frac{\partial
(2d^{2}+p_{0}^{2}+p^{2})}{\partial \mu _{\text{eff}}}.
\end{equation}%
In \autoref{Fig:Compressibility}, $\partial n/\partial \mu _{\text{eff}}$ is plotted
versus $n$ for the nearest neighbor hopping model ($t^{\prime }=0$) and for $%
U=10.$ Interestingly, the compressibility changes sign in the magnetically
ordered phase \cite{fresardw1992,igoshev2013} where $\mu _{\text{eff}}$ has a maximum. With increasing density $n$ towards half filling, $|\beta| $ becomes larger and decreases the energy of the occupied band (compare \autoref{Fig:zfactorsU13.5}). This has to be counteracted by also reducing $\mu
_{\text{eff}}$ to ensure the correct electron filling, causing the compressibility to turn negative. We indicated the
portion of the phase diagram where negative compressibility occurs by adding
a dot into the colored circles marking the ordering wave vector in \autoref%
{Fig:Phasediagram_t2=0} and \autoref{Fig:Phasediagram_t2=-0.2}. A negative
compressibility signals a transition to a spatially modulated density
distribution or phase separation. The simultaneous presence of two ordering fields, one
magnetic, the other nonmagnetic, at generally different ordering vectors,
requires a numerical effort beyond the scope of the present work.

\subsection{Fluctuations around the paramagnetic mean field}

We have calculated the spin and charge susceptibilities in the paramagnetic
phase from the fluctuations of the slave-boson fields around the saddle point
as described in \app{Chapter:App:Fluc} and \app{Chapter:App:Correlation} to provide a general stability analysis. The divergence of the
static spin (charge) susceptibility at some wave vector indicates the
appearance of magnetic (charge) order with a spatial period given by this
wave vector. 
This will be used to determine the magnetic phase boundary of the paramagnet,
which turns out to be a numerically more efficient way to identify the appearance of
magnetic order, compared to the magnetic mean field analysis presented in the
previous subsection. It is reassuring that both methods provide consistent
results.

Notice that first order phase transitions cannot be identified via Gaussian fluctuations around a paramagnetic saddle point. This is because a local minimum of the paramagnetic free energy ($p=0$) like, e.g., shown in \autoref{Fig:MF_n1} is metastable and the global minimum is out of reach of the quadratic expansion of the action.

\begin{figure}[t!]
\includegraphics[width=0.48\textwidth]{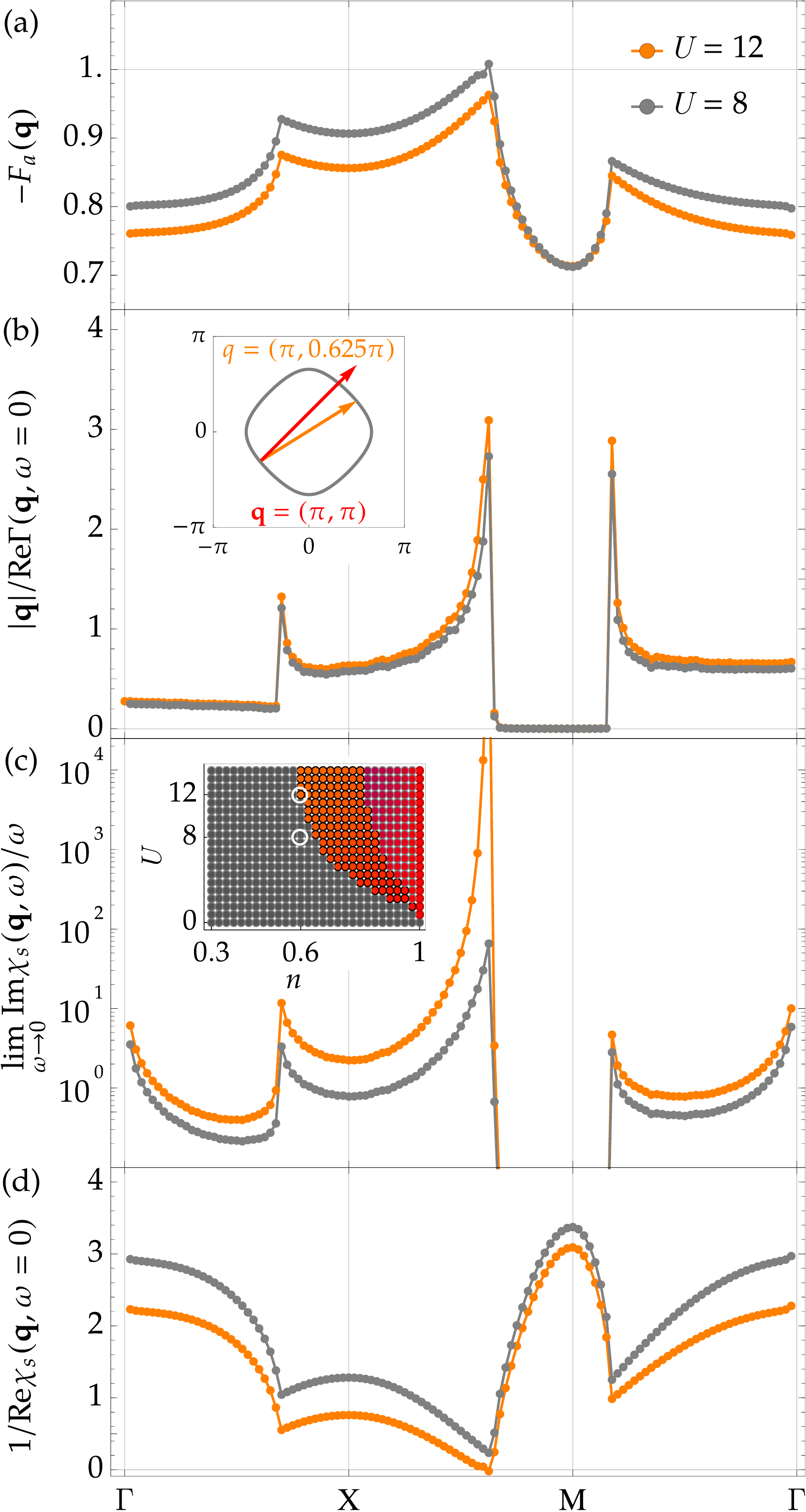}
\caption{Spin susceptibility and Landau factors at filling $n=0.6$, $t^{\prime }=0$, $T=0$ and
$\protect\eta =0.001$ for $U=8$ (PM) and $U=12$ [phase boundary from PM to $(\pi,Q)$-magnetic order] shown on the high symmetry path $\Gamma$--X--M--$\Gamma$.
We show an excerpt of the phase diagram (compare \autoref{Fig:Phasediagram_t2=0}) in the inset of (c), where the two chosen interaction values are highlighted. 
(a) Landau interaction function $-F_{a}(\bs q)$. The phase transition is indicated by $F_{a}(\bs Q)=-1$.
(b) Landau damping function plotted as $|\bs q|/\Gamma (\bs q,\protect\omega=0)$.
Its signal is overall drastically reduced for momenta larger than the diameter of the Fermi surface [compare inset (b)]. 
(c) Imaginary part of the spin-susceptibility and in the zero frequency limit plotted as
$\lim_{\omega \to 0}\Im \protect\chi _{s}(\bs q,\protect\omega)/\omega$.
Its signal is overall drastically reduced for momenta larger than the diameter of the Fermi surface [compare inset (b)]. 
(d) Inverse real part of the spin susceptibility. A root of $1/\Re \chi_s(\bq,\omega=0)$ indicates a magnetic instability.}
\label{Fig:SpinFluctuations}
\end{figure}

\subsubsection{Spin susceptibility}

\paragraph{Phenomenological form of susceptibility}

The functional behavior of the dynamical spin susceptibility at low
frequencies can be represented in terms of auxiliary functions $F_{a}(\bs q)$
and $\Gamma (\bs q,\omega )$. In the static limit we define 
\end{subequations}
\begin{subequations}
\begin{equation}
\chi _{s}(\bs q,\omega =0)=\frac{\chi _{0}(\bs q,\omega =0)}{1+F_{a}(\bs q)}
\label{eq:dynamic_chi_s}
\end{equation}%
where $F_{a}(\bs q)$ is a generalization of the well-known Landau parameter
to finite wave vectors and $\chi _{0}(\bs q,\omega =0)$ is the
 ``non-interacting'' quasiparticle susceptibility (see \app
{Chapter:App:Correlation}), which carries a hidden influence of the
interaction through its dependence on the mean field parameters. At small
but finite frequency the leading dynamical addition is given by the Landau
damping term in the denominator, parametrized by a function $\Gamma (\bs %
q,\omega )$ 
\begin{equation}
\chi _{s}(\bs q,\omega )=\frac{\chi _{0}(\bs q,\omega =0)}{1+F_{a}(\bs %
q)-i\omega /\Gamma (\bs q,\omega )}.
\end{equation}%
These quantities are shown in \autoref{Fig:SpinFluctuations}a and \autoref{Fig:SpinFluctuations}b at $n=0.6$, $t^{\prime
}=0$ for interactions $U=8$ and $U=12$ on the high symmetry path $\Gamma $--X--M--$\Gamma $ in the Brillouin zone. The corresponding susceptibility
will be discussed below. The imaginary part $\Im \Gamma = \mathcal{O}(\omega )$ is
negligible at small $\omega $. Around the M point, the wavevector $\bq\approx
(\pi ,\pi )$ is larger than the diameter of the Fermi surface, as shown in
the inset of \autoref{Fig:SpinFluctuations}b and\ therefore the imaginary part of $%
\chi _{s}$ and consequently of $1/\Gamma $\ is zero (compare \autoref%
{Fig:SpinFluctuations}b). Rather than plotting $\Gamma$, we therefore show 
$|\bq|/\Gamma (\bs q,\omega =0)$. The limiting behavior of $|\bq|/\Gamma (\bs %
q,\omega =0)\rightarrow const.$ as $\bq\rightarrow 0$ is demonstrated. The
upper panel shows $F_{a}(\bs q)$ along $\Gamma$--X--M--$\Gamma $. The
curve for $U=12$ is seen to reach $F_{a}(\bs q)=-1$, signaling a phase transition into a magnetically ordered
state characterized by the wave vector $\bs Q$, which is discussed below.

\paragraph{Magnetic instability}
In \autoref{Fig:SpinFluctuations} the imaginary (c) and real (d) part of the spin
susceptibility at $n=0.6$ and $t^{\prime }=0$ are shown for two different
interactions. For $U=8$ we find a stable paramagnet for any wavevector and
for $U=12$ a magnetic instability appears at the incommensurate ordering
vector $\bs Q\approx(0.625\pi ,\pi )$.

A magnetic phase transition is indicated if $\chi _{s}(\bs q,\omega =0)$
diverges at some ordering vector $\bs q=\bs Q$. It is numerically more
viable to investigate $1/\Re \chi _{s}(\bs q,\omega =0)$, a sign change of $%
1/\Re \chi _{s}$ indicates a divergence of $\Re \chi _{s}$. This represents
the most precise criterion to define a magnetic instability. The imaginary
part can be evaluated numerically only at finite $\eta $ and $\omega $,
since $\Im \chi _{s}(\bs q,\omega +i\eta )\propto \omega $ in the limit $%
\eta \rightarrow 0$. In \autoref{Fig:SpinFluctuations}c we show $%
\lim_{\omega \rightarrow 0}[\Im \chi _{s}(\bs q,\omega +i\eta )/\omega ]$ as
a function of $\bq$ exhibiting a diverging peak at $\bq=\bs Q$, as $U$ increases
towards the critical value of $U_{c}\approx 12$. The growth of peaks at other ordering vectors can be explained by the enhancement of the density of states at the Fermi level and do not indicate a magnetic instability.

To determine the
paramagnetic phase boundary from the divergence of the static spin
susceptibility, we steadily increase the interaction $U$ and look for the
first appearance of a zero of $1/\chi _{s}(\bs Q,\omega =0)$ as shown in %
\autoref{Fig:SpinFluctuations}d by example. 
Following this procedure, the phase boundaries to the paramagnet obtained by the magnetic
mean field analysis shown in \autoref{Fig:Phasediagram_t2=0} and \autoref{Fig:Phasediagram_t2=-0.2} are reproduced
consistently.

Identifying the onset of magnetic instabilities from a study of the spin
susceptibility as compared to solving the saddle point equations of the spiral magnetic mean field ansatz is more general. In contrast to
the latter, which is restricted to the assumed form of the
order (spin spiral), the divergence of the susceptibility signals the
emergence of magnetic order of any kind with spatial periodicity described
by the wavevector $\bs Q$. The fluctuation approach is, however, not suited
to determine the type of magnetic order beyond the boundary of the
paramagnetic regime.

\paragraph{Critical exponent}

We determine the critical exponent $\alpha $ at magnetic instabilities of
the paramagnet where the spin susceptibility diverges as 
\end{subequations}
\begin{equation}
\chi _{s}(\bs Q,0)\propto (n_{c}-n)^{-\alpha },\text{ \ \ }n<n_{c},
\end{equation}%
when $n\rightarrow n_{c}$. We find a critical exponent of $\alpha =1$ for
phase transitions towards the commensurate antiferromagnet $\bs Q=(\pi ,\pi )$ which occupies an extended domain in the phase diagram for $t^{\prime
}\neq 0$ as shown in \autoref{Fig:Phasediagram_t2=-0.2}. For incommensurate
magnetic instabilities, the critical exponent is found as $\alpha \approx 1/2$,
as demonstrated in \autoref{Fig:ncrit_fit} for $t^{\prime }=0$ and $U=12$,
which shows the inverse spin susceptibility at $\bs Q\approx (0.625\pi ,\pi
) $ as function of the filling $n$.

\subsubsection{Charge susceptibility}
\label{sec:Chargesusceptibility}

We also considered the possibility of charge order in the Hubbard model as indicated
by a divergence of $\chi _{c}(\bs Q,0)$ . In the paramagnetic regime, we
did not find any charge instabilities for $U\geq 0$, which confirms the
magnetic phase diagrams shown in \autoref{Fig:Phasediagram_t2=0} and \autoref%
{Fig:Phasediagram_t2=-0.2}. However, we cannot exclude a combination of spin
and charge order in the magnetically ordered regime, because the
investigation would require fluctuations around a magnetic saddle point,
which is outside the scope of our present work. Such an analysis would
certainly be of interest, especially in the regime of negative
compressibility.\newline
\begin{figure}[t]
	\includegraphics[width=0.48\textwidth]{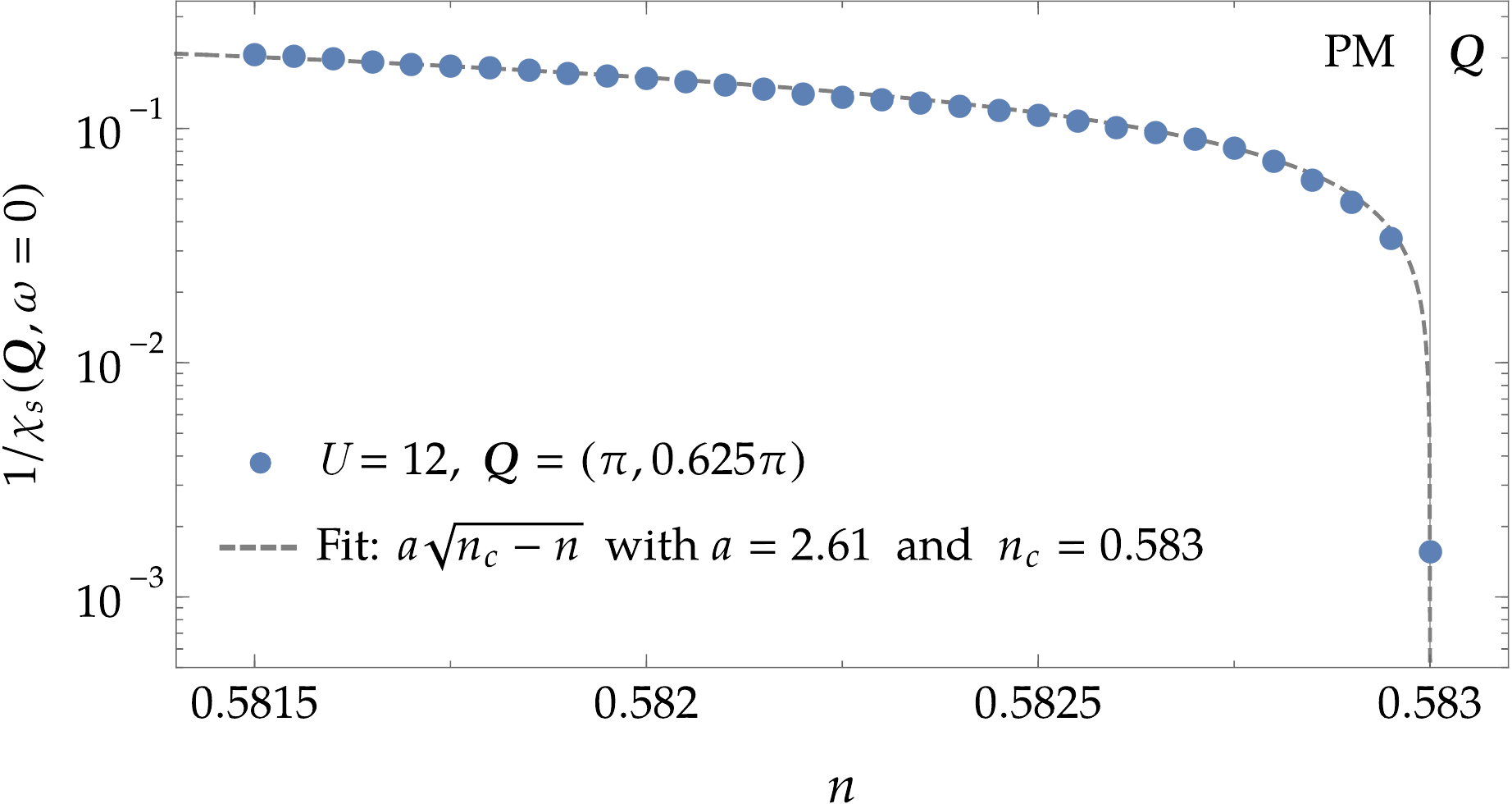}
	\caption{Inverse static spin susceptibility (blue plot markers) at wave vector $(\protect\pi,0.625 \protect\pi)$ versus $n$ at $U=12$
		in the paramagnetic regime.
		The gray dashed line shows a fit of $1/\chi_s(\bs Q, \omega=0)$ to the square root $\sqrt{n_c-n}$,
		indicating a critical exponent of $\alpha=1/2.$
		The vertical line marks the magnetic phase transition at the critical doping $n_c$.}
	\label{Fig:ncrit_fit}
\end{figure}

\begin{figure}[t]
	\includegraphics[width=0.48\textwidth]{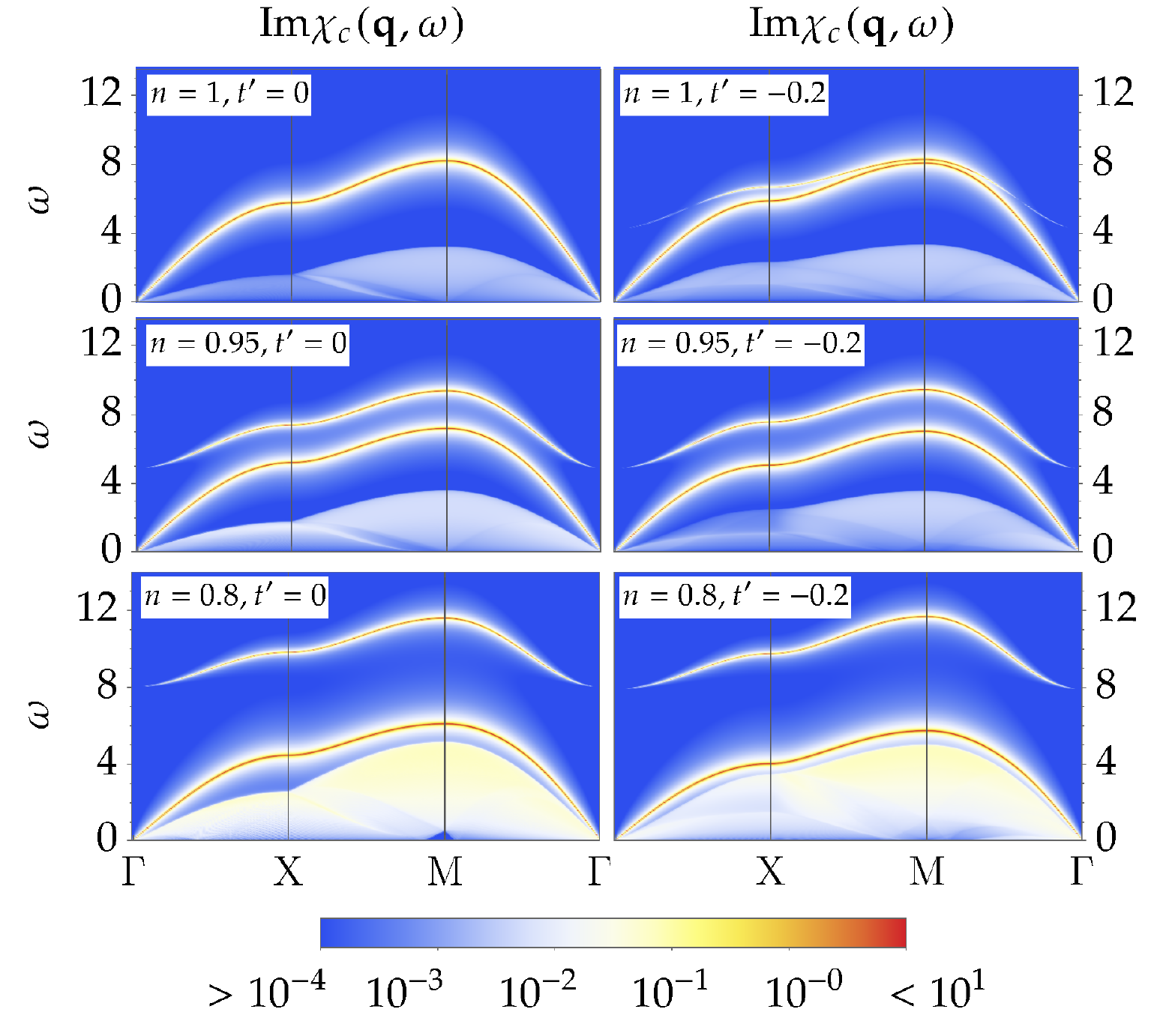}
	\caption{(a) Imaginary part (color coded) of the bare susceptibility $\protect%
		\chi _{0}$ (left column) and charge susceptibility $\protect\chi _{c}$
		(right column) in the frequency $\protect\omega-$wavevector $\bq$ plane
		at $n=0.6$, and $t^{\prime }=0$ for various interaction strengths. For $U>0$, the $\chi_s$ features a collective mode and the upper Hubbard band in addition to the particle hole excitation spectrum.
		(b) Imaginary part of the charge susceptibility $\protect%
		\chi _{c}$ versus frequency $\protect\omega$ at wavevector $\bq=(\pi,0)$ at $%
		n=0.6$, $U=10$ and  $U=0$ and at $t^{\prime }=0$.}
	\label{Fig:ChargeSusceptibility}
\end{figure}
\begin{figure}[t]
	\includegraphics[width=0.48\textwidth]{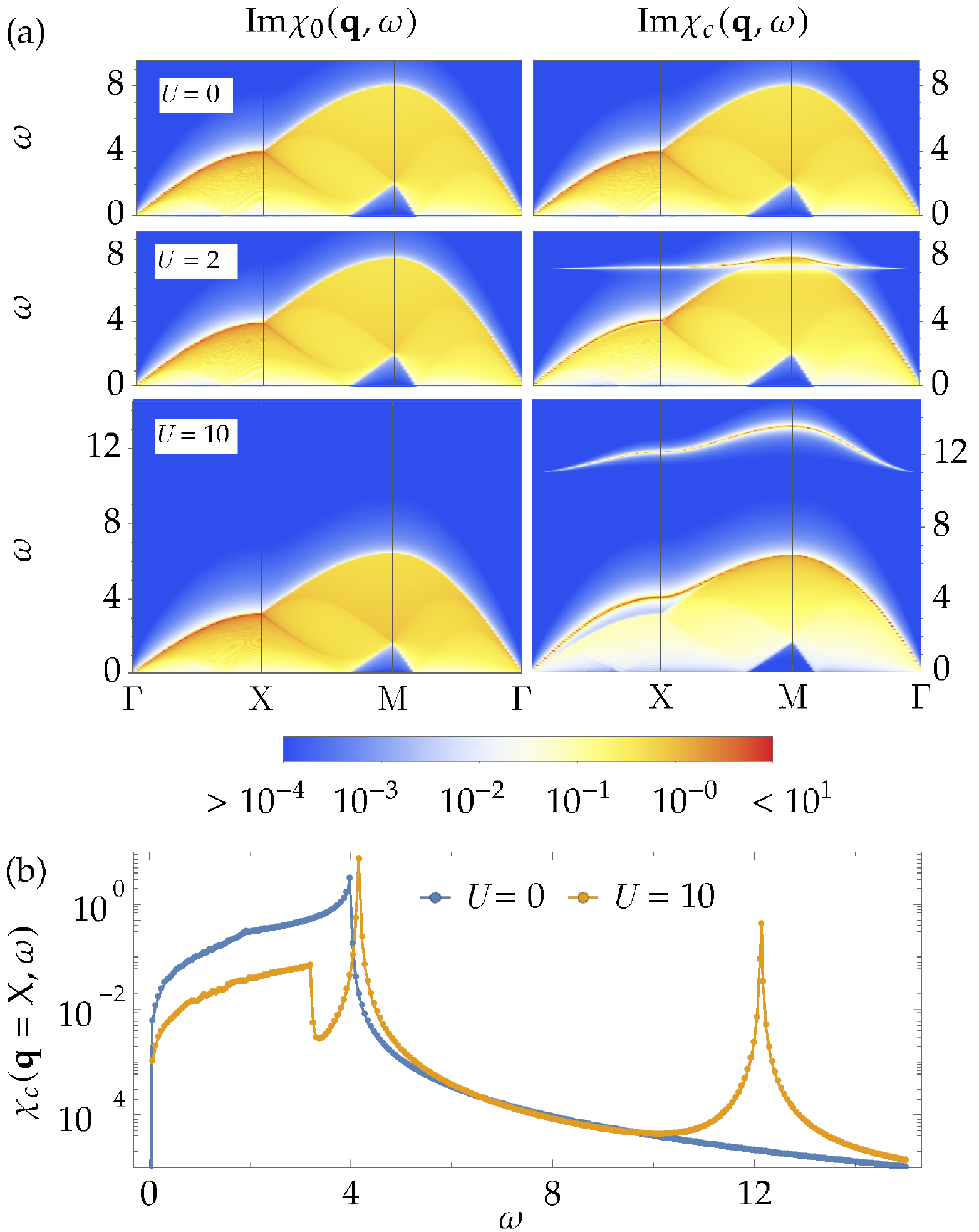}
	\caption{Imaginary part (color coded) of the charge susceptibility $\protect%
		\chi _{c}$ in the frequency $\protect\omega-$wavevector $\bq$ plane for
		$U=10$ and various dopings $n$ at $t^{\prime }=0$ (left column) and $t^{\prime }=-0.2$ (right
		column).
		With increasing $n$ towards half filling, the spectral weight of the particle hole continuum is shifted towards the collective mode. The upper Hubbard band vanishes at half filling for $t'=0$ and is kept for the long range hopping model $t'=-0.2$.}
	\label{Fig:Xc_U10_t2}
\end{figure}

In \autoref{Fig:ChargeSusceptibility}a we present in the left column the
``bare'' susceptibility $\chi _{0}$ (corresponding to the bubble diagram in
mean field approximation, and therefore dependent on interaction) as
function of momentum $\bq$ and frequency $\omega $ and compare it with the
the charge susceptibility $\chi _{c}$ (right column of \autoref%
{Fig:ChargeSusceptibility}a). The chosen set of parameters ($n=0.6,t^{\prime
}=0,U=0,2,10$) lies within the paramagnetic regime of the phase diagram %
\autoref{Fig:Phasediagram_t2=0}. The bare susceptibility is determined by
the paramagnetic mean field band structure, given by the spin degenerate
eigenvalue $\epsilon _{\bk}=-2z_{0}^{2}\left[ \cos (k_{1})+\cos (k_{2})\right]
-\mu _{\text{eff}}$ (compare \app{Chapter:App:MF}). The slave-boson
renormalization $z_{0}(U)$ depends on the interaction and is normalized to $%
z_{0}(0)=1$, the resulting bandwidth is given by $W(U)=8z_{0}^{2}(U)$.
Hence, for vanishing interaction, we have $W=8$ and accordingly the width of
the excitation spectrum is equal to the bandwidth, as can be seen in the top
panel of \autoref{Fig:ChargeSusceptibility}a, moreover it is $\chi
_{c}=\chi _{0}$ for $U=0$. Increasing the interaction has two effects.
First, the excitation widths of $\chi _{0}$ are reduced, matching the
renormalized bandwidths $W(2)\approx 7.8$ and $W(10)\approx 6.5$. Second, $%
\chi _{c}$ exhibits the emergence of two excitation gaps, splitting the
charge susceptibility into three regimes \cite{PhysRevB.95.165127}. There is a particle-hole
excitation continuum for $\omega <W$, where $\chi _{c}$ resembles $\chi _{0}$
and also scales with the bandwidth. The second regime, which may be
identified with the upper Hubbard band, features a sharp energy momentum
relation and is separated from the first regime by a gap, which approaches $%
U $ in the limit of large interactions (upper excitation band). This is due
to the fact that $\chi _{c}$ as a fluctuation quantity goes beyond the band
structure picture of the mean field and allows excitations which result in
the creation of new doubly occupied sites at the cost of the interaction $U$%
. The feature we identify with excitations into the the upper Hubbard band
is seen to vanish for $\bq\rightarrow 0$ .\ Third, a collective mode feature
situated between the continuum and the upper Hubbard band emerges, which may
be identified as a collective density mode as appears in a Fermi liquid for
sufficiently large repulsive interaction. At half-filling only one
collective mode is visible. This is different for the longer range hopping
model ($t^{\prime }=-0.2$) for which both excitation features are present
even at half-filling as shown in \autoref{Fig:Xc_U10_t2}. The structure of
the charge excitation spectrum as a function of frequency at $\bq=(\pi ,0)$
(X point) is shown in more detail at doping $n=0.6$ , $t^{\prime }=0$ and
for $U=10$ and $U=0$ in \autoref{Fig:ChargeSusceptibility}b. The comparison
shows how the interaction (i) shifts spectral weight from the lower to the
upper Hubbard band and (ii) generates a collective mode at the upper edge of
the lower Hubbard band. The reason for the appearance of two excitation
bands lies in the different dynamics of the fermionic and bosonic degrees of
freedom. As shown in \app{Chapter:App:Correlation} the charge
susceptibility is determined by inverse matrix elements of the $4\times 4$ charge block of the fluctuation matrix $%
\mathcal{M}_{\mu \nu }(\bs q,\omega)=\mathcal{M}_{\mu \nu }^{B}(\bs q,\omega)+\mathcal{M}_{\mu \nu
}^{F}(\bs q,\omega)$, $\mu,\nu \in (e,d_1,d_2,\beta_0)$. In opposite to the spin sector, $\mathcal{M}_{\mu \nu }^{B}(\bq,\omega)$ explicitly depends on the frequency because the slave-boson field $d=d_1+id_2$ is complex valued. Our results are in full agreement with the detailed analysis of collective charge modes in the Hubbard model presented in \cite{PhysRevB.95.165127}.

\subsection{Dynamical conductivity}

\begin{figure}[tbp]
\includegraphics[width=0.48\textwidth]{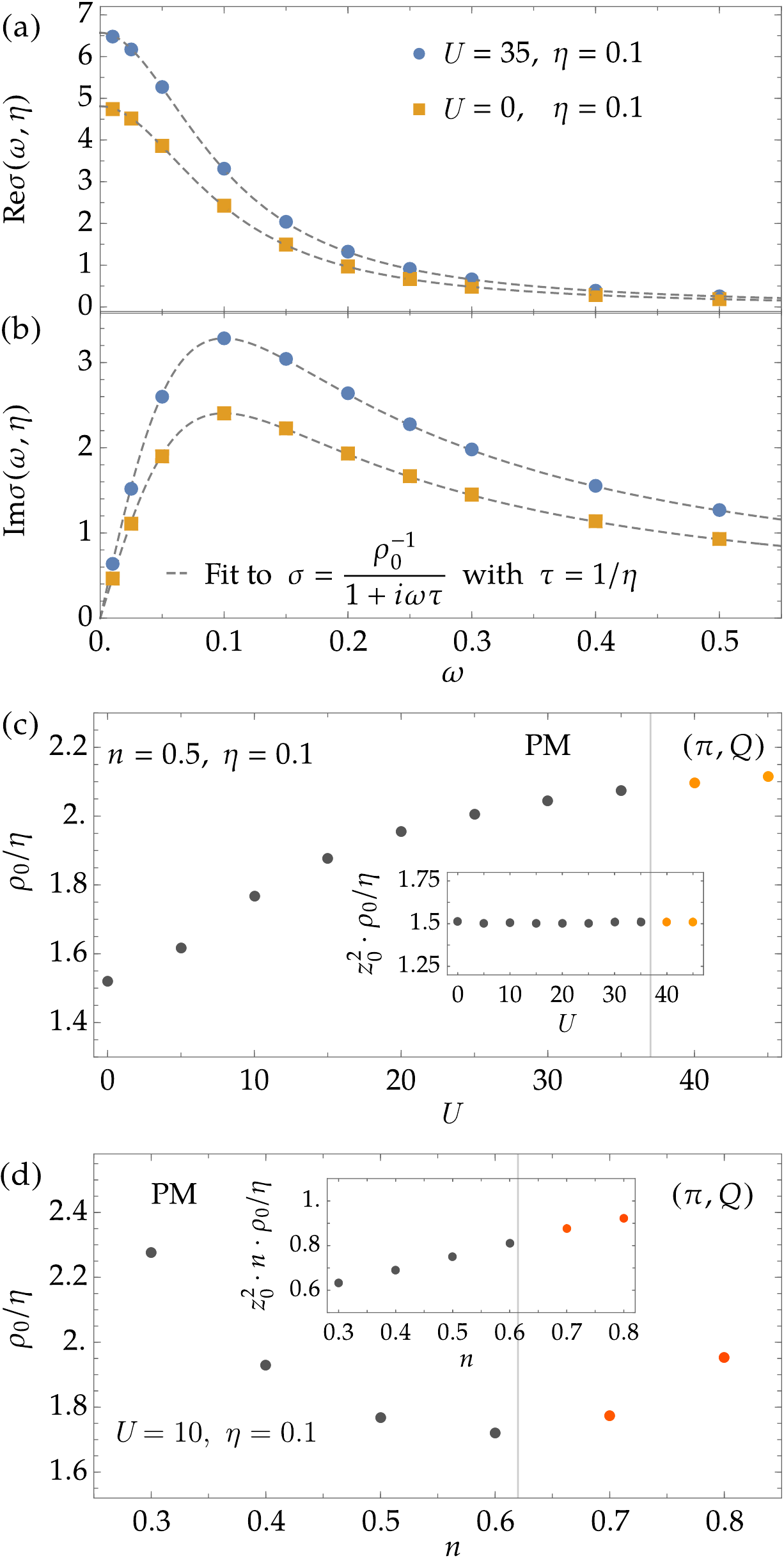}
\caption{(a,b) Real and imaginary part of the dynamical conductivity for $\protect%
\eta =0.1$, $T=0$, quarter filling $n=0.5$ for the two interactions $U=0$
and $U=35$ at $t^{\prime }=0$ . In both cases we see a Drude-like behavior
according to \eq{Model:Drude} as indicated by the dashed gray lines.
(c) DC-resistivity $\protect\rho_{0}$ versus interaction $U$ for
filling $n=0.5$. The inset shows the scaling of $\protect\rho_{0}$ with $%
1/z_{0}^2$. The vertical grid lines indicate phase transitions, the respective
phases are denoted in the upper part of the plot and the color of the plot makers display the value of the ordering vector
(compare \autoref{Fig:Phasediagram_t2=0}).
(d) DC-resistivity $\protect\rho_{0}$ versus doping $n$ for $U=10$. The
inset shows the approximate scaling of $\protect\rho_{0}$ with $1/(z_{0}^2
n) $. The vertical grid lines indicate phase transitions, the respective
phases are denoted in the upper part of the plot and the color of the plot makers display the value of the ordering vector 
(compare \autoref{Fig:Phasediagram_t2=0}).}
\label{Fig:Conductivity}
\end{figure}

We studied the dynamical conductivity $\sigma (\omega +i\eta)$ according to %
\autoref{section:conductivity}. In %
\autoref{Fig:Conductivity}, results for the real (a) and imaginary (b) part of the dynamical conductivity are
presented for the nearest neighbor hopping model ($t^{\prime }=0$), at
quarter filling $n=0.5$ and for two values of interaction, $U=0$ and $U=35$%
.\ The parameter $\eta =1/\tau =0.1$ is kept finite and is
identified with the inverse scattering time of the Drude model, which fits our data. For $\eta \rightarrow 0$, the
DC-conductivity goes to infinity, because our model does not include a momentum
dissipation mechanism (no umklapp scattering, no phonons). One may interpret 
$\eta $ as an effective scattering parameter accounting for impurity
scattering, while the limit $\eta \rightarrow 0$ corresponds to a perfect,
impurity-free crystal.

\autoref{Fig:Conductivity}c shows the DC-resistivity $\rho _{0}=1/\sigma _{0}$ as
function of the interaction $U$ at filling $n=0.5$. The inset demonstrates
that $z_{0}^{2}\rho _{0}$ is nearly independent of $U$, reflecting the
scaling of $\rho _{0}$ with the effective mass $\rho _{0}\propto m^{\ast }/m$
, which is given by $m^{\ast }/m=1/z_{0}^{2}$ in the one band Hubbard model.
The density dependence of $\rho _{0}$ at $t^{\prime }=0$ and $U=10$\ is
shown in \autoref{Fig:Conductivity}d. The inset shows the scaling of $\rho _{0}$
with density and effective mass according to Drude's formula, requiring $%
z_{0}^{2}n\rho _{0}$ to be nearly independent of density, which happens to
be satisfied only approximately.

\subsection{Spin and charge structure factors}

\begin{figure}[tbp]
\includegraphics[width=0.48\textwidth]{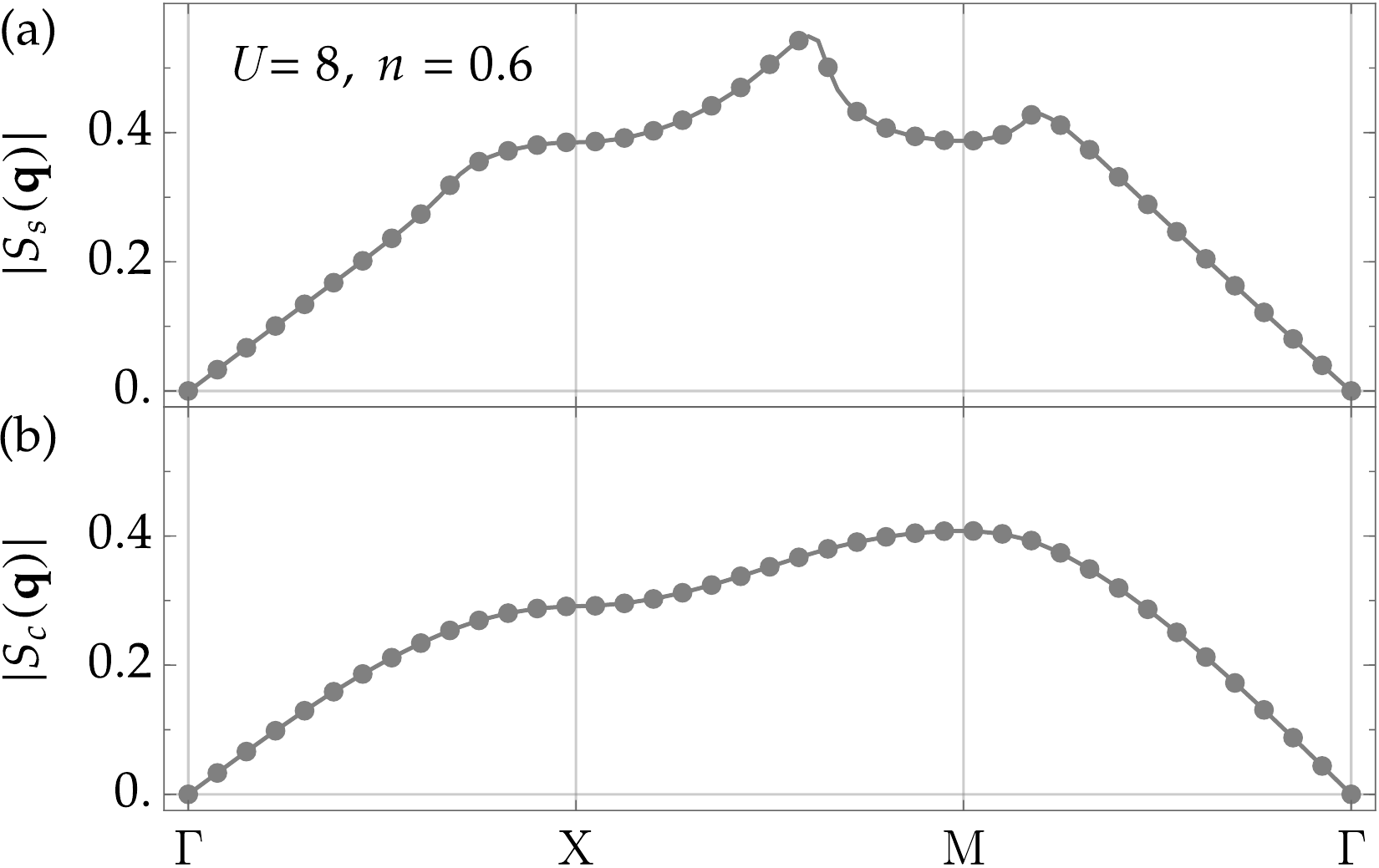}
\caption{Spin structure factor $S_{s}$ (a) and charge structure
factor $S_{c}$ (b) for $n=0.6$, $t^{\prime }=0$, $T=0$, broadening 
$\protect\eta =0.01$ for interaction $U=8$ on the high symmetry path $\Gamma$--X--M--$\Gamma$.}
\label{Fig:Structurefactor}
\end{figure}

The spin and charge structure factors at $T=0$ are obtained as 
\begin{equation}
S_{s,c}(\bs q)=-\int_{0}^{\infty }\frac{d\omega }{\pi }\Im \chi _{s,c}(\bs %
q,\omega +i\eta ).
\end{equation}%
The structure factors at $n=0.6$ , $t^{\prime }=0$, and $U=8$ are\ shown
along the path $\Gamma$--X--M--$\Gamma$ in the Brillouin zone in \autoref%
{Fig:Structurefactor}. 

Similar to $\Im \chi _{s}(\bs q,\omega \rightarrow 0)$ the spin
structure factor is enhanced at $\bq=(\pi ,0.625\pi )$,
reflecting the upcoming magnetic instability at larger $U$. Due to the
integration over $\omega $ the structure factors do not necessarily have to
resemble the corresponding susceptibilities in one distinct frequency range.

\subsection{Comparison with DMET-Results}\label{sec:Results:DMET}
Zheng et al.~computed the ground-state of the
Hubbard model on the square lattice in 2D \cite{PhysRevB.93.035126} by
employing DMET using clusters of up to 16 sites.
They report competition between inhomogeneous charge, spin, and pairing
states at low doping. In the following, we compare their results with our
results from slave-boson theory.

\subsubsection{Results at half filling}

\begin{figure}[tbp]
\includegraphics[width=0.48\textwidth]{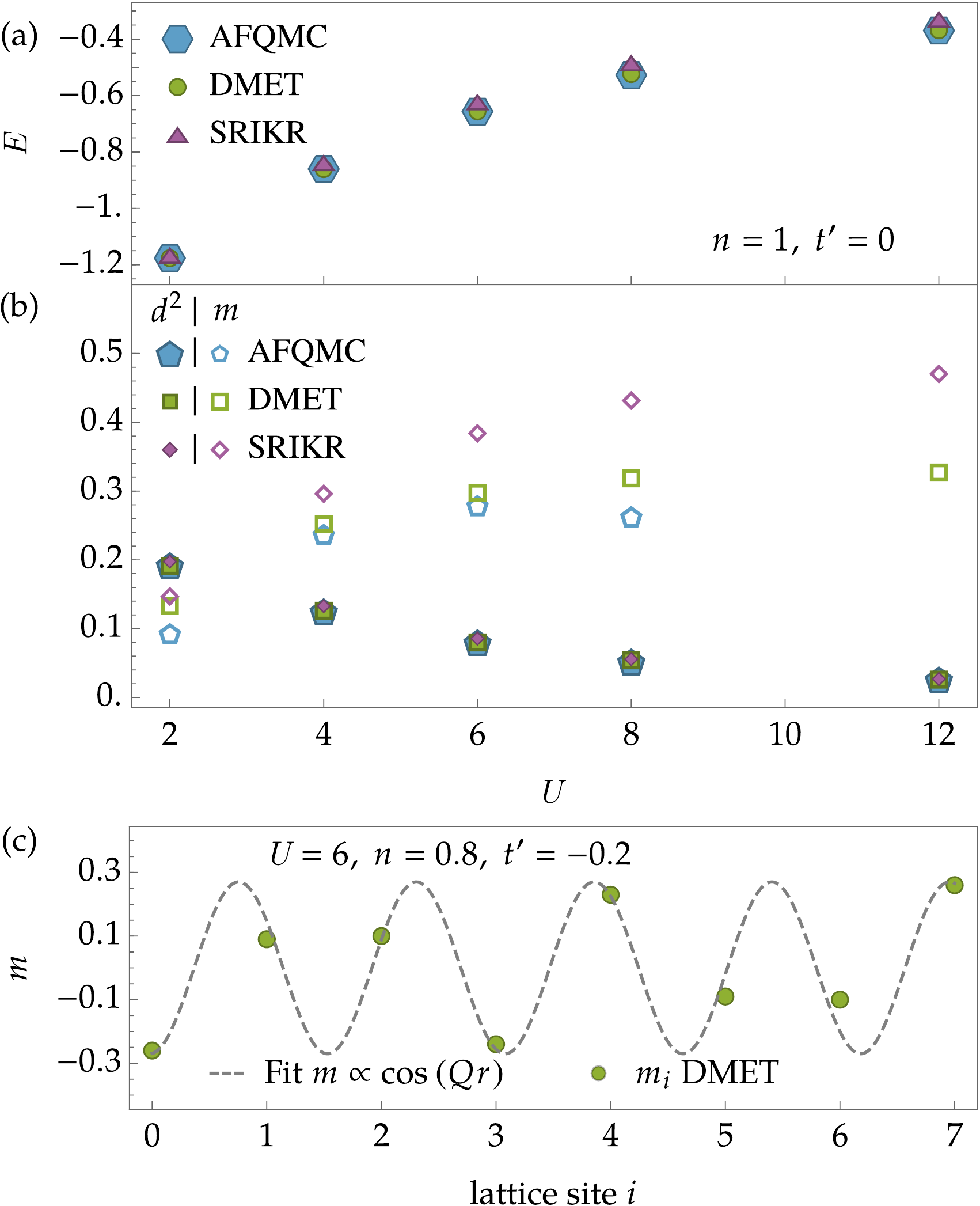}
\caption{Energy per site (a), double occupancy $d^2$ and magnetization (b) for half filling, $t'=0$ and different $U$. We compare the results obtained from AFQMC and DMET by Zheng et al.~\cite{PhysRevB.93.035126} with our SRIKR slave-boson results;
(c) Local magnetization as a function of the lattice site [assuming $(\pi,Q)$ order on a $2\times 8$ cluster] for $n=0.8$, $t'=-0.2$ and $U=6$ within DMET~\cite{PhysRevB.93.035126}. SRIKR yields $\bs Q\approx(\pi,0.71\pi)$, implying a spin profile of $m\propto \cos(Qr)$ which fits very well to the DMET data.}
\label{Fig:CompareDMET}
\end{figure}

\autoref{Fig:CompareDMET} (a) and (b) compare the energy per site, double occupancy and
staggered magnetization of the AFM obtained by Auxiliary-Field Quantum Monte Carlo (AFQMC), DMET and SRIKR for $t^{\prime }=0$ and different $U$ at half filling.
While we find very good agreement for the double occupancy, the magnetization deviates considerably for increasing interaction.
For $U\rightarrow \infty $ we find the
fully magnetized Neel state with $m_{\text{SRIKR}}=1/2$ within the SRIKR slave-boson analysis whereas the magnetization saturates at $m_{\text{DMET}}\approx 0.33$ within DMET,
close to the exact Heisenberg value in 2D which is given by $0.307$ according to Quantum Monte Carlo (QMC) 
calculations \cite{PhysRevB.56.11678}.
This overestimation of the magnetization coincides with an increased energy per site in SRIKR compared to the other methods for large $U$. 
We expect the magnetization
to be decreased by fluctuation corrections to the magnetic mean field, which
are however beyond the scope of the present work.

\subsubsection{Results for finite doping}
\begin{figure}[tbp]
	\includegraphics[width=0.48\textwidth]{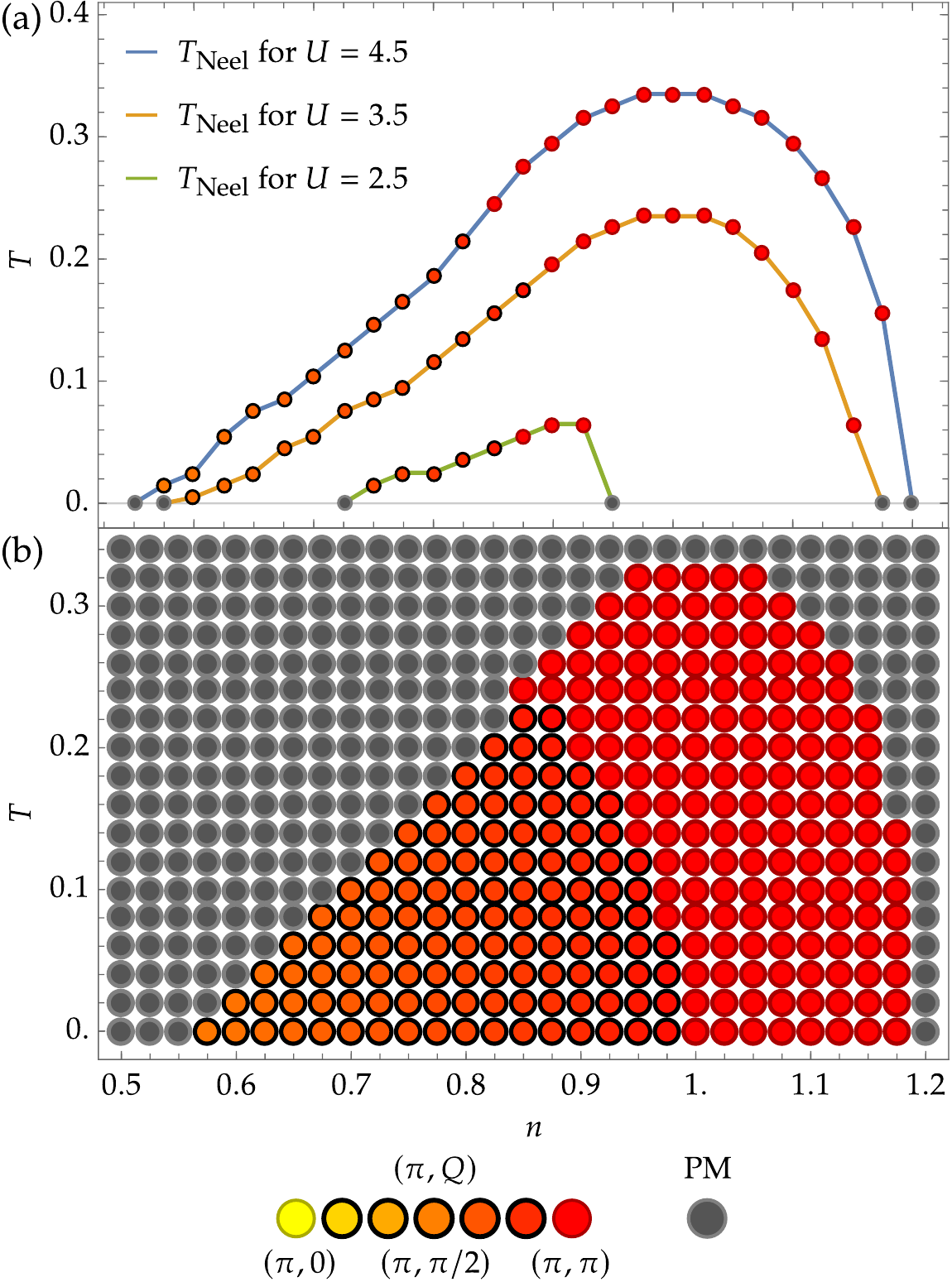}
	\caption{(a) Transition temperature into the magnetically ordered phase versus filling $%
		n $ for $U=2.5,3.5,4.5$ and $t^{\prime }=-0.2$.
		The ordering vectors on the line of the phase transition are indicated by the color of the plot markers.
		All presented transitions from the magnetic to the non magnetic phase are of second order; 
		(b) Temperature dependent magnetic phase diagram for the Hubbard model
		for $U=4.5$ and $t^{\prime }=-0.2$. It features two distinct phases, namely
		the PM (gray) and the ($\protect\pi ,Q$) phase denoted by black circles
		filled with coloring from red to yellow. The ordering vector within
		one phase regime changes continuously with $U$ and $n$ visualized by the
		color scheme as indicated in the plot legend. The AFM is denoted by a red circle.}
	\label{Fig:Phasediagram_nT_U4,5}
\end{figure}
The domain of the $n$--$U$ SRIKR phase diagram exhibiting $(\pi,Q)$ magnetic order, is in good agreement with the DMET data given in Ref.~\cite{PhysRevB.93.035126}. This is exemplary shown in \autoref{Fig:CompareDMET}, where the spin spiral with ordering vector $\bs Q\approx(\pi,0.71\pi)$ found by SRIKR is fitted to the spin profile according to DMET on a $2\times 8$ cluster.
However, in the $(Q,Q)$ domain, the ordering cannot be matched. Coincidingly, there are increasing inconsistencies between DMET clusters of size $2\times 8$ and $4\times 4$ which could be due to more severe finite size effects in the case of $(Q,Q)$ order compared to $(\pi,Q)$ order.

Moreover, we find the general trend, that points in parameter space which
feature a negative compressibility within SRIKR, show highly
inhomogeneous charge and/or superconducting orders according to DMET, while points with a positive compressibility are approximately
homogeneous in that regard.

\section{Results at finite temperature}\label{sec:Results_finiteT}
The slave boson-mean field theory may be extended to finite temperature,
provided $T$ is not too high. Although in the limit of infinite temperature
the free energy is found to approach the correct limit of $F=-NT\ln 4$, the
equipartition of slave bosons expected in this limit is not obtained.
Rather, one finds, e.g., at half filling and for particle-hole symmetric
spectrum, that $d=e=0$, for any $U>0$ with $T\rightarrow \infty$. We expect the slave boson mean field
theory to be applicable up to temperatures of the order of the band width $W$%
. \ In this section, we discuss the temperature dependence of the
slave-boson mean field and fluctuation results.

\begin{figure}[t!]
	\includegraphics[width=0.5\textwidth]{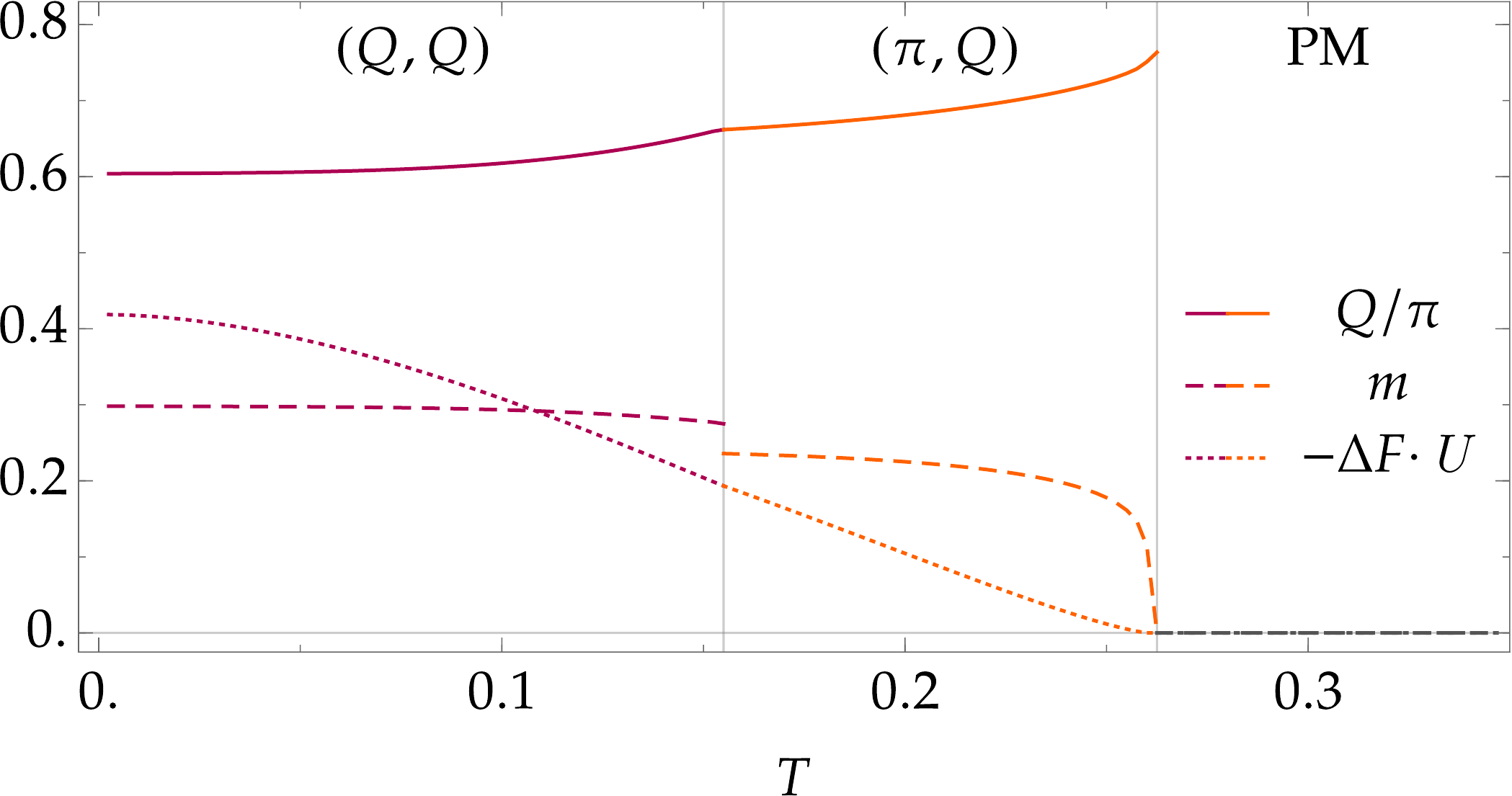}
	\caption{Ordering wave vector component $Q$, magnetization $m=p_0p$ and relative difference
		of the magnetic free energy at ordering vector $\boldsymbol Q$ and paramagnetic free energy 
		versus temperature $T$ for $U=10$, filling $%
		n=0.8$ and $t^{\prime }=-0.2$. 
		The vertical grid lines indicate phase transitions, the respective
		phases are denoted in the upper part of the plot. Near $T=0.155$ occurs a first order phase
		transition from $(Q,Q)$ to $(\protect\pi,Q)$, indicated by the discontinuity in the magnetization $m$.}
	\label{Fig:Q_T}
\end{figure}

\subsection{Magnetic mean field phase diagram}

\autoref{Fig:Phasediagram_nT_U4,5}b shows the temperature dependent
slave-boson mean field phase diagram at $U=4.5$ \ and $t^{\prime }=-0.2$.
In the presented temperature range we have $T\lesssim W(U)$, the
renormalized bandwidth. The paramagnetic second order phase boundary coincides with
results obtained from a temperature dependent fluctuation analysis of
magnetic instabilities. At stronger interaction the transition into the $%
(\pi ,\pi )-$state becomes a first order transition, which is presumably an
artifact of the mean field approximation. We determined the transition
temperature signaling the instability of the paramagnetic phase by first
finding the root of the inverse susceptibility $1/\chi _{s}(\bs Q,0)$
as a function of temperature defined by $T_{c}(\bs Q)$\ and then determining the maximum $%
T_{c}=\max_{\bs Q}\{T_{c}(\bs Q)\}$. The transition temperatures into the\ ordered
phase so determined as a function of doping are shown in \autoref{Fig:Phasediagram_nT_U4,5}a, for $t^{\prime }=-0.2$ and $U=2.5,3.5,4.5$. Our results also show that
a change in temperature leads to a continuous variation of the ordering
vector $Q$ and can induce a first order phase transition between a $(Q,Q)$
and $(\pi ,Q)$ ordering as illustrated in \autoref{Fig:Q_T}, for $n=0.8$, $%
t^{\prime }=-0.2$, and $U=10$,\ where also the magnetization and the free
energy are shown. For not too small $U\gtrsim 3$\ the Neel temperature has
its maximum around half filling and decreases with (hole- or electron)
doping.

\subsection{Critical Exponent}

\begin{figure}[t]
\includegraphics[width=0.48\textwidth]{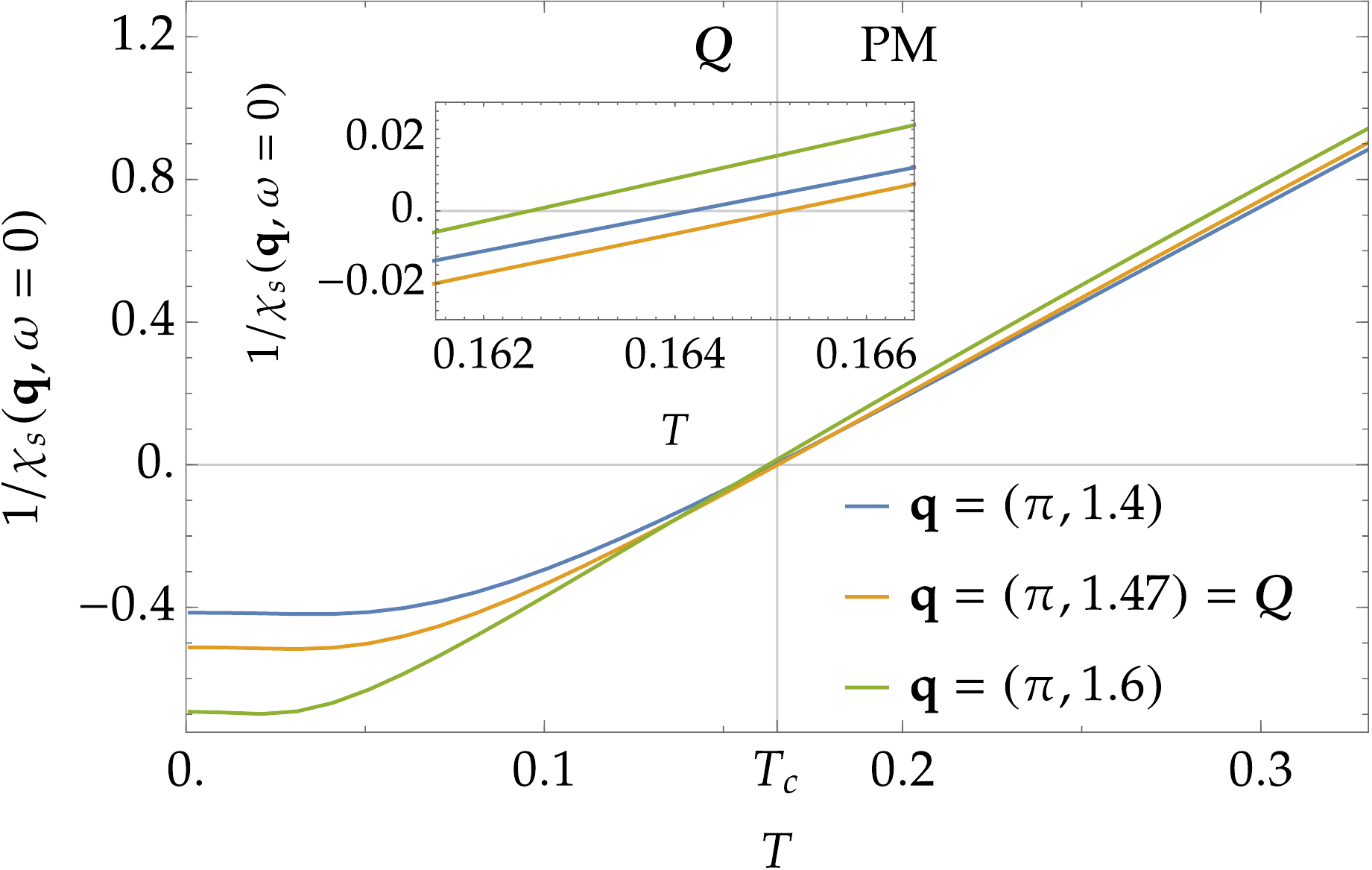}
\caption{Inverse static spin susceptibility for $U=12$, $t^{\prime }=0$, filling $n=0.6$
for different $\bq$ as function of temperature in the zero frequency limit. 
The inset shows the vicinity of
$1/\chi_s(\bq,\omega=0)$ around the critical temperature $T_{c}=0.165$.
The vertical grid line indicates the magnetic phase transition.}
\label{Fig:Tc_incomm}
\end{figure}

Furthermore we present the critical exponent $\gamma $ at the phase
transition defined as 
\begin{equation}
\chi _{s}(\bs Q,0)\propto (T-T_{c})^{-\gamma },
\end{equation}%
where $\bs Q$ is the ordering vector determined at $T_{c}$ featuring the
lowest free energy in mean field approximation. \autoref{Fig:Tc_incomm}
shows $\chi _{s}(\bs Q,0)^{-1}$ around the phase transition, which is
situated at the sign change of the reciprocal susceptibility for $\bq=\bs Q$
and two neighboring ordering vectors. Note that $\bq=\bs Q$ features the
highest $T_{c}$. The reciprocal susceptibility $\chi _{s}(\bs Q,0)^{-1}$
scales linearly in $T$ as shown by the comparison with the straight line in
the inset, resulting in a critical exponent of $\gamma =1$.

\subsection{Dynamical conductivity}

\begin{figure}[tbp]
\includegraphics[width=0.48\textwidth]{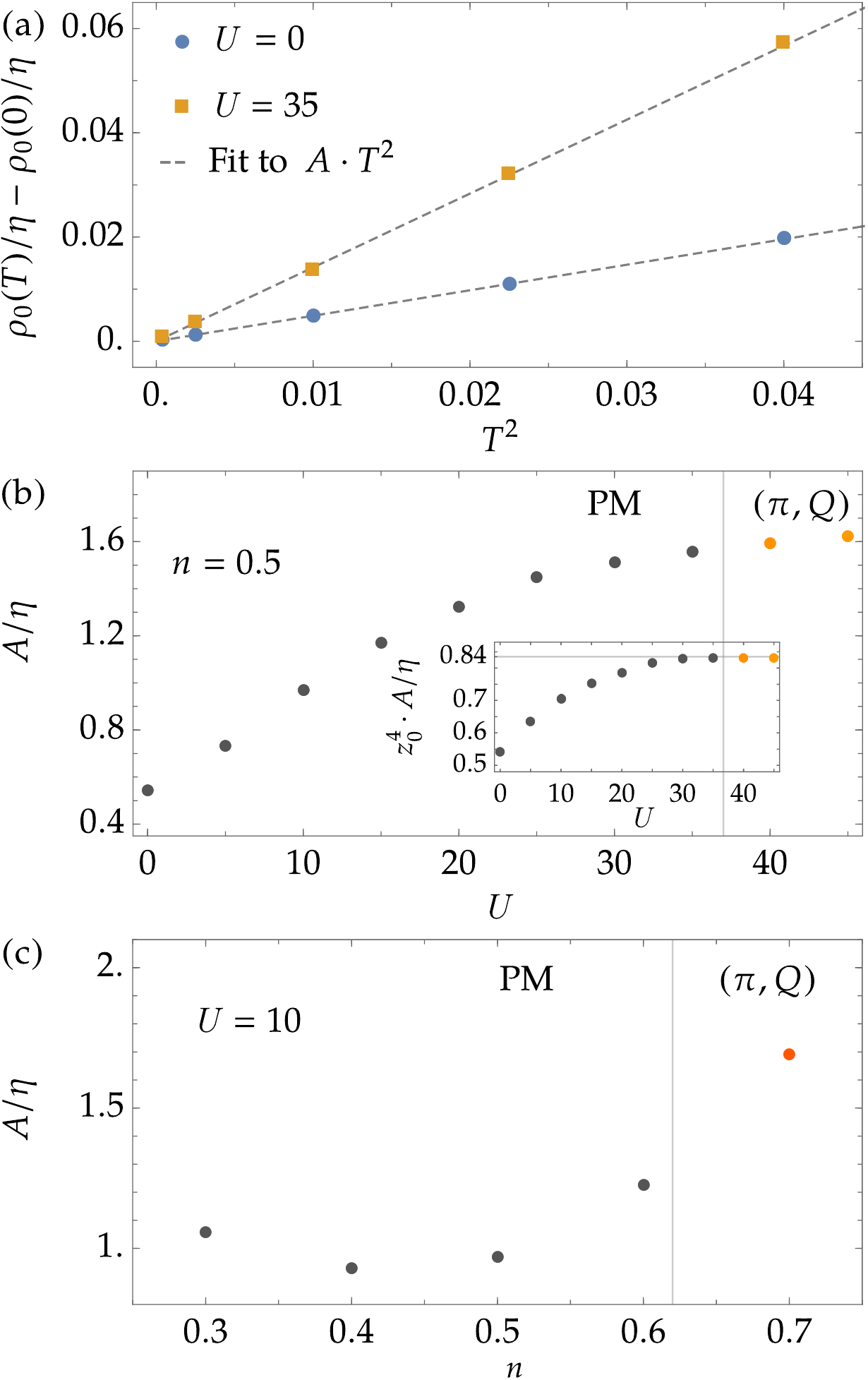}
\caption{(a) Temperature dependence of the resistivity $[\protect\rho (T)-%
\protect\rho (T=0)]$ versus $T^2$ at filling $n=0.5$, $t^{\prime }=0$ for $U=0$ (blue plot marker) and $U=35$ (orange plot marker).
(b) Coefficient $A$ versus interaction $U$ for
filling $n=0.5$ and $t^{\prime }=0$. The Inset shows the normalized coefficient $Az_{0}^{4}$, for large interactions the Kadowaki-Woods ratio
$A \propto (m/m^*)^2$ is fulfilled.
The vertical grid line indicates the magnetic phase transition. The color of the plot makers display the value of the ordering vector (compare \autoref{Fig:Phasediagram_t2=0}).
(c) Coefficient $A$ versus doping $n$ for $U=10$, $t^{\prime
}=0$. The vertical grid line indicates the magnetic phase transition. The color of the plot makers display the value of the ordering vector (compare \autoref{Fig:Phasediagram_t2=0}).}
\label{Fig:rho_T}
\end{figure}

The temperature dependence of the DC resistivity $\rho (T)=1/\sigma (T)$ at $%
n=0.5$, two values of interaction $U=(0,35)$ and $t^{\prime }=0$\ is shown
in \autoref{Fig:rho_T}a. For $T\ll W$, $\rho (T)$ follows the
behavior 
\begin{equation}
\rho (T)=\rho _{0}+A\cdot T^{2},
\end{equation}%
where $\rho _{0}$ and $A$ are temperature independent functions of filling,
interaction and hopping parameters (for $\rho _{0}$\ see the discussion
given above). For large $U$, we find that the coefficient $A$
of the quadratic term is proportional to $(m^{\ast }/m_{0})^{2}\propto
1/(z_{0})^{4}$, reminiscent of what is observed in heavy fermion compounds
(Kadowaki-Woods relation), as shown in \autoref{Fig:rho_T}b at $n=0.5$ and $%
t^{\prime }=0$. The density dependence of $A$ is weak, see \autoref%
{Fig:rho_T}c.

\subsection{T--U phase diagram at half filling}

\begin{figure}[t!]
\includegraphics[width=0.48\textwidth]{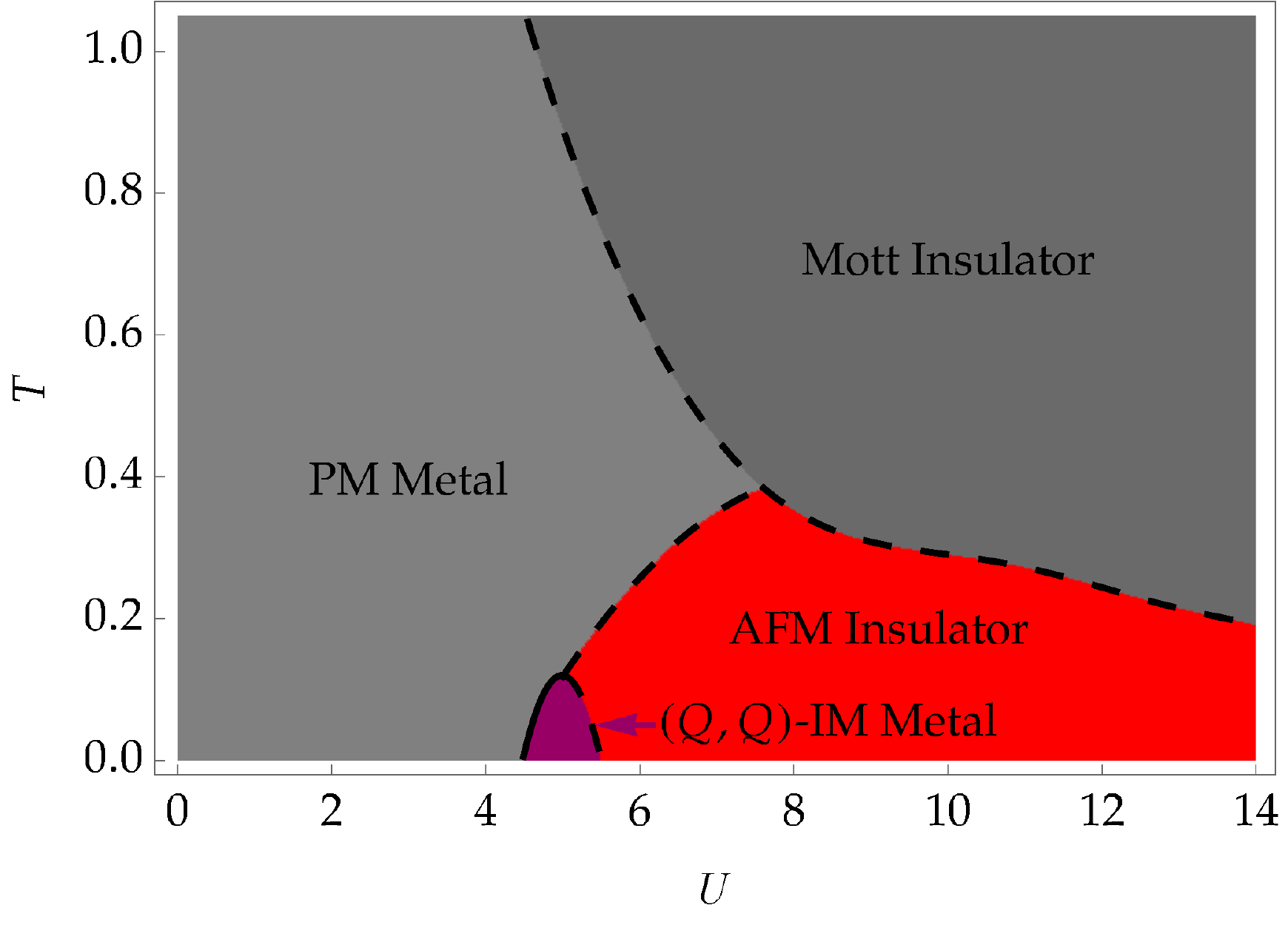}
\caption{Phasediagram in the $T$--$U$-plane at half filling for $t^{\prime}=1/\sqrt{3}$.
It features four distinct phases: a metallic PM (gray), a Mott insulator (dark gray),
a metallic magnetic phase (incommensurate ordering vector $Q \approx 0.57\pi$ (purple) and the antiferromagnetic insulating phase (red).
The ordering vector is nearly constant for all parameter points in the incommensurate magnetic phase.
The solid (dashed) lines indicate a phase transition of second (first) order. SB mean field yields no critical 
point at high $T$ which terminates the transition line of the Mott and paramagnetic metallic phase, as predicted by DMFT.}
\label{Fig:T-U_phasediagram_n1}
\end{figure}

The phase diagram in the temperature-interaction plane at half-filling is
shown in \autoref{Fig:T-U_phasediagram_n1}, at $t^{\prime }=1/\sqrt{3}$. For
given lower temperature, $T\lesssim 0.38$ and increasing $U$ the metallic
paramagnet is entering an insulating antiferromagnetic phase and eventually
a paramagnetic Mott insulator. both transitions are of first order. At low
temperatures $T\lesssim 0.13$ a narrow region of metallic magnetic $(Q,Q)-$%
phase, $Q\approx 0.57\pi$ is found between the paramagnet and the
antiferromagnetic\ insulator. The phase transition from the paramagnet to
the metallic magnetic $(Q,Q)-$phase is of second order. At high temperature, 
$T\geq 0.38$, the paramagnetic metal crosses over directly into the Mott
insulator phase by way of a first order transition. A comparison with the
results obtained for the two sublattice frustrated model ($t^{\prime }=1/%
\sqrt{3}$) of Rozenberg, Kotliar and Zhang using DMFT \cite{PhysRevB.49.10181}
(see Fig.~43 in \cite{georges1996}) shows remarkable similarity at not too
high temperatures even quantitatively. Only the behavior at very high
temperature\ is not captured correctly in the slave boson MFA, in that the
first order phase separation line between metal and Mott insulator does not
terminate at a critical point at about $T_{\text{crit}}\approx 1.5$, as it should,
but continues up to infinite temperature. This is a consequence of the fact
that in MFA the slave boson occupation numbers do not assume equilibrated
values for $T\rightarrow \infty $.

\bigskip

\bigskip

\bigskip

\bigskip


\section{Summary}

In this paper we presented a detailed derivation of the SRIKR slave boson
formalism (\autoref{sec2} and Appendix). It is shown within a path integral
representation that the atomic limit is exactly recovered. The mean field
theory of \ spiral magnetic states is derived. The spin and charge
correlation functions in the paramagnetic state are expressed in terms of
the fluctuation amplitudes. We showed that the $\alpha$ constraint which fixes the number of bosons per 
lattice site can be enforced exactly not only on MF level, but also within the fluctuation calculation. 
This reduces the dimension of the fluctuation matrix $\mathcal{M}_{\mu\nu}$ by 
two and simplifies the calculation of the charge susceptibility compared to the formalism presented in the previous literature. 

In \autoref{sec:Results:ZeroT}, results for zero temperature are presented. The solution of
the mean field equations of spiral magnetic states are used to construct a
phase diagram in the interaction $U$ -- density $n$ plane. A number of
different phases are found, characterized by the ordering wavevector $%
\bs Q$, classified in two types $(Q,Q)$ and $(\pi ,Q)$, with $Q$
varying continuously within a given phase. The two types of phases are
separated by first order transitions. We considered two hopping models:
nearest neighbor hopping only ($t^{\prime }=0$) and additional next nearest
neighbor hopping ($t^{\prime }=-0.2$). The $z$-factors renormalizing the
hopping have been calculated and discussed in their dependence on density
and interaction.\ The magnetization, the free energy gain of the ordered
state and the ordering wave vector component $Q$ have also been evaluated.
We presented the renormalized band structures in the paramagnetic and the
various magnetically ordered phases. At half-filling the Mott-Hubbard
transition in the paramagnetic phase, signaled by a vanishing of the $z-$%
factor, is preempted by the formation of magnetic antiferromagnetic order
for which the $z$-factor stays finite. Examples of the Fermi surfaces in the
various phases were presented as well. The compressibility in the
paramagnetic phase is found to become negative in a region of the phase
diagram around $n=1$, signaling an instability towards charge separation or
charge order. A sign change of the compressibility happens in the
magnetically ordered phase. The calculation of the charge susceptibility in
the magnetically ordered phase is beyond the scope of our present work.

We calculated the spin susceptibility in the paramagnetic phase. The static
spin susceptibility is parametrized in terms of a Landau interaction
function $F^{a}(\bs q)$, found to vary in the interval $[-1,0]$, with $%
F^{a}(\bs Q)=-1$ signaling the transition to a magnetic state with
ordering vector $\bs Q$. We determined the phase boundary to the
magnetically ordered phase by finding the zeros of $\min_{\bq%
}\{1/\chi (\bq,0)\}$, the results being fully consistent with what
was found from the magnetic mean field study. The dynamic spin
susceptibility is parametrized in terms of a Landau damping function $\Gamma
(\bq,0)$, found to vary as $\Gamma \propto |\bq|$ in the limit 
$|\bq|\rightarrow 0$.\ At the phase transition the static spin
susceptibility at the ordering wave vector is found to diverge as $\chi (%
\bs Q,0)\propto (n-n_{c})^{-\alpha }$ where $n_{c}$ is the critical
doping. Surprisingly, the exponent $\alpha $ turned out to depend on whether
the magnetic state was commensurate, where $\alpha \approx 1$, or incommensurate,
for which $\alpha \approx 1/2$.\ 

We calculated the charge excitation spectrum finding an interesting structure
to be interpreted as two collective modes induced by interaction on top of
the particle-hole continuum. The higher frequency mode has the character of
an excitation into the upper Hubbard band. The mode in between the continuum
and the latter mode resembles the zero sound mode of a Fermi liquid. These
modes show a considerable dependence on density, interaction and on the
range of hopping. The charge response function is employed to calculate the
dynamical conductivity. We employed a finite imaginary part of the
frequency, $\eta $, to be interpreted as an impurity scattering induced
relaxation rate. The real and imaginary parts of the conductivity are found
to assume Drude form, renormalized by interaction. The DC resistivity as a
function of $U$ is shown to be proportional to the inverse effective mass in
good approximation $\rho _{0}\propto z_{0}^{-2}$. As a function of density $n
$ the relation $\rho _{0}\propto 1/(z_{0}^{2}n)$, expected to hold for the
Drude conductivity is obeyed only approximately. The spin and charge
structure factors were also calculated.

In \autoref{sec:Results_finiteT} we presented results at finite temperature. Stable solutions
of the mean field equations have been found for temperatures less than the
renormalized band width. We determined the magnetic phase diagram in the
temperature $T$ -- doping $n$ plane at fixed interaction $U$. We found the phase boundaries separating the magnetically ordered phases from the
paramagnetic phase and also separating different ordered states. A continuous
change of the ordering wave vector as the temperature and doping are varied
is presented. The static spin susceptibility at fixed $U$ and $n$ and at the
ordering vector $\bs Q$\ is found to diverge at the transition as $\chi
(\bs Q,0)\propto (T-T_{c})^{-\gamma }$, where $T_{c}$ is the critical
temperature and $\gamma \approx 1$. The temperature dependent DC
resistivity is shown to follow a quadratic dependence $\rho (T)=\rho
_{0}+AT^{2}$. The coefficient $A$ is found to be proportional to $(m^{\ast
}/m)^{2}\propto z_{0}^{-4}$, reminiscent of the Kadowaki-Woods relation
found for heavy fermion compounds. Finally, we established a phase diagram
in the temperature $T$ - interaction $U$ plane at half filling and choosing
a next-nearest neighbor hopping parameter $t^{\prime }=1/\sqrt{3}$. The general
features of the phase diagram agree very well with results obtained by other
methods. The only exception is the behavior at higher temperatures, where
the slave-boson mean field approximation shows a first order metal-insulator
transition instead of a phase boundary ending at a critical point. 

The results presented above show that the SRIKR slave-boson method is a
powerful alternative to other approximate methods in the interacting fermion
problem, such as DMFT, Functional
Renormalization Group Method (FRG), and purely numerical methods such as QMC, Density Matrix Renormalization Group (DMRG), or
DMET, to name a few prominent examples. Our
method is not limited to local quantum fluctuations (like DMFT), but can describe
long-range ordered phenomena. It is not limited to low to intermediate
interaction (like FRG), but works for arbitrarily strong interaction, it does not
suffer from a ``sign problem'' limiting its application to sufficiently high
temperatures (like QMC), but works at low temperatures up to the bandwidth limit,
it is not restricted to small systems (like DMRG and DMET), but works in the
thermodynamic limit. The detailed comparison of our results with those of a
recent DMET study presented in \autoref{sec:Results:DMET} demonstrates an impressive
degree of compatibility as far as the fine-structure of the phasediagram at $%
T=0$ is concerned .


\section*{Acknowledgments}
The work in W\"urzburg is funded by the Deutsche Forschungsgemeinschaft (DFG, German Research Foundation) through Project-ID 258499086 - SFB 1170 and through the W\"urzburg-Dresden Cluster of Excellence on Complexity and Topology in Quantum Matter -- \textit{ct.qmat} Project-ID 39085490 - EXC 2147. Titus Neupert acknowledges support from the Swiss National Science Foundation (grant number: 200021\_169061) and from the European Union’s Horizon 2020 research and innovation program (ERC-StG-Neupert-757867-PARATOP).
Peter W\"olfle acknowledges support through a Distinguished Senior Fellowship of Karlsruhe Institute of Technology.

The authors David Riegler and Michael Klett contributed equally to the method development.

\bibliography{Literatur_SB}
\newpage
\clearpage
\appendix

\section{slave-boson formalism on operator level}\label{Chapter:App:Operatorlevel}
The slave-boson formalism was originally introduced by Kotilar and Ruckenstein~\cite{kotliar_new_1986} (KRSB) as a strong coupling mean field theory for a unified treatment of magnetism, metal-to-insulator 
transitions and Kondo physics. The method was later generalized to be manifestly spin rotation invariant~\cite{woelfle_spin_rotation_1989,woelfle_spin_1992} (SRIKR) and applied to charge and spin structure factors in the 
Hubbard model by means of bosonic fluctuations around the saddle point solution~\cite{Li1991,woelfle_spin_1997}. 
This note provides a detailed summary of spin rotation invariant slave-boson mean field formalism with 
fluctuations in a general notation for the Hubbard model, which can be generalized to models with one interacting and an arbitrary number of non-interacting orbitals~\cite{PAM_wuerzburg}. We include a guide how to numerically implement the mean field equations and
a section how to derive response functions from the fluctuation matrix. Moreover, we present the exact evaluation of the atomic limit in the path integral representation. 

The general idea of slave-boson formalism is to define a set of bosonic operators $e_i$, $p_{0,i}$, $\bs p_i=(p_{1,i},p_{2,i},p_{3,i})$ and $d_i$, labeling empty, singly and doubly occupied lattice sites $i$, 
respectively for the interacting orbital. Spin rotation invariance requires to introduce four bosonic fields to represent a singly occupied site in comparison to two fields in the original Kotliar-Ruckenstein description.
Furthermore, one needs to introduce two auxiliary fermionic fields $f_{i,\downarrow} , f_{i,\uparrow}$, referred to as pseudofermions, which correspond to the quasiparticle degrees of freedom. A set of additional constraints, allows an exact mapping from the original fermionic creation and annihilation operators $(c^\dagger,c^\nodag)$ to the slave-boson and pseudofermion fields, 
where the Hubbard interaction becomes bilinear, whereas hopping terms adapt a non bilinear form in bosonic operators. This way, the problem is investigated from a strong coupling perspective compared to conventional fermionic mean field theory. 

The empty, singly and doubly occupied states are created by
\begin{subequations}\label{App:States}
	\begin{align} 
		\ket{0}_i&\equiv\ed_i \ket{vac},\\
		\ket{\sigma}_i&\equiv\sum\limits_{\sigma'}\pd_{i,\sigma \sigma'}\fd_{i,\sigma'} \ket{vac},\\
		\ket{2}_i&\equiv\dd_i \fd_{i,+\frac12} \fd_{i,-\frac12} \ket{vac} =\dd_i \fd_{i,\uparrow} \fd_{i,\downarrow} \ket{vac}
	\end{align}
\end{subequations}
at each lattice site $i$ where $\sigma \in \pm \frac12$ corresponds to the spin of the fermionic operators. The matrix operator $\pd_{i,\sigma \sigma'}$ will be defined in the following section. The occurring fermionic $f_{i,\sigma}$ and bosonic $b_{\alpha,i} \in \left\{e_i,p_{0i},\bs p_i, d_i\right\}$ fields 
fulfill the usual (anti-) commutation relations
\begin{subequations}
	\begin{align}
		\Big{\{}f^\nodag_{i,\sigma} \ , f^\dagger_{j,\sigma'}\Big{\}}
		&= \delta_{\sigma \sigma'}\ \delta_{ij}, \\
		\Big{\{}f^\nodag_{i,\sigma} \ , f^\nodag_{j,\sigma'}\Big{\}}
		&=\Big{\{}f^\dagger_{i,\sigma} \ , f^\dagger_{j,\sigma'}\Big{\}}=0, \\
		\left[b^\nodag_{\alpha,i} \ ,  b^\dagger_{\beta,j}\right]&=\delta_{\alpha \beta}\ \delta_{ij}, \\
		\left[b^\nodag_{\alpha,i} \ ,  b^\nodag_{\beta,j}\right]
		&=\left[b^\dagger_{\alpha,i} \ ,  b^\dagger_{\beta,j} \right]=0.
	\end{align}
\end{subequations}
The site index $i$ will be dropped for readability in the following, it is implied that the all equations without an additional index $ i $ hold for every lattice site.

\subsection{Construction of the p-Matrix}\label{app:p-matrix}
While the empty and doubly occupied states transform like scalars, the singly occupied state $\ket{\sigma}$ needs to transform like a spinor under spin rotation, consequently $\pd_{i,\sigma \sigma'}$ 
represents an element of a  $2\times 2$ matrix. The total spin of the singly occupied state is $S=\frac{1}{2}$ and consists of a  pseudofermionic ($S_f=\frac{1}{2}$) and a bosonic component. The possible bosonic 
spins are $S_b=0$ and $S_b=1$ yielding a scalar boson field $p_0$ and a vector boson field $\bs p=(p_x,p_y,p_z)$ where $x,y,z$ are the Cartesian components.
The spin operator for $S_b=1$ is given by
\begin{gather}
	\hat{\bs S}=
	\begin{pmatrix}
		\underline{S}^{x}\\
		\underline{S}^{y}\\
		\underline{S}^{z}
	\end{pmatrix},
\end{gather}
where
\begin{subequations}
	\begin{align}
		\underline{S}^{x}&=  \frac{1}{\sqrt{2}}
		\begin{pmatrix}
			0&1&0\\
			1&0&1\\
			0&1&0
		\end{pmatrix}, \\
		\underline{S}^{y}&=  \frac{1}{\sqrt{2}}
		\begin{pmatrix}
			0&-\i &0\\
			\i &0&-\i \\
			0&\i &0
		\end{pmatrix}, \\  
		\underline{S}^{z}&=  
		\begin{pmatrix}
			1&0&0\\
			0&0&0\\
			0&0&-1
		\end{pmatrix}.
	\end{align}
\end{subequations}
For a spin rotational invariant representation, we choose the $\bs p^\dagger$ operator to create a bosonic state ($S_b=1$) which is polarized in the $x, y $ and $z$ direction respectively with the magnetic quantum number $m=0$
\begin{equation}
	\begin{gathered}
		\underline{S}^{i} \ket{\chi_i} =0\\
	\bs	\chi_x= \frac{1}{\sqrt{2}}
		\begin{pmatrix}
			-1\\0\\1         
		\end{pmatrix},
		\qquad
	\bs	\chi_y= \frac{1}{\sqrt{2}}
		\begin{pmatrix}
			\i \\0\\\i          
		\end{pmatrix},\qquad
	\bs	\chi_z= 
		\begin{pmatrix}
			0\\1\\0         
		\end{pmatrix}.
	\end{gathered}
\end{equation}
This basis is orthonormal on the spin Hilbert space $\bra{\chi_i}\chi_j\rangle=\delta_{ij}$. The relative phases of $\bs \chi_i$ are not arbitrary because they are related
by spin rotation $\e^{-i \bs \phi \hat{\bs S}  }$ and chosen such that
\begin{subequations}
	\begin{align}
	\bs	\chi_y=\e^{-i \frac{\pi}{2} \underline{S}^z  } \bs \chi_x, \\
		\bs \chi_z=\e^{-i \frac{\pi}{2} \underline{S}^x  } \bs \chi_y, \\
		\bs \chi_x=\e^{-i \frac{\pi}{2} \underline{S}^y  } \bs \chi_z .
	\end{align}
\end{subequations}
In order to add the spin of the boson and the pseudofermion, it is convenient to use the basis of eigenstates of the $\underline{S}^z$ operator, which can be found as superposition 
of the spinors $\bs \chi_x, \bs \chi_y$ and $\bs \chi_z$ yielding the ladder operators

\begin{subequations}\label{App:Construction:p1m}
\begingroup
\allowdisplaybreaks
	\begin{align}
		\pn_{1,1}&\equiv-\frac{1}{\sqrt{2}}\left(\pn_x+\i  \pn_y\right),\\
		\pn_{1,0}&\equiv \pn_z \ , \\
		\pn_{1,-1}&\equiv\frac{1}{\sqrt{2}}\left(\pn_x-\i  \pn_y\right) \ .
	\end{align}
\endgroup
\end{subequations}
Consequently a state with total spin of $S=\frac{1}{2}$ composed of a $S_f=\frac{1}{2}$ pseudofermion $f_{\sigma'}$ and a $S_b=1$ slave-boson $p_{1,m_1}$ is given by \citep{woelfle_spin_1992}
\begin{widetext}
	\begin{align}\label{App:Mapping:AddSpin}
		\begin{split}
			\ket{\sigma}_{S={\frac{1}{2}}}&=
			\sum_{\sigma'=\pm \frac{1}{2}} C\left( S_b=1, S_f =\frac{1}{2}; m_b=\sigma-\sigma', m_f=\sigma' \Big{|} S =\frac{1}{2}; \sigma \right)
			\ \pd_{1,m_b} \fd_{\sigma'}\ket{vac}= \sum_{\sigma'=\pm \frac{1}{2}} \left(p^\dagger_{S=1}\right)_{\sigma\sigma'} \fd_{\sigma'}\ket{vac} 
		\end{split}
	\end{align}
\end{widetext}
with $\sigma=\pm \frac{1}{2}$ and the Clebsch-Gordon coefficients
\begin{align}
	C\left( 1,\frac{1}{2}; \sigma \mp \frac{1}{2},\pm\frac{1}{2}\Big{|}\frac{1}{2},\sigma\right)=\mp \sqrt{\frac{3\mp2\sigma}{6}}.
\end{align}
As \eq{App:Mapping:AddSpin} implies, we can write the bosons in a convenient matrix notation $\underline{p}_{~S=1}$, which reads
\begin{equation}
	\underline{p}^\dagger_{~S=1}=
	\begin{pmatrix}
		-\sqrt{\frac{1}{3}}\pd_{1,0} & \sqrt{\frac{2}{3}} \pd_{1,1} \\
		-\sqrt{\frac{2}{3}} \pd_{1,-1} & \sqrt{\frac{1}{3}} \pd_{1,0}
	\end{pmatrix},
\end{equation}
using the basis
\begin{equation}
	\fd_{\sigma'}=
	\begin{pmatrix}
		\fd_\uparrow\\
		\fd_\downarrow\\
	\end{pmatrix}
\end{equation}
for the pseudofermions. To obtain the full matrix, one has to take contributions of the scalar field $p_0$ as a superposition into account, which only acts diagonal on the spin subspace. 
Inserting \eq{App:Construction:p1m}, one finds for the full matrix
\begin{equation}
	\underline{p}^\dagger=
	\begin{pmatrix}
		a p_0^\dagger+b\pd_z & b(\pd_x-\i \pd_y) \\
		b(\pd_x+\i \pd_y) & a p_0^\dagger-b \pd_z
	\end{pmatrix}.
\end{equation}
The coefficients $a$ and $b$ are not arbitrary, but have to be chosen such that the normalization
\begin{equation}
	\sum_{\sigma' \sigma''}\bra{vac}f^\nodag_{\sigma''} p^\nodag_{\sigma'' \sigma} \pd_{\sigma \sigma'} \fd_{\sigma'} \ket{vac}=1 
\end{equation}
is fulfilled for $\sigma=\pm \frac{1}{2}$ which implies $ 3b^2 +a^2 = 1 $. The ratio $a/b$ is a free parameter. It can be chosen $a=b=1/2$ which finally yields
\begin{subequations}
	\begin{align}
		\underline{p}^\dagger &=\frac{1}{2}\sum_{\mu=0}^3 \pd_\mu \underline{\tau}^\mu=\frac{1}{2}
		\begin{pmatrix}
			\pd_0+\pd_z & \pd_x-\i \pd_y \\ 
			\pd_x+\i \pd_y & \pd_0-\pd_z
		\end{pmatrix},\\
		\underline{p}^\nodag &=\frac{1}{2}\sum_{\mu=0}^3 p_\mu \underline{\tau}^\mu=\frac{1}{2}
		\begin{pmatrix}
			\pn_0+p_z & \pn_x-\i \pn_y \\ 
			\pn_x+\i p_y & \pn_0-\pn_z
		\end{pmatrix}
	\end{align}
\end{subequations}
where $\underline{\tau}^\mu$ is the vector of the Pauli matrices, including the identity matrix $\underline{\tau}^0\equiv \underline{\mathbb{1}}_2$. The commutator of these matrix operators is given by
\begin{equation}\label{App:Construction:Commutator}
	\left[p_{\sigma_1 \sigma_2}^{} , p_{\sigma_3 \sigma_4}^\dag \right]= \frac{1}{2} \delta_{\sigma_1 \sigma_4} \delta_{\sigma_2 \sigma_3}.
\end{equation}

\subsection{slave-boson representation and time reversal properties}
The original, fermionic operators $c^\dagger_\sigma, c^\nodag_\sigma$, are mapped to the slave-boson operators by
\begin{subequations}\label{App:Mapping:zf}
	\begin{align}
	c^\dagger_\sigma &\equiv \sum_{\sigma'} z^\dag_{\sigma\sigma'} \fd_{\sigma'},\\
		c^\nodag_\sigma &\equiv \sum_{\sigma'} \ff_{\sigma'} z^{\nodag}_{\sigma'\sigma},
	\end{align}
\end{subequations}
with
\begin{subequations}\label{App:Mapping:z}
	\begin{align}
		z_{\sigma\sigma'}&=e^\dag p_{\sigma\sigma'}+\tilde p^\dag_{\sigma\sigma'} d,\\
		z^\dagger_{\sigma\sigma'}&= p^\dag_{\sigma\sigma'}e+d^\dag\tilde p^\nodag_{\sigma\sigma'}, 
	\end{align}
\end{subequations}
and
\begin{subequations}
	\begin{equation}
		\tilde p_{\sigma \sigma'}=\frac{1}{2}\left(p_0\tau^0_{\sigma \sigma'}- \sum_{\mu=1}^3 p^\mu_{\sigma \sigma'} \right)
	\end{equation}
	or equivalently
	\begin{equation}\label{App:Mapping:pTilde}
		\tilde{p}_{\sigma \sigma'} = 4\sigma \sigma' p_{-\sigma' -\sigma} \qquad \sigma \in \Bigg{\{}\frac{1}{2},-\frac{1}{2}\Bigg{\}}.
	\end{equation}
\end{subequations}
Note, that In this notation $-\sigma$ corresponds to a spin flip.

\eq{App:Mapping:zf} is straight forward to understand when acting on an empty or doubly occupied site. For single occupation, there are two states because of the spin, which makes the situation slightly more complicated and it is necessary to define $\tilde{\bs p}$ in order to obtain the expected result
\begin{align}\label{App:Mapping:2}
\begin{split}
c^\dagger_{\sigma}\ket{\sigma'}
&=\sum_{\sigma_1 \sigma_2}z^\dag_{\sigma\sigma_1} \fd_{\sigma_1}\pd_{\sigma' \sigma_2}\fd_{\sigma_2} \ket{vac}\\
&=2\sigma\delta_{-\sigma \sigma'}\dd f^\dagger_\uparrow f^\dagger_\downarrow \ket{vac}=2\sigma \delta_{-\sigma \sigma'}\ket{2}.
\end{split}
\end{align}

\eq{App:Mapping:zf} fulfills the expected behavior under time reversal. Fermionic operators need to fulfill 
\begin{subequations}\label{App:Mapping:timereversal}
	\begin{align}
		\hat{T}c^\nodag_\uparrow\hat{T}^{-1}&=c^\nodag_{\downarrow},\\
		\hat{T}c^\nodag_{\downarrow}\hat{T}^{-1}&=-c^\nodag_{\uparrow}
	\end{align}
\end{subequations}
where $\hat{T}$ is the time reversal operator. Since $p_0$ annihilates a spin singlet and $\bs p$ a spin triplet, we expect $p_0$ to be even, $\hat{T}p_0\hat{T}^{-1}=p_0$ and $\bs p$ to be 
odd $\hat{T}\bs p \hat{T}^{-1}=-\bs p$ under time reversal. Moreover, the operator $\hat{T}$ is anti-unitary $\hat{T}i\hat{T}^{-1}=-i$.
The properties of the other slave-boson fields under time reversal can now be determined demanding that Eq.~\eqref{App:Mapping:timereversal} holds within slave-boson formalism
\begin{subequations}
	\begin{align}
		\hat{T}p_0\hat{T}^{-1}&=p_0,\\
		\hat{T}\bs p \hat{T}^{-1}&=-\bs p,\\
		\hat{T}e\hat{T}^{-1}&=e,\\
		\hat{T}d\hat{T}^{-1}&=d,\\
		\hat{T}f_\uparrow\hat{T}^{-1}&=f_{\downarrow},\\
		\hat{T}f_{\downarrow}\hat{T}^{-1}&=-f_{\uparrow}.
	\end{align}
\end{subequations}

\subsection{Constraints in slave-boson formalism}\label{App:constraints}
In order to have an exact mapping of original fermionic operators to slave-boson operators, one needs to enforce the following constraints to recover from the extended Fock space to the physical Hilbert space
\begin{subequations}
	\begin{align} \label{App:Constraints:Constraint1}
		1&= \ed e+\dd d + \ppuu\\ 
		\label{App:Constraints:Constraint2}
		\fd_{\sigma'} \ff_{\sigma\vphantom'} &=2\sum_{\sigma_1}\pd_{\sigma_1 \sigma} \pn _{\sigma' \sigma_1} + \delta_{\sigma \sigma'} \dd d.
	\end{align}
	\eq{App:Constraints:Constraint2} can be rewritten in terms of $p_\mu$'s to four scalar equations by expanding in Pauli matrices including the identity matrix, i.e. applying $\sum_{\sigma\sigma'}\tau^\mu_{\sigma\sigma'}$ 
	on both sides of the equation
	\begin{align}
		\sum_{\sigma}\fd_{\sigma} f^\nodag_{\sigma}&= \ppuu+2\dd d \label{App:Constraints:Constraint21} \\
		\sum_{\sigma \sigma'} \bs \tau_{\sigma \sigma'} \fd_{\sigma'}f^\nodag_{\sigma\vphantom'}&= \pd_0 \bs p + \bs \pd p_0-\i \bs \pd \times \bs p\;. \label{App:Constraints:Constraint22}
	\end{align}
\end{subequations}
These constraints are enforced on each lattice site. The first constraint \eq{App:Constraints:Constraint1} makes sure that every site is occupied by exactly one slave-boson. The second constraint 
\eq{App:Constraints:Constraint21} matches the number of pseudofermions and slave-bosons according to \eq{App:States}. The third constraint \eq{App:Constraints:Constraint22} relates the spin of the pseudofermions
and slave-bosons which are not independent as \eq{App:Mapping:AddSpin} indicates. It states that a spin flip in pseudofermions can be recast as a spin flip in the slave-bosons. 
Since such a recast spin flip has to obey the previous assignment of $p$-bosons and pseudofermions one has to employ the third constraint .

The necessity of the constraints can be seen mathematically by calculating the anti-commutator $\{c^\nodag_{\sigma\vphantom'}  , c^\dagger_{\sigma'}\}=\delta_{\sigma\sigma'}$ 
in slave-boson formalism, which is only recovered correctly when applying all of the constraints.
It is sufficient to verify the commutator on the physical subspace. This way, one can exploit, that two bosonic annihilation operators to the very right side of an equation annihilate any 
state because of \eq{App:Constraints:Constraint1}. Such an ordering can be achieved by using Eq.~\eqref{App:Mapping:pTilde} and the commutator Eq.~\eqref{App:Construction:Commutator}. 
Moreover, the pseudofermions can be replaced by slave-boson operators by means of the second constraint Eq.~\eqref{App:Constraints:Constraint2}. It turns out that only terms which are bilinear in bosonic operators remain
\begin{subequations}
	\begin{equation}\label{App:Constraints:CalcCommutator}
		\begin{gathered}
			\Big{\{}c^\nodag_{\sigma\vphantom'} \ , c^\dagger_{\sigma'}\Big{\}}
			= \sum_{\sigma_1 \sigma_2} \left(f^\nodag_{\sigma_1}z^\nodag_{\sigma_1 \sigma}z^\dagger_{\sigma'\sigma_2}f^\dagger_{\sigma_2}+ z^\dagger_{\sigma'\sigma_2}f^\dagger_{\sigma_2}f^\nodag_{\sigma_1}z^\nodag_{\sigma_1 \sigma} \right)\\
			=\delta_{\sigma \sigma'}\left(e^\dagger e + d^\dagger d \right)
			+2\sum_{\sigma_1}\left(4\sigma \sigma' p^\dagger_{-\sigma\sigma_1}p^\nodag_{\sigma_1 -\sigma'}+ p^\dagger_{\sigma'\sigma_1}p^\nodag_{\sigma_1\sigma}\right).
		\end{gathered}
	\end{equation}
	The second term in Eq.~\eqref{App:Constraints:CalcCommutator} can be further decomposed
	\begin{equation}
		\begin{gathered}
			\sum_{\sigma_1}p^\dagger_{\sigma'\sigma_1}p^\nodag_{\sigma_1\sigma}
			=\frac{1}{4}\sum_{\mu=0}^3 p^\dagger_\mu p^\nodag_\mu \delta_{\sigma\sigma'}\\
			+\frac{1}{4}\sum_{\mu=1}^3\tau^\mu_{\sigma'\sigma}\left(p^\dagger_\mu  p^\nodag_0+p_0^\dagger p^\nodag_\mu \right)
			+\frac{i}{4}\sum_{\mu\mu'\nu=1}^3\epsilon^{\ \mu\mu'\nu}\tau^\nu_{\sigma\sigma'} p^\dagger_\mu p^\nodag_{\mu'}
		\end{gathered}
	\end{equation}
	Now, making use of 
	\begin{equation}
		\tau^\mu_{\sigma'\sigma}=-4\sigma \sigma' \tau^\mu_{-\sigma -\sigma'}, \quad \mu \in \{1,2,3\}
	\end{equation}
	all terms containing Pauli matrices vanish in Eq.~\eqref{App:Constraints:CalcCommutator} which yields
	\begin{align}
		\Big{\{} c^\nodag_{\sigma \vphantom'} \ , c^\dagger_{\sigma'}\Big{\}}=\delta_{\sigma\sigma'}\left(e^\dagger e+d^\dagger d+ \sum_{\mu=0}^3 p^\dagger_\mu p^\nodag_\mu \right)=\delta_{\sigma\sigma'}
	\end{align}
\end{subequations}
and leads to the expected result by once more using the first constraint. Consequently, the fermionic character of the fields is preserved in slave-boson formalism.

Note, that pseudofermions can be replaced by slave-bosons every time they appear bilinear with \eq{App:Constraints:Constraint2}. However, a combination of pseudofermions on different sites cannot be replaced.

The constraints can be enforced by means adequate projection operators. We define
\begin{subequations} \label{App:Path:Projectors}
	\begin{align} 
		A&\equiv e^\dagger e+d^\dagger d + \sum_{\mu=0}^3 p^\dagger_\mu p^\nodag_\mu-1\\ 
		B_0&\equiv \sum_{\mu=0}^3 p^\dagger_\mu p^\nodag_\mu+2 d^\dagger d -\sum_{\sigma} f^\dagger_{\sigma} f^\nodag_{\sigma}\\ 
		\bs B&\equiv p^\dagger_0 \bs p + \bs p^\dagger p^\nodag_0-\i \bs p^\dagger \times \bs p -\sum_{\sigma \sigma'} \bs \tau_{\sigma \sigma'} f^\dagger_{\sigma'}\ff_{\sigma}
	\end{align}
\end{subequations}
and need to enforce $A=B_0=\bs B=0$ in order to fulfill the constraints. To do so, we define the following projection operators:
\begin{subequations}
	\begin{align}
		\mathcal{P}_\alpha&=\frac{1}{2\pi T}\int_{-\pi T}^{\pi T}\e^{i\alpha A/T}d\alpha =\delta_{A,0} \\
		\mathcal{P}_{\beta_0}&=\frac{1}{2\pi T}\int_{-\pi T}^{\pi T}\e^{iB_0 Y/T}d\beta_0 =\delta_{B_0,0} \\
		\mathcal{P}_{\bs \beta}&= \lim_{N \rightarrow \infty}\frac{1}{(2\pi N T)^3}\iiint_{-\pi N T}^{\pi N T}\prod_{i=1}^{3} \e^{i \bs \beta_i B_i/T}d\beta_i =\delta_{\bs B,0} \\
		\mathcal{P}&\equiv \mathcal{P}_\alpha \mathcal{P}_{\beta_0} \mathcal{P}_{\bs\beta}.
	\end{align}\label{App:Constraints:Projectors}
\end{subequations}
Note that, since $\bs Z$ contains operators which are not number operators, its eigenvalues may have non integer values. Therefore the integral has to be extended to infinity to project out all unphysical states. 
The partition function of the physical subspace for a Hamiltonian $H$ is then given by
\begin{equation}
	Z_{\text{eff}}=\tr\left[e^{-H/T}\mathcal{P}\right]. 
\end{equation}
The constraints commute with the slave-boson representation of fermionic creation (annihilation) operators
\begin{subequations}
	\begin{align}
		\left[\sum_{\sigma'} z^\dag_{\sigma\sigma'} \fd_{\sigma'}\ ,\ A\right]= \left[\sum_{\sigma'} z^\dag_{\sigma\sigma'} \fd_{\sigma'}\ ,\ B_0\right]= \left[\sum_{\sigma'} z^\dag_{\sigma\sigma'} \fd_{\sigma'}\ , \ \bs B\right]=0\\
		\left[\sum_{\sigma'} \ff_{\sigma'} z^{\nodag}_{\sigma'\sigma}\ , \ A\right]= \left[\sum_{\sigma'} \ff_{\sigma'} z^{\nodag}_{\sigma'\sigma} \ , \ B_0\right]= \left[\sum_{\sigma'} \ff_{\sigma'} z^{\nodag}_{\sigma'\sigma} \ , \ \bs B\right]=0.
	\end{align}
\end{subequations}
 Consequently, the constraints commute with any reasonable Hamiltonian in second quantisazion
 \begin{equation}
  \left[ H,A\right]= \left[ H,B_0\right]= \left[ H,\bs B\right]=0
 \end{equation}
and with the time evolution operator, i.e. a state on the physical subspace cannot propagate into an unphysical state.

\subsection{Operators in slave-boson formalism}\label{App:Operators}
This section summarizes important fermionic operators and their representation in slave-boson formalism.
\subsubsection{Spin density operator}
\begin{subequations}
	The spin density operator in fermionic language is given by
	\begin{equation}
		\hat{\bs S}=\frac{1}{2} \sum\limits_{\sigma \sigma '} c^{\dagger}_\sigma \boldsymbol \tau^\nodag_{\sigma \sigma '}c^{\nodag}_{\sigma '}
	\end{equation}
	Within slave-boson formalism, one finds
	\begin{align}\label{App:Operators:Spinoperator}
		\begin{split}
			\hat{\bs S}&= \frac{1}{2} \sum\limits_{\sigma\sigma ' \sigma_1 \sigma_2} z^{\dagger}_{\sigma \sigma_1} f^{\dagger}_{\sigma_1} \boldsymbol \tau^\nodag_{\sigma \sigma '} z^{\nodag}_{\sigma_2 \sigma '} f^{\nodag}_{\sigma_2} \\
			& =  \sum\limits_{\sigma \sigma ' \sigma_1} p^{\dagger}_{\sigma \sigma_1} \boldsymbol \tau^\nodag_{\sigma \sigma '}   p^{\nodag}_{\sigma_1 \sigma '}\\
			&=\frac{1}{4}\sum_{\mu\mu'=0}^3 \pd_{\mu'}\pn_{\mu}  \sum_{\sigma\sigma'\sigma_1}\tau^\mu_{\sigma_1\sigma'}\bs\tau^\nodag_{\sigma\sigma'}\tau^{\mu'}_{\sigma\sigma_1} \\
			&=\frac{1}{2}\left( p^{\dagger}_0\check{{\bs p}}+ \check{\bs p}^{\dagger} p_0 - i \check{\bs p}^{\dagger}\times  \check{{\bs p}}\right)
		\end{split}
	\end{align}
	with
	\begin{equation}
		\check{\bs p}=(p_1,-p_2,p_3)^T.
	\end{equation}
	It is easy to verify that this representation fulfills the spin algebra $\left[\hat{S}_i, \hat{S}_j \right]=i \epsilon_{ijk}\hat{S}_k$.
\end{subequations}
\subsubsection{Density operator}
\begin{subequations}
	The fermionic density operator of the interacting electrons is defined by
	\begin{equation}
		\hat{n}=\sum_{\sigma}c^\dagger_{\sigma} c^\nodag_{\sigma}.
	\end{equation}
	Replacing the $p$-bosons with the second constraint Eq.~\eqref{App:Constraints:Constraint2} yields 
	\begin{align}\label{App:Operators:Numberoperator}
		\hat{n}&=\sum_ {\sigma}  \fd_{\sigma}\ff_{\sigma},
	\end{align}
	i.e. the number of original fermions matches the number of pseudofermions. Using the constraints, it can also be written by means of slave-bosons only
	\begin{equation}\label{App:Operators:NumberoperatorBoson}
		\hat{n}=1+d^\dagger d-e^\dagger e.
	\end{equation}
\end{subequations}
\subsubsection{Hubbard interaction operator}
\begin{subequations}
	The Hubbard interaction is defined by
	\begin{equation}
		\hat{U}=Uc^\dagger_\uparrow c^\nodag_\uparrow c^\dagger_\downarrow c^\nodag_\downarrow.
	\end{equation}
	It is translated to slave-boson formalism by using the transformation property of the density operator Eq.~\eqref{App:Operators:Numberoperator}
	\begin{align}
		\begin{split}
			\hat{U}=U f^\dagger_\sigma f^\nodag_{\sigma} f^\dagger_{-\sigma} f^\nodag_{-\sigma}
			=Ud^\dagger d.
		\end{split}\label{App:Operators:Hubbard}
	\end{align}
	The Hubbard interaction becomes bilinear, which is why a consecutive mean field treatment is well adapted for strong coupling.
\end{subequations}\\

\section{Path Integral formulation of slave-boson formalism}\label{Chapter:App:Pathintegral}
Our goal is to derive the partition function which will be used to calculate thermodynamic quantities on mean field level and correlation functions by means of fluctuations around the mean field solution. It is given by the path integral over coherent states with imaginary time propagation \cite{negele1988quantum}
\begin{subequations}
	\begin{equation}
		Z= \int \mathcal{D}\left[f^*, f\right]\mathcal{D}\left[\psi^*,\psi\right] \e^{-\mathcal{S}\left[(f^*,f),(\psi^*,\psi)\right]}
	\end{equation}
	where
	\begin{equation}
		\mathcal{S}\left[(f^*,f),(\psi^*,\psi)\right]=\int_0^{\frac{1}{T}}  \mathcal{L}\left[(f^*_\tau,f^\nodag_\tau),(\psi^*_\tau,\psi^\nodag_\tau)\right] \ d\tau
	\end{equation}
\end{subequations}
is the action and $\mathcal{L}$ the Lagrangian. In the path integral, the operators are replaced by their coherent state eigenvalues, which are complex (Grassmann) numbers for the slave-bosons (pseudofermions) represented by $\psi_\tau$ ($f_\tau)$ at imaginary time $\tau$.
Moreover, $T$ is the temperature and $\mathcal{D}\left[\psi^*, \psi\right]$  ($\mathcal{D}\left[f^*, f\right]$) represents the integration over all field configurations.

The constraints can be enforced by means of the projectors defined in \eq{App:Constraints:Projectors}. Since the they commute with the Hamiltonian on operator level, the physical subspace is recovered with the following
effective Lagrangian, featuring time independent Lagrange multipliers.
\begin{widetext}
	\begin{subequations}
		\begin{gather}
			Z_{\text{eff}}=\lim_{N\rightarrow \infty}\frac{1}{(2\pi T)^2}\frac{1}{(2\pi NT)^3}\int_{-\pi T}^{\pi T}\d\alpha \int_{-\pi T}^{\pi T}\d\beta_0 
			\iiint_{-\pi N T}^{\pi N T}\d^3\beta \int\mathcal{D}[f^*, f]\mathcal{D}[\psi^*,\psi] \e^{-\mathcal{S}_{\text{eff}}\left[(f^*,f),(\psi^*,\psi),\alpha,\beta_0,\bs \beta\right]}\\
			\text{with} \quad \mathcal{L}_{\text{eff}}=\mathcal{L}+\i (\alpha A+\beta_0 B_0 +\bs \beta \bs B)
		\end{gather}
	\end{subequations}
\end{widetext}

\subsection{Effective Lagrangian in momentum space}
A general one band Hubbard Hamiltonian is give by
\begin{widetext}
	\begin{subequations} \label{App:Def:Hamilton}
	\begin{gather}
				H =\sum_{ij} \sum_{\sigma\sigma'} \cd_{i,\sigma} t_{ij} \, \cc_{j,\sigma} 
			-\mu_0\sum_{i}\hat{n}_{i}
			+U\sum_{i}c_{i,\uparrow }^\dagger c^\nodag_{i,\uparrow }c^\dagger_{i,\downarrow }c_{i,\downarrow }^\nodag \label{App:Hamilton_model} 
			\\=\sum_{\bk} \bs c^\dagger_{\bk} \underline{\mathcal{H}}_\bk \bs c^\nodag_\bk -\mu_0\sum_{i}\hat{n}_{i}
			+U\sum_{i}c_{i,\uparrow }^\dagger c^\nodag_{i,\uparrow }c^\dagger_{i,\downarrow }c_{i,\downarrow }^\nodag.\label{App:Hamiltonian_model_momentumspace}
		\end{gather}
		\end{subequations}
\end{widetext}
The tensor $t_{ij}$ may contain arbitrary hopping amplitudes. Moreover, we define the density operator $ \hat{n}_{i} = \sum_\sigma c^\dagger_{i,\sigma} c^\nodag_{i,\sigma}$, $\mu_0$ is the chemical potential and $U$ is the on-site Hubbard interaction strength. 

By Fourier transformation, the Hamiltonian can be rewritten as in \eq{App:Hamiltonian_model_momentumspace}
where  $\bs c_\bk\equiv(c_{1,\bk, \uparrow},c_{1,\bk, \downarrow})^T$ is to be understood as a $2$-dimensional spinor  and $\underline{\mathcal{H}}_\bk$ is the $2\times 2$ bare hopping matrix.

The Hamiltonian can be rewritten using the slave-boson representation given in \eq{App:Mapping:zf} and the representation of the operators given by \eq{App:Operators:Numberoperator} and \eq{App:Operators:Hubbard}. The effective Lagrangian within path integral formulation after Fourier transformation of $z^\dagger_i,\ f^*_i,\ z^\nodag_j$ and $\ f^\nodag_j$ is given by
\begin{widetext}
	\begin{equation}
		\begin{gathered}
			\mc L_{\text{eff}}[f,\psi]= \mc L_\mathrm{F}[f,\psi]+ \mc L_\mathrm{B}[\psi]	= \sum_{\bk_1,\bk_2} \bs f_{\bk_1}^\dag\left( \partial \tau+ \underline{H}_{\bk_1,\bk_2}[\psi]\right)\bs f_{\bk_2}^\nodag
			\\
			+ \sum_{i} \Bigg[ d^*_{i}(\partial_\tau + U)d^\nodag_{i} + e^*_{i}\partial_\tau e^\nodag_{i} + \bs p^*_{i}\partial_\tau \bs p^\nodag_{i}
			+i\alpha_{i}\left(e^*_{i}e^\nodag_{i}+p^*_{0,i}p^\nodag_{0,i}+\bs p^*_{i}\cdot \bs p^\nodag_{i}+d^*_{i}d^\nodag_{i}-1\right)
			\\
			-i\beta_{0,i}\left(p^*_{0,i}p^\nodag_{0,i}+\bs p^*_{i}\cdot\bs p^\nodag_{i}+2d^*_{i}d^\nodag_{i}\right)
			-i\bs\beta_{i}\cdot\left(p_{0,i}^* \bs p_{i}^\nodag+\bs p^*_{i} p_{0,i}^\nodag-\i\bs p^*_{i}\times \bs p_{i}^\nodag\right)\Bigg].
		\end{gathered}
	\end{equation}
\end{widetext}
 Above,  $\bs f^\dagger_\bk \equiv \left( f^*_{\bk,\uparrow}, f^*_{\bk,\downarrow} \right)$ represents 
 the collection of pseudo-fermionic fields and $\ul H_{\bk_1,\bk_2}[\psi]$ is defined as a slave-boson dependent hopping matrix
	\begin{widetext}
	\begin{equation}
		\begin{gathered}
			\ul H_{\bk_1,\bk_2}[\psi] \equiv
			 -\mu_0 \underline{\mathbb{1}}_{2}\delta^\nodag_{\bk_1,\bk_2}+ \sqrt{\frac{1}{N}}  \ \Big(\ul \beta \Big)^T_{\bs k_1-\bs k_2} 
		 +\frac{1}{N}\sum_\bk \left(\underline{z}^\dagger\right)^T_{\bs k-\bs k_1}
			\underline{\mathcal{H}}_{\bs k}
			\left(\underline{z}^\nodag\right)^T_{\bs k-\bs k_2}  \\
			\end {gathered} \label{App:Hkk}
		\end{equation}
	\end{widetext}
		where $\mathcal{H}_{\bs k}$ is the bare hopping matrix of the Hamiltonian without $\mu_0$ as defined in \eq{App:Hamilton_model} and $N$ the number of lattice sites moreover we define
		\begin{equation}
			\underline{\beta}\equiv\sum_{\mu=0}^3 \beta_\mu \ttau^{\mu} \label{App:Pathintegral:Beta}
		\end{equation}
		to enforce the pseudofermionic part of the constraints. \eq{App:Hkk} means descriptively, that every matrix element which will be multiplied with a pseudofermion is renormalized with a respective matrix element of $\underline{z}$ compared to the bare hopping matrix $\underline{\mathcal{H}}$, however the chemical potential $\mu_0$ is not renormalized. Moreover, the hopping matrix is complemented with Lagrange multipliers to enforce the fermionic parts of the constraints.

	\subsection{Gauge fixing}\label{Chapter:App:Gauge}
	In this section, it will be shown that by a gauge transformation, the phases of the $e,p_1,p_2$ and $p_3$ fields can be removed, which greatly simplifies the Lagrangian for the following calculations. The effective Lagrangian for the one band Hubbard model is given by
	\begin{widetext}
		\begin{subequations}
			\begin{gather}
				\mc L_{\text{int}} = \sum\limits_{ij} \sum\limits_{\sigma \sigma' \sigma_1 \sigma_2} \left(z^\dagger_i\right)_{ \sigma \sigma_1} f^*_{i, \sigma_1} t_{ij} f^\nodag_{j, \sigma_2} z^\nodag_{j, \sigma_2 \sigma} 
				+  \sum\limits_{i} \sum\limits_{\sigma, \sigma'} f^*_{i,\sigma} \left( \delta_{\sigma \sigma'} \left( - \mu_0 + \beta_{i,0} \right)  \right) f^\nodag_{i,\sigma'} \label{App:Gauge:L1}\\
				+i\sum_i\left[\alpha_{i}\left(e^*_{i}e^\nodag_{i}+2\sum\limits_{\sigma_1\sigma}\left(p^\dagger_i\right)_{\sigma_1\sigma}p^\nodag_{i,\sigma\sigma_1}+d^*_{i}d^\nodag_{i}-1\right)
				-i\sum\limits_{\mu=0}^3 \beta_{\mu,i} \sum_{\sigma\sigma'}\left( 2\sum_{\sigma_1} \left(p^\dagger_i\right)_{ \sigma_1 \sigma'} \tau^\mu_{\sigma' \sigma} p^\nodag_{i, \sigma \sigma_1}+d^* \tau^\mu_{\sigma' \sigma}d
				+f^*_{\sigma}\tau^\mu_{\sigma' \sigma}f^\nodag_{\sigma'}\right)\right]\label{App:Gauge:L2}\\
				+  \sum_{i}\left[ d^*_{i}(\partial_\tau + U)d^\nodag_{i}
				+e^{*}_i \partial_\tau e^{\nodag}_i + 2\sum\limits_{\sigma_1\sigma}\left(p^\dagger_i\right)_{\sigma_1\sigma}\partial_\tau p^\nodag_{i,\sigma\sigma_1}+\sum_\sigma f^*_\sigma \partial_\tau f^\nodag_\sigma\right]\label{App:Gauge:L3}.
			\end{gather}
		\end{subequations}
	\end{widetext}
	
	Now we rewrite the fields in radial description by means of their absolute value and a phase. The following transformations are applied on each site $i$ 
	independently, the site index $i$ will be dropped for readability
	\begin{subequations}
		\begin{align}
			e&=e^{i\theta}|e|,\\
			e^*&=e^{-i\theta}|e|, \\
			d&=e^{i\Phi}|d|, \\
			d^*&=e^{-i\Phi}|d|, \\
			p_{\sigma \sigma'} &\equiv \sum_{\sigma_1}e^{i\chi_{0}} U_{\sigma \sigma_1 } q_{\sigma_1 \sigma'}, \\
			\left(p^\dagger\right)_{\sigma \sigma'} &\equiv \sum_{\sigma_1} e^{-i \chi_{0}} q_{\sigma \sigma_1} \left(U^\dagger\right)_{\sigma_1 \sigma'},\\
			U_{\sigma \sigma'}&\equiv e^{i\sum\limits_{\alpha=1}^3 \chi_{\alpha} \tau^{\alpha}_{\sigma \sigma' }},
		\end{align}
	\end{subequations}
	where $q^\nodag_{\sigma\sigma'}$ is defined as the phaseless $p-$matrix
	\begin{equation}\label{eq:q_Matrix}
		q_{\sigma \sigma'} \equiv \frac{1}{2} \sum_{\mu = 0}^{3} q_\mu \tau^{\mu}_{\sigma \sigma'},   \qquad q_\mu \in \mathbb{R}.
	\end{equation}
	With these definitions and \eq{App:Mapping:pTilde} which also holds for the $q_{\sigma \sigma'}$-matrix, we can calculate the transformation properties of the matrix $ \tilde{p} $ ,
	\begin{subequations}
		\begin{align}
			\tilde{p}_{\sigma\sigma'}  
			&= e^{i \chi_{0}} \tilde{q}_{\sigma\sigma''} \left(U^\dagger\right)_{\sigma'' \sigma'},\\
			\left(\tilde{p}^\dagger\right)_{\sigma\sigma'}& = e^{-i \chi_{0}} U_{\sigma \sigma''} \tilde{q}_{\sigma''\sigma'},
		\end{align}
	\end{subequations}
	which can be shown, using the identity
	\begin{equation}
		\begin{gathered}\label{App:Gauge:PauliExp}
			U_{\sigma \sigma'}=\delta_{\sigma \sigma'} \cos \chi +i\sum_{\mu=1}^3 \tau^{\mu}_{\sigma \sigma'} n_{\mu}\sin \chi\\
			\text{with} \quad \chi \equiv \sqrt{\chi_1^2+\chi_2^2+\chi_3^2}\,  \qquad n_\mu \equiv\frac{\chi_{\mu}}{\chi} .
		\end{gathered}
	\end{equation}
	Now, we apply the following  SU$(2)\ \otimes  \ $U$(1)$ gauge transformation for the pseudofermions and U$(1) $ gauge transformation for the $ d $ and $ p $ bosons 
	\begin{subequations}\label{eq:gauge}
		\begin{align}
			f_{\sigma} &\rightarrow  e^{-i \chi_{0}} f_{\sigma'} \left(U^\dagger\right)_{\sigma' \sigma}, \\
			f^*_{\sigma} &\rightarrow  e^{i \chi_{0}}  U^\nodag_{ \sigma \sigma'} f^*_{\sigma'},\\
			d &\rightarrow e^{i(\theta + 2\chi_{0})} d, \\
			p_{\sigma \sigma'} &\rightarrow e^{i\theta} p_{\sigma \sigma'}, \\
			\tilde{p}_{\sigma \sigma'} &\rightarrow e^{i\theta} \tilde{p}_{\sigma \sigma'}.
		\end{align}
	\end{subequations}
	Since the Jacobi determinant of this unitary transformation is equal to one, the fields in the effective Lagrangian can simply be replaced by the gauge fields, leaving the partition function invariant.
	
	Now we look at the transformation properties of the Lagrangian term by term, beginning with \eq{App:Gauge:L1}. For the hopping term, all phases, except for the phase of the d-field are gauged away
	\begin{subequations}
		\begin{align}
			\left(z^\dagger\right)_{\sigma \sigma'} f^*_{\sigma'}& \rightarrow  q^\nodag_{\sigma \sigma'} |e| f^*_{\sigma'} + d^* \tilde{q}^\nodag_{\sigma \sigma'} f^*_{\sigma'}, \\
			f^\nodag_{\sigma'}  z^\nodag_{\sigma' \sigma} & \rightarrow f^\nodag_{\sigma'} |e| q^\nodag_{\sigma' \sigma} +  f^\nodag_{\sigma'} \tilde{q}^\nodag_{\sigma' \sigma} d.
		\end{align}
	\end{subequations}
	The pseudo-fermionic onsite terms remain invariant.
	
	Next we investigate the constraints (\eq{App:Gauge:L2}). For the first constraint, all fields except for the d-field simply lose their phase information
	\begin{align}
		\begin{split}
			i\alpha\left(e^* e^\nodag+2\sum\limits_{\sigma_1\sigma}p^\dagger_{\sigma_1\sigma}p^\nodag_{\sigma\sigma_1}+d^* d^\nodag-1\right)\\
			\rightarrow \quad  i\alpha\left(\big{|}e^2\big{|}+2\sum\limits_{\sigma_1\sigma}q^\nodag_{\sigma_1\sigma}q^\nodag_{\sigma\sigma_1}+d^* d^\nodag-1\right).
		\end{split}
	\end{align}
	The second constraint Eq.~\eqref{App:Constraints:Constraint2} in the new variables reads
	\begin{align}
		\begin{gathered}\label{App:Gauge:Constraint2}
			\sum_{\sigma_2 \sigma_3}e^{i \chi_0} U^\nodag_{\sigma \sigma_2} f^*_{\sigma_2 } e^{-i \chi_0} f^\nodag_{\sigma_3} \left(U^\dagger\right)_{\sigma_3 \sigma'}  \\
			=2\sum\limits_{\sigma_1\sigma_2\sigma_3} q_{\sigma_1 \sigma_2} \left(U^\dagger\right)_{\sigma_2 \sigma'}U^\nodag_{\sigma \sigma_3} q^\nodag_{\sigma_3 \sigma_1}  + \delta^\nodag_{\sigma \sigma'} |d|^2.
		\end{gathered}
	\end{align}
	It needs to be expanded in the unitary rotated basis of Pauli matrices in order to obtain four scalar equations which simplify the Lagrangian in the new gauge. Applying 
	$\sum_{\sigma\sigma'}U^\nodag_{\sigma'  \sigma_4} \tau^\mu_{\sigma_4\sigma_5} \left(U\right)^\dagger_{\sigma_5 \sigma}$ 
	with $\mu \in \lbrace0,1,2,3\rbrace$ 
	on both sides of \eq{App:Gauge:Constraint2} yields
	the transformation properties of the second constraint. After tracing out the Pauli matrices associated with the slave-bosons in the new gauge, one finds
	\begin{widetext}
		\begin{align}
			\begin{gathered}
				-i\sum\limits_{\mu=0}^3 \beta_{\mu} \sum_{\sigma\sigma'}\left( 2\sum_{\sigma_1} \left(p^\dagger\right)_{\sigma_1 \sigma'} \tau^\mu_{\sigma' \sigma} p^\nodag_{ \sigma \sigma_1}+d^* \tau^\mu_{\sigma' \sigma}d
				+f^*_{\sigma}\tau^\mu_{\sigma' \sigma}f^\nodag_{\sigma'}\right)\\
				\rightarrow \sum\limits_{\sigma \sigma'} f^*_{\sigma} \left( i\beta_{0} +  \sum\limits_{\mu=1}^3 i\beta_{\mu} \tau^\mu_{\sigma' \sigma} \right) f^\nodag_{\sigma'} 
				- i\beta_{0} \left( \sum_{\mu=0}^3 q_\mu^2 +2d^* d^\nodag\right)
				-\sum_{\mu=1}^3 i\beta_{\mu}2q_{\mu}q_{0}.
			\end{gathered}
		\end{align}
	\end{widetext}
	The $p-$fields again loose their phase information while the rest remains invariant. The cross product $\bs p^\dagger \times \bs p^\nodag$ which occurred in the 
	vector constraint \eq{App:Constraints:Constraint22} vanishes as the phases are removed.
	
	Now we investigate the time derivative terms of the Lagrangian \eq{App:Gauge:L3}. Note, that total derivatives like $|e|\partial_\tau |e|$ vanish because of the periodic boundary conditions of the path integral
	\begin{widetext}
		\begin{equation}
			\begin{gathered}
				d^*(\partial_\tau + U)d^\nodag+e^* \partial_\tau e^{\nodag} + 2\sum\limits_{\sigma_1\sigma}\left(p^\dagger\right)_{i,\sigma_1\sigma}\partial_\tau p^\nodag_{\sigma\sigma_1}+
				\sum_\sigma f^*_\sigma \partial_\tau f^\nodag_\sigma\\
				\rightarrow 
				i(\dot{\theta}+2\dot{\chi_0}+U)d^* d+i\dot{\theta}\big{|}e^2\big{|}
				+\sum_{\sigma\sigma_1\sigma_2\sigma_3}\left[-2i(\dot{\chi}+\dot{\theta})q^{\nodag}_{\sigma_1 \sigma}q^{\nodag}_{\sigma \sigma_1}+\dot{U}^{\nodag}_{\sigma_3\sigma_1}\left(U^{\dagger}\right)_{\sigma_1\sigma_2}q^\nodag_{\sigma_2\sigma}q^\nodag_{\sigma\sigma_3}\right]\\
				+\sum_{\sigma\sigma_1 \sigma_2}\left[f^*_\sigma \partial_\tau \ff_\sigma-i\dot{\chi_0}f^*_\sigma \ff_\sigma
				+f^*_{\sigma_1 }(\dot{U}^{\dagger})_{\sigma_2 \sigma}U^{\nodag}_{\sigma \sigma_1} \ff_{\sigma_2}\right].\\
			\end{gathered}
		\end{equation}
	\end{widetext}
	The time derivative of the unitary matrix can be evaluated with Eq.~\eqref{App:Gauge:PauliExp}. The terms containing time derivatives can be further simplified
	\begin{widetext}
		\begin{subequations}
			\begin{align}
				\sum_{\sigma\sigma_1\sigma_2\sigma_3}\dot{U}^{\nodag}_{\sigma_3\sigma_1}\left(U^{\dagger}\right)_{\sigma_1\sigma_2}q^\nodag_{\sigma_2\sigma}q^\nodag_{\sigma\sigma_3}&=i q_0 \sum_{\mu=1}^3 q_{\mu}\left( n_{\mu}\dot{\chi}+\dot{n}_{\mu}\sin\chi \cos\chi -\sum_{ij}\epsilon^{\mu i j} \dot{n}_i n_j \sin^2\chi\right)\\
				\sum_{\sigma}\left(\dot{U}^\dagger\right)_{\sigma_2\sigma}U^\nodag_{\sigma\sigma_1}&=-i\sum_{\mu=1}^3\left( n_{\mu}\dot{\chi}+\dot{n}_{\mu}\sin\chi \cos\chi -\sum_{ij}\epsilon^{\mu i j} \dot{n}_i n_j \sin^2\chi\right)\tau^\mu_{\sigma_2\sigma_1}.
			\end{align}
		\end{subequations}
	\end{widetext}
	
	Using all previous results, terms containing the phase factors $\chi_0$, $\bs \chi$ and $\theta$ can be absorbed in the Lagrange multipliers by
	\begin{widetext}
	\begingroup
	\allowdisplaybreaks
		\begin{subequations}
			\begin{gather}
				\left(i\alpha+i\dot{\theta}\right)\rightarrow i\alpha\\
				\left(i\beta_0+i\dot{\chi}_0\right)\rightarrow i\beta_0\\
				\left(i\beta_\mu-i\left( n_{\mu}\dot{\chi}+\dot{n}_{\mu}\sin\chi \cos\chi -\sum_{ij}\epsilon^{\mu i j} \dot{n}_i n_j \sin^2\chi\right)\right)\rightarrow i\beta_\mu.
			\end{gather}\label{App:Gauge:absorb}
		\end{subequations}
	\endgroup
	\end{widetext}
	The Lagrange multipliers are now formally time dependent and are considered as Lagrange multiplier fields. 
	The resulting Lagrangian in the new gauge is much simplified, since all bosonic fields except for the $d-$field are real valued
	\begin{widetext}
		\begin{equation}
			\begin{gathered}\label{App:Gauge:LR}
				\mc L \rightarrow \sum\limits_{ij} \sum\limits_{\sigma \sigma' \sigma_1 \sigma_2}\left(z^\dagger_i\right)_{ \sigma \sigma_1} f^*_{i, \sigma_1} t_{ij} f^\nodag_{j, \sigma_2} z^\nodag_{j, \sigma_2 \sigma'} 
				+  \sum\limits_{i} \sum\limits_{\sigma, \sigma'} f^*_{i,\sigma} \left( \delta_{\sigma \sigma'} \left( - \mu_0 + \beta_{0,i} \right) +  \sum\limits_{\mu=1}^3 \beta_{\mu,i} 
				\tau_{\sigma' \sigma}^\mu \right) f^\nodag_{i,\sigma'} \\
				+\sum_i\left[d^*_{i}(\partial_\tau + U)d^\nodag_{i}+i\alpha_{i}\left(\big{|}e^2_{i}\big{|}+\sum_{q=0}^3 q_{\mu,i}^2+d^*_{i}d^\nodag_{i}-1\right)
				-i\beta_{0,i}\left(\sum_{\mu=0}^3q_{\mu,i}^2+2d_i^* d^\nodag_i\right)-i\sum\limits_{\mu=0}^3 \beta_{\mu,i}2q_{\mu,i} q_{0,i}\right].\\
			\end{gathered}
		\end{equation}
	\end{widetext}
	In the following notation, we will go back to the $p$-fields notation rather than $q$ and it is implied that these fields are phaseless, but identically to their original definition in terms of physical interpretation
	since only redundant information has been removed by the gauge transformation.
	After Fourier transformation of the hopping and on-site terms in Eq.~\eqref{App:Gauge:LR}, including the non-interacting part of the Lagrangian one finds
	\begin{widetext}
		\begin{equation}
			\begin{gathered}\label{App:Gauge:LK}
				\mc L_{\text{eff}}[f,\psi]= \mc L_\mathrm{F}[f,\psi]+ \mc L_\mathrm{B}[\psi]
				\rightarrow \sum_{\bk_1,\bk_2} \bs f_{\bk_1}^\dag\left( \partial \tau+ \underline{H}_{\bk_1,\bk_2}[\psi]\right)\bs f_{\bk_2}^\nodag
			\\				+ \sum_{i} \Bigg[ d^*_{i}(\partial_\tau + U)d^\nodag_{i} 
				+i\alpha_{i}\left(\big{|}e^2_{i}\big{|}+p^2_{0,i}+\bs p^2_{i}+d^*_{i}d^\nodag_{i}-1\right)
				-i\beta_{0,i}\left(p^2_{0,i}+\bs p^2_{i}+2d^*_{i}d^\nodag_{i}\right)
				-i\bs\beta_{i}\cdot 2 p_{0,i}^\nodag \bs p_{i}^\nodag\Bigg].
			\end{gathered}
		\end{equation}
	\end{widetext}
	Note that when calculating the partition function in this gauge, one needs to replace the integration measure $d\psi^* d\psi$ by the radial expression $d\psi^2$ for the real valued fields.
	It turns out that the removal of the phase variables is necessary in order to have a well defined path integral as will be discussed for the atomic limit later on. Whenever a 
	physical field, e.g a fermion field, is represented by a product of two (complex valued) slave-boson fields,  an additional degree of freedom is necessarily introduced,
	namely the relative phase of the two fields. The final result should not depend on the choice of this phase, consequently these spurious phases have to be removed by fixing the gauge
	to avoid double counting in the path integral.	
	
	\subsection{Spin interaction}\label{Chapter:App:Path:Spin}
	In the new gauge, the spin density vector takes a much simpler form, since $p$ is a real field. Using \eq{App:Operators:Spinoperator}, one finds
	\begin{subequations}
		\begin{equation}\label{App:Gauge:Spinoperator}
			\hat{\bs{S}} \rightarrow  \check{\bs p} p_0
		\end{equation}
		with
		\begin{equation}
			\check{\bs p}=(p_1,-p_2,p_3)^T. 
		\end{equation}
	\end{subequations}
	Therefore, it is very convenient to add spin interaction terms to the Lagrangian. Note, that since the cross product has been gauged away, the spin density vector in pseudofermions fields is now equivalent to the spin density vector in 
	original fermions within path integral formalism according to \eq{App:Constraints:Constraint22}
	\begin{equation}\label{App:Path:Spin:pseudofermion}
	\hat{\bs S} \rightarrow \frac12 c^*_{\sigma}\bs \tau_{\sigma\sigma'} c^\nodag_{\sigma'}= \frac12 f^*_{\sigma}\bs \tau_{\sigma\sigma'} f^\nodag_{\sigma'}. 
	\end{equation}
	\subsubsection{External magnetic field}
	An external magnetic field, coupling to the spin density vector can expressed as purely bosonic term
	\begin{equation}\label{App:Path:Spin:B}
		\hat{B}:= \bs B\sum_i \hat{\bs S}_i\rightarrow \bs B\sum_i\check{\bs p}_i p_{0,i},
	\end{equation}
	or alternatively represented with pseudofermions by means of \eq{App:Path:Spin:pseudofermion}
	\subsubsection{Spin-Spin interaction}\label{Chapter:App:Path:SpinSpin}
	A spin-spin interaction of the form
	\begin{subequations}
		\begin{equation}
			\hat{J}:= J \sum_{<ij>}\hat{\bs S}_i\hat{\bs S}_j
		\end{equation}
		can also be represented by slave-bosons
		\begin{align}\label{App:Path:Spin:J}
			\hat{J}\rightarrow \sum_{<ij>}J\sum_\mu\check{p}_{\mu,i} p_{0,i} \check{p}_{\mu,j} p_{0,j} 
		\end{align}
	\end{subequations}

	\subsection{Atomic limit}\label{Chapter:App:Atomic}
	In the following, we will calculate the exact partition function for the slave-boson Lagrangian in the atomic limit within path integral formulation. Thermodynamics dictates the result to be
	\begin{equation}
		Z=1+\e^{-U/T+2\mu/T}+2\e^{\mu/T}
	\end{equation}
	since we consider only one interacting orbital at one site in the atomic limit. We apply the Lagrangian after the gauge transformation given by \eq{App:Gauge:LR} and rewrite it in terms of matrices
	\begin{subequations}
	\begin{equation}
		\mathcal{L}= i\alpha \big{|}e^2\big{|} +d^* (U+\partial_\tau +\i\alpha +2\i\beta_0) d -\i\alpha +  \bs p^T  \underline{\mathcal{B}}\ \bs p +\bs \fd  \underline{\mathcal{F}} \bs \ff
	\end{equation}
	with
	\begin{align}
		\begin{split}
			\bs p &\equiv
			\begin{pmatrix}
				p_0 \\ p_1 \\ p_2 \\ p_3
			\end{pmatrix},
			\quad 
			\underline{\mathcal{B}}\equiv
			\begin{pmatrix}
				i(\alpha+\beta_0) &i\beta_1& i\beta_2  &i\beta_3 \\
				i\beta_1 & i(\alpha+\beta_0) & 0 & 0 \\
				i\beta_2 &0 & i(\alpha+\beta_0) & 0\\
				i \beta_3 & 0 & 0 & i(\alpha+\beta_0)
			\end{pmatrix},
			\\
			\bs f &\equiv 
			\begin{pmatrix}
				f_{\uparrow} \\
				f_{\downarrow}  
			\end{pmatrix},
			\quad 
			\underline{\mathcal{F}}\equiv 
			\begin{pmatrix}
				\partial_\tau - \mu -\i (\beta_0 +\beta_3) & -\i \beta_1-\beta_2 \\
				-\i \beta_1+\beta_2 & \partial_\tau - \mu -\i (\beta_0 -\beta_3)
			\end{pmatrix}.
		\end{split}
	\end{align}
	\end{subequations}
	
	The effective Lagrangian in the atomic limit only contains bilinears and consequently the fields can be integrated out analytically. With the knowledge of 
	generalized Gaussian (Grassmann) Integrals, one finds for a bilinear Lagrangian
	\begin{subequations}
		\begin{equation}
			\mathcal{L}_0= a^* \partial_\tau a^\nodag+\epsilon  a^*a^\nodag  
		\end{equation}
		that the partition function is given by
		\begin{equation}\label{App:Atomic:Z0}
			Z_0=\left(1-\zeta \e^{-\epsilon/T}\right)^{-\zeta} \qquad
			\begin{cases}
				a \text{ bosonic, } \zeta=1 \\
				a \text{ fermionic, } \zeta=-1.
			\end{cases}
		\end{equation}
	\end{subequations}
	\eq{App:Atomic:Z0} also holds for real fields where the time derivative vanishes because of the periodic boundary conditions of the path integral. Even though the Lagrange multipliers are formally
	time dependent in the fixed gauge, it is sufficient to enforce the constraints only at one time slice, since physical states cannot propagate out of the physical subspace, which means that we can choose them to be time independent.
	To integrate out the fields, one needs to diagonalize the matrices $\mathcal{ B}$ and $\mathcal{F}$, whose eigenvalues
	are given by
		\begin{align}
		\begin{split}
			\mathcal{B}_1&=\i \alpha+ i\beta_0\\
			\mathcal{B}_2&=\i \alpha+ i\beta_0\\
			\mathcal{B}_3&=\i \alpha+ i\beta_0 +\i |\bs \beta | \\
			\mathcal{B}_4&=\i \alpha+ i\beta_0 -\i |\bs \beta |\\
			\mathcal{F}_1&=\partial_\tau -\i \beta_0-\mu_0+\i |\bs \beta | \\
			\mathcal{F}_2&=\partial_\tau -\i \beta_0-\mu_0-\i |\bs \beta | 
		\end{split}
		\end{align}
	where $|\bs \beta|\equiv\sqrt{\beta_1^2+\beta_2^2+\beta_3^2}$. Integrating out the fermionic Grassmann fields $\bs f$ and the bosonic fields $e,d,\bs p$ with \eq{App:Atomic:Z0}, one finds
	\begin{widetext}
		\begin{equation}
			\begin{split}
				Z=\lim_{N\rightarrow \infty}\frac{1}{(2\pi T)^2}\frac{1}{(2\pi N T)^3}\int_{-\pi T }^{\pi T }\d\alpha \int_{-\pi T}^{\pi T}\d\beta_0 \iiint_{-\pi N T}^{\pi N T}\d^3\beta
				\ \e^{i\alpha /T}\left(1-\e^{-\i \alpha /T}\right)^{-1}\left(1-\e^{-(i\alpha+2i\beta_0+U)/T}\right)^{-1}\times\\
				\left(1-\e^{-\i (\alpha+\i \beta_0)/T}\right)^{-2}\left(1-\e^{-\i( \alpha+\i \beta_0+\i |\bs\beta|)/T}\right)^{-1}\left(1-\e^{-(\i \alpha+\i \beta_0-\i |\bs\beta|)/T}\right)^{-1}\left(1+\e^{(i\beta_0+\i |\bs\beta|+\mu)/T}\right)\left(1+\e^{(i\beta_0-\i |\bs\beta|+\mu)/T}\right)
			\end{split}
		\end{equation}
	\end{widetext}
	and is left with the integrals over the Lagrange multipliers. The $\alpha$-integral can be mapped on a complex contour integral by making use of the fact that the projectors (\eq{App:Path:Projectors}) are invariant
	when adding an imaginary part to the Lagrange multiplier $\alpha \rightarrow \bar{\alpha}+\i \tilde{\alpha}$. The substitution 
	\begin{equation}
		\xi\equiv \e^{-\i \alpha/T} \qquad \int d\alpha\rightarrow T\oint \frac{i}{\xi}d\xi
	\end{equation}
	leads to a contour integral around the origin with radius $\e^{\tilde{\alpha}}$ Since $\tilde{\alpha}$ can be chosen arbitrary small, the integral is determined by the residuum at the origin $\xi=0$ which can be found 
	by expanding the integrand as a geometric series. The $\beta_0$ integral can be carried out in the same way, which finally yields
	\begin{equation}
		\begin{gathered}
			Z=1+\e^{-(U-2\mu)/T}+2\e^{\mu/T}\\
			+\e^{\mu/T} \lim_{N\rightarrow \infty}\frac{1}{(2\pi N)^3}\iiint_{-\pi N }^{\pi N }\d^3\tilde\beta  \left(2\cos(2| \tilde{\bs\beta}|)+4\cos(|\tilde{\bs\beta}|)\right),
		\end{gathered}
	\end{equation}
	where $\tilde{\bs\beta}=\bs \beta/T$. The remaining integral is equal to zero in the limit, since it is of the order $\mathcal{O}(\tilde{\bs\beta}^2)$ while being suppressed by $1/N^3$ by the normalization. 
	Consequently the path integral description yields the same result as expected
	\begin{equation}
		Z=1+\e^{-(U-2\mu)/T}+2\e^{\mu/T}.
	\end{equation}
	Note that if the atomic limit is calculated before the gauge transformation discussed in \autoref{Chapter:App:Gauge}, one finds the false result $1+\e^{-(U-2\mu)/T}+4\e^{\mu/T}$ for the partition function. This is because of over counting introduced by the cross product $\bs p^* \times \bs p^\nodag$ due to spurious fields if the gauge is not fixed.

	\section{Paramagnetic Mean field}\label{Chapter:App:MF}
	We now investigate the paramagnetic mean field solution of the Lagrangian \eq{App:Gauge:LK}. As approximation, the spacial and time dependent slave-boson
	fields are replaced by a static, uniform expectation value $\psi_i \rightarrow \langle \psi \rangle$ with $\partial\tau \langle \psi \rangle=0$.
	Since the Hamiltonian is hermitian, the eigenvalues of the pseudofermionic part of the Lagrangian only depend on $\langle \psi \rangle \langle \psi^* \rangle$ which is also true for the bosonic part. 
	Consequently, $\langle \psi \rangle$ and $\langle \psi^*\rangle $ have the same saddle point equations which means that $\langle \psi \rangle $ is real, as we would expect.
	
	Since the Lagrange multipliers cannot be integrated out analytically, they will also be included in the mean field. As we have seen, the Lagrange multipliers can be chosen complex since the projectors are invariant under $\alpha \rightarrow \alpha+\i \bar{\alpha}$.
	In order to find a real Free valued energy, we assign them to be purely imaginary and uniform such that the constraints are enforced exactly at saddle point of the mean filed equations.
		
	The paramagnetic mean field is further defined with a vanishing expectation value of the spin density vector 
	\eq{App:Gauge:Spinoperator}, which is found by $\langle \bs p \rangle \equiv 0$. Consequently, it is also $\langle \bs \beta\rangle =0$,
	because otherwise the bands would not be spin degenerate and the pseudofermionic representation of the spin density vector would not yield a vanishing expectation value.
	
	All paramagnetic mean field assumptions are summarized by
		\begin{align}
		\begin{split}
			\bs p_i &\rightarrow \langle \bs p \rangle :=0,\\
			\bs \beta_i &\rightarrow \langle \bs \beta\rangle  :=0,\\
			d_i &\rightarrow \langle d\rangle  \in \mathbb{R}_0^+, \ \partial_\tau \ \langle d\rangle  :=0,\\
			p_{0,i}&\rightarrow \langle p_0 \rangle \in \mathbb{R}_0^+, \\
			e_{i}&\rightarrow \langle e \rangle \in \mathbb{R}_0^+, \\
			i\alpha_{i}&\rightarrow \langle\alpha\rangle \in  \mathbb{R}, \\
			i\beta_{0,i}&\rightarrow \langle\beta_0\rangle \in  \mathbb{R}.
			\end{split}
		\end{align}
	In the following, the brackets $\langle \rangle$ will be droped for readability. 
		\subsection{Non interacting limit}
	Because of the constraints, there is a considerable freedom in choice of the slave-boson representation, leaving the exact solution unchanged, but having an immense impact on the mean field solution.  
	We choose the following renormalization \cite{woelfle_spin_1992_Errata} 
	\begin{subequations}\label{App:Meanfield:Noninteracting:z}
		\begin{align}
			\zz \rightarrow (e^\dag\underline{L} M\underline{R} \ \pp+\tilde\pp^\dag \underline{L}M\underline{R} d), \label{eq:z_0}
		\end{align}
		with
		\begin{align}
			\underline{L} &= \left((1-d^\dag d)\ttau_0-2 \pp^\dag \pp\right)^{-1/2}, \\
			M &= \left(1 + d^\dag d +e^\dag e +\sum\limits_{\mu} p^\dag_\mu p_\mu^\nodag \right)^{1/2}, \\
			\underline{R} &= \left((1-e^\dag e)\ttau_0-2 \tilde{\pp}^\dag \tilde{\pp}\right)^{-1/2}.
		\end{align}
	\end{subequations}
	\eq{App:Meanfield:Noninteracting:z} can be expanded in a power series and it appears that all additional terms compared to the bare definition of the slave-boson representation in 
	\eq{App:Mapping:z} exhibit two annihilators to the very right of the equation. Consequently these terms annihilate every state on the physical subspace enforced by the constraints and the exact solution remains unchanged.

	For the paramagnetic mean field, we find
	\begin{align}
		\underline{z}_0&=\frac{p_0(e+d)}{\sqrt{2(1-d^2-p_0^2/2)(1-e^2-p_0^2/2)}}\ttau^0\equiv z_0 \ttau^0. \label{App:Meanfield:z0}
	\end{align}
	One can infer from \eq{App:Hkk}, that hopping terms between different sites of the interacting orbital are renormalized by $t\rightarrow z_0^2 t$. In the limit of no interaction, there should not be a renormalization effect on the band structure, consequently we demand $z_0=1$ for $U=0$ which is true for any occupation $e, p_0, d$
	because of the following statistical argument:
	
	Without interaction, orbitals are occupied randomly by a probability $0\leq x \leq 1$. Consequently the probability that a site is doubly occupied is given by 
	$d^2=x^2$. The probability that a site is singly occupied is $p_0^2=2x(1-x)$ taking spin degeneracy into account. It follows $e^2=1-p_0^2-d^2=(1-x)^2$. Inserting these results into \eq{App:Meanfield:z0}
	yields $z_0=1$ as demanded.
	
	\subsection{Free Energy}
	The Free energy is given by
	\begin{equation}\label{App:Meanfield:Free:def}
		F=-T\ln Z+\mu_0  \mathcal{N},
	\end{equation}
	where $\mathcal{N}$ is the total number of electrons in the system.
	\begin{subequations}
		The Lagrangian in the paramagnetic mean field given by
		\begin{equation}
			\begin{gathered}
				\mc L_{0}= \sum_\bk \bs f^\dagger_\bk\left[\partial_\tau +\underline{H}_{\bk}[\psi]\right] \bs f_\bk\\
				+N \left[ U d^2+\alpha(e^2+p_{0}^2+d^2-1)-\beta_{0}(p_{0}^2+2d^2)\right],
			\end{gathered}
		\end{equation}
		with the mean field renormalized hopping matrix 
		\begin{equation}\label{App:Meanfield:Hamiltonian}
			\begin{gathered}
				\underline{H}_{\bk}[\psi]\equiv 
				 \underline{z_0}_{}
			 \underline{\mathcal{H}}_{\bs k}
			 \underline{z_0}_{} +
			 (\beta_0-\mu_0) \underline{\mathbb{1}}_2,
			\end{gathered}
		\end{equation}
	\end{subequations}
	The pseudofermions in the mean field Lagrangian can be integrated out with \eq{App:Atomic:Z0}. The slave-boson dependent spin degenerate eigenvalues of the matrix 
	$\underline{H}_{\bk}[\psi]$ are labeled by $\epsilon_{\bs k}$ in the following.
	The mean field free energy per lattice site is then found to be
	\begin{equation}
		\begin{gathered}
			f_0\equiv
			\frac{F_0}{N}=-T\frac{2}{N}\sum_{\bk} \log\left(1+\e^{-\epsilon_{\bk}/T}\right) \\
			\ +U d^2+\alpha(e^2+p_{0}^2+d^2-1)-\beta_{0}(p_{0}^2+2d^2)+ \mu_0 n,
		\end{gathered}
	\end{equation}
	where $n=\mathcal{N}/N$ is the total electron filling and $N$ is the number of lattice sites. Spin interactions like \eq{App:Path:Spin:B} or \eq{App:Path:Spin:J} do not change the paramagnetic mean field, however impact the fluctuations around the saddle point.
	\subsection{Saddle point equations}
	In order to find the mean field solution for the ground state, we need to minimize the free energy with respect to the fields $e,p_0,d$, while enforcing the constraints, which can be recovered by deriving the Free energy by the respective Lagrange parameter. The resulting saddle point equations are given by
	\begin{subequations}\label{App:Meanfield:Saddlepoint}
		\begin{align}
			\frac{\partial f_0}{\partial e}&=\frac{2}{N}\sum_{\bs k}n_F(\epsilon_{\bs k})\frac{\partial \epsilon_{\bs  k}}{\partial e} +2\alpha e=0, \label{App:Meanfield:Saddlepoint:e}\\
			\frac{\partial f_0}{\partial p_0}&=\frac{2}{N}\sum_{\bs k}n_F(\epsilon_{\bs k})\frac{\partial \epsilon_{\bs  k}}{\partial p_0} +2p_0(\alpha -\beta_0)=0, \label{App:Meanfield:Saddlepoint:p_0} \\
			\frac{\partial f_0}{\partial d}&=\frac{2}{N}\sum_{\bs k}n_F(\epsilon_{\bs k})\frac{\partial \epsilon_{\bs  k}}{\partial d} +2d(U+\alpha-2\beta_0)=0 , \\
			\frac{\partial f_0}{\partial \alpha}&=e^2+p_0^2+d^2-1=0,\label{App:Meanfield:Saddlepoint:alpha} \\
			\frac{\partial f_0}{\partial \beta_0}&=\frac{2}{N}\sum_{\bs k}n_F(\epsilon_{\bs k})\frac{\partial \epsilon_{\bs  k}}{\partial \beta_0} -2d^2-p_0^2=0, \label{App:Meanfield:Saddlepoint:beta0}  \\
			\frac{\partial f_0}{\partial \mu_0}&=-\frac{2}{N}\sum_{\bs k}n_F(\epsilon_{\bs k})+n=0,  \label{App:Meanfield:Saddlepoint:mu} 
		\end{align}
	\end{subequations}
	where $n_F(\epsilon_{\bs k})=(1+\exp(\epsilon_{\bs k}/T))^{-1}$ is the Fermi-Dirac distribution. The last equation has to be enforced additionally to ensure the correct electron filling, instead of fixing the chemical potential.
	\subsection{Reduction of mean field equations}\label{App:Meanfield:Numerical}
	It turns out that instead of solving the six saddle point equations given above, one can reduce the system to a set of only two independent equations. To do so, we substitute $\beta_0 = -\mu_{\text{eff}}+\mu_0$ and find
	\begin{equation}
		\begin{gathered}
			f_0=-T\frac{2}{N}\sum_{\bk} \log\left(1+\e^{-\epsilon_{\bk}/T}\right)
			+U d^2 + \mu_{\text{eff}}(p_{0}^2+2d^2) \\ \ +\alpha(e^2+p_{0}^2+d^2-1) + \mu_0 (n-p_{0}^2-2d^2),
			\label{App:Meanfield:Saddlepoint:f0mueff}
		\end{gathered}
	\end{equation}
	which means effectively, that we fix the filling by a purely bosonic constraint with Lagrange parameter $\mu_0$, since the eigenvalues $\epsilon_\bk$ now only depend on $ \mu_{\text{eff}}$ rather than $\beta_0$ and $\mu_0$. 
	
	We then exploit the the two constraints which only couple to bosonic degrees of freedom, i.e. the constraint which ensures, that there is only one boson per site associated with $\alpha$ and the constraint which fixes the total number of particles associated with $\mu_0$ by setting
	\begin{subequations}\label{App:Meanfield:Saddlepoint:constraints}
	\begin{align}\label{App:Meanfield:Saddlepoint:constraint1}
		1&=e^2+p_0^2+d^2 \\
		\label{App:Meanfield:Saddlepoint:constraint2}
		n&=p_0^2+2d^2\\
		\mu_{\text{eff}}&=\mu_0-\beta_0
	\end{align}
	\end{subequations}
This way, the redundant degrees of freedom $\alpha$, $\mu_0$ and two arbitrary slave-boson fields (we choose $p_0$, and $e$ without loss of generality) are removed from the mean field equations.
The mean field solution is given by the saddle point of the free energy
\begin{subequations}
 \begin{align}
 f_0 \big{|}_\eqref{App:Meanfield:Saddlepoint:constraints} &=
 -T\frac{2}{N}\sum_{\bk} \log\left(1+\e^{-\epsilon_{\bk}/T}\right)
			+U d^2 + n\mu_{\text{eff}} \\
			z_0^2 \big{|}_\eqref{App:Meanfield:Saddlepoint:constraints} &=
			\frac{2\left(\sqrt{d^2-n+1}+d\right)^2\left(2d^2-n\right)}{n(n-2)}.
 \end{align}
\end{subequations}
We are left to determine
	\begin{equation}
	 \frac{\partial f_0}{\partial d}\Bigg{|}_\eqref{App:Meanfield:Saddlepoint:constraints} = \frac{\partial f_0}{\partial \mu_{\text{eff}}}\Bigg{|}_\eqref{App:Meanfield:Saddlepoint:constraints}=0,
	\end{equation}
which we do by minimizing $f_0$ with respect to $d$ and maximizing with respect to $\mu_{\text{eff}}$ between each minimization step. To do so, we used gsl$\_$multimin.h in our numerical evaluation, which is available in the GNU Scientific Library

	On mean field level, $\mu_{\text{eff}}$ employs the role of the chemical potential on the constrained subspace.
	The original chemical potential is recovered by evaluating
	\begin{equation}
		\mu_0=\frac{1}{2 \mf p_0}\frac{2}{N}\sum_{\bs k}n_F(\epsilon_{\bs k})\frac{\partial \epsilon_{\bs  k}}{\partial p_0}\Bigg{|}_{\mf\psi,\eqref{App:Meanfield:Saddlepoint:constraint1}}+\mu_{\text{eff}}
	\end{equation}
	where $\mf \psi$ represents the slave-bosons at the saddle point solution. Note, that the previous equation is to be understood such, that only \eqref{App:Meanfield:Saddlepoint:constraint1} is applied to reduce the degrees of freedom to eliminate the $e$ field.
	
	There is an ambiguity whether to define the electronic compressibility via $\mu_0$ or $\mu_\text{eff}$. We choose the definition 
	\begin{equation}
	 	n^2 \kappa_T =\partial n /\partial \mu_\text{eff},
	\end{equation}
	because in this description, redundant fields have been removed, and not only $f_0$ but also $\delta f_0$ satisfies the constraints exactly at the saddle point solution.
		However, $\mu_0$ plays an important role for Gaussian fluctuations around the saddle point, which infinitesimally violate the constraints, i.e. $\delta f_0$ must not satisfy the constraints in that case.
	
	Analogously, one can formally calculate 	
	\begin{equation}
		\mf \alpha=-\frac{1}{2 \mf e}\frac{2}{N}\sum_{\bs k}n_F(\epsilon_{\bs k})\frac{\partial \epsilon_{\bs  k}}{\partial e}\Bigg{|}_{\mf\psi},
	\end{equation}
	albeit $\alpha$ has no physical meaning, neither on mean field, nor on fluctuation level.
	
	\subsection{Solution at half filling}\label{App:Meanfield:Numerical}
	For a half filled system the set of mean field equations read
	\begin{align}
	 0 &= \frac{2}{N}\sum_{\bs k}n_F(\epsilon_{\bs k}) - 1,\\
	 0 &= \left(16d-64d^3 \right)\frac{2}{N}\sum_{\bs k}n_F(\epsilon_{\bs k})\frac{\partial \epsilon_{\bs  k}}{\partial z_0^2} +2Ud.
	\end{align}
	In the limit of large system size and zero temperature one can solve these equations analytically for a square lattice with only nearest neighbor hopping 
	\begin{equation}
	 d = \begin{cases}
	      0 &\text{for} \quad U > \frac{128t}{\pi^2} \\
              \frac{1}{16}\sqrt{64-\frac{\pi^2U}{2t}} \, &\text{else},
	\end{cases}
	\end{equation}
	where $t$ is the hopping amplitude between the neighboring sites and $U > 0$. Notice, that the solution is defined with the global minimum of $f_0\big{|}_\eqref{App:Meanfield:Saddlepoint:constraints}$ with respect to $d$.
	For large $U$, all sites are occupied by exactly one electron, the system is now in the insulating Mott state. The chemical potential $\mu_0$ for large $U$ is then given by   
	\begin{equation}
	\lim_{\delta\to 0}\mu_0(\delta) =
	 \begin{cases}
	      0 &\text{for} \quad \delta > 0, \\
              U/2 \  &\text{for} \quad \delta = 0, \\
              U \  &\text{for} \quad \delta < 0, 
         \end{cases}
	\end{equation}
	where we defined $n=1-\delta$ \cite{woelfle_spin_1992}.

	\section{Fluctuations around the saddle point}\label{Chapter:App:Fluc}
	Now we consider fluctuations around the saddle point solution, which allow to calculate correlation functions such as the spin or charge susceptibility and provide a stability analysis.
	Since the first order variation of the action vanishes at the saddle point, the fluctuations are determined by the variation of the action to the second order.
	To do so, we apply apply the following Fourier transformation in space and time
	\begin{subequations}\label{App:Fluctuations:Fouriertransform}
		\begin{align}
			\psi_\mu(\bs x_i,\tau)&=\sqrt{\frac{T}{N}}\sum_{\bs q}\sum_{\omega_n}e^{i \bs q \bs x_i} e^{-i \omega_n \tau}\psi_\mu(\bs q, i\omega_n) \\
			\psi^*_\mu(\bs x_i,\tau)&=\sqrt{\frac{T}{N}}\sum_{\bs q}\sum_{\omega_n}e^{-i \bs q \bs x_i} e^{i \omega_n \tau}(\psi_\mu(\bs q,i \omega_n))^* \\
			\omega_n&=2\pi nT \qquad n \in \mathbb{Z}
		\end{align}
	\end{subequations}
	where $\omega_n$ is the bosonic Matsubara frequency and $\mu$, $\nu$ label the fields which are subject to fluctuations. In \autoref{Chapter:App:Gauge} it has been shown, that all slave-boson fields except for the $d$-field are real valued. We decompose it into its real and 
	imaginary part $d=d_1+id_2$ where $d_1$ and $d_2$ are independent fields. 
	Using \eq{App:Fluctuations:Fouriertransform}, we see that
	\begin{equation}
		(\psi_\mu(\bs q, i\omega_n))^*=\psi_\mu(-\bs q, -i\omega_n)
	\end{equation}
	holds for the fields in momentum space. The second variation of the action is given by
	\begin{subequations}\label{App:Fluctuations:Kernel}
		\begin{equation}
			\delta\mc S^{(2)}=\sum_{q}\sum_{\mu\nu}\delta \psi_\mu(-\bs q,-i\omega_n) \mathcal{M}_{\mu\nu}(q) \delta\psi_\nu(\bs q,i\omega_n)\\
		\end{equation}
	\end{subequations}
	with $\sum_q\equiv  \sum_{\bs q}\sum_{n}$ and $q\equiv (\bs q,i\omega_n)^T$. 
	where
	\begin{equation}
	 			\mathcal{M}_{\mu\nu}(\bs q,\omega_n)
			\equiv\frac12 \frac{\delta^2 \mathcal{S}(\psi)}{\delta \psi_\mu(-\bs q,-i\omega_n)\delta \psi_\nu(\bs q,i\omega_n)}
	\end{equation}
	defines the fluctuation matrix which satisfies $\mathcal{M}_{\mu\nu}(\bs q,i\omega_n)=\mathcal{M}_{\nu\mu}(-\bs q, -i\omega_n)$. 
		
Since the fluctuations are calculated by means of functional
derivatives, they violate the constraints which are exactly enforced only at the saddle point. Such a
violation is actually necessary in order to resolve correlations
and evaluate whether the system will relax back to the paramagnetic mean field solution or whether it features an instability. Since the Lagrange multipliers are part of the effective field theory, one needs to consider the fluctuation of $\beta_0$ which couples to the charge density (i.e. necessary to calculate charge fluctuations) and $\bs \beta$ which couples to the spin density vector (i.e. necessary to calculate spin fluctuations) as well.
However, $\alpha$ does not couple to any physical degree of freedom and its fluctuations would yield bosonic occupations per lattice site unequal to one, which can be associated with a violation of the Pauli principle. This needs to be avoided by replacing an arbitrary slave-boson field (we choose $p_0$ w.l.o.g) via \eq{App:Constraints:Constraint1}
	\begin{equation}
	 p_0=\sqrt{1-\bs p^2 -|d|^2 - e^2},
	\end{equation}
i.e. fluctuating on the subspace where the $\alpha$ constraint is exactly fulfilled.
Thereby, we reduce the number of independent fields by two and guarantee, that the derivatives are evaluated in the physical subspace. Moreover, 
since the Lagrangian only depends on $\beta_0$ and $\bs \beta$,
	but not on $\beta_0^*$ or $\bs \beta^*$, it is sufficient to fluctuate the real part of the Lagrange multiplier fields.

	We choose the following basis of real fields for the fluctuations: 
	$\psi_{1}=e$, $\psi_{2}=d_1$, $\psi_{3}=d_2$, $\psi_{4}=\beta_0$, $\psi_{5,6,7}=p_{1,2,3}$ and $\psi_{8,9,10}=\beta_{1,2,3}$. Notice, that we actually fluctuate with respect to $\beta_0=\mu_0-\mu_{\text{eff}}$ rather than $\mu_{\text{eff}}$
	because fluctuations in $\beta_0$ are induced by the interaction while fluctuations in $\mu_0$ are due to the external particle bath and are not featured in the expansion of Lagrangian.
	
	We numerically confirmed, that it does not matter which slave-boson field is eliminated via the $\alpha$-constraint. The results for the charge and spin susceptibility remain invariant, while single matrix elements of $\mathcal{M}_{\mu,\nu}$ are of course subject to change, depending on the substitution. Moreover, as elaborated in \autoref{sec:Results:ZeroT}, we find consistent magnetic phase boundaries comparing the divergence of the paramagnetic spin susceptibility and the results of a spiral magnetic field defined in \autoref{Chapter:App:MaME}.
	
 	Despite fluctuations in $\alpha$ erroneously have been considered in the slave-boson literature so far, our results are mostly in good agreement with previous slave-boson studies, 
 	as the term $M = \left(1 + d^\dag d +e^\dag e +\sum\limits_{\mu} p^\dag_\mu p^\nodag_\mu \right)^{1/2}$ in \eq{App:Meanfield:Noninteracting:z}, which causes the largest deviation, has been correctly set to $\sqrt{2}$ in earlier works.
		
	\subsection{Integration of the pseudofermions}
	We consider the partition function
	\begin{gather}
		Z=\int\mathcal{D}[f^*, f]\mathcal{D}[\psi^*,\psi] \e^{-S_{\text{eff}}\left[(f^*,f),(\psi^*,\psi)\right]}\label{App:Fluctuations:Integration:PF}
	\end{gather}
	with the action according to the Lagrangian given in \eq{App:Gauge:LK}. Note, that the Lagrange multipliers are included in the integration measure, since they and act as effective bosonic fields $\psi$ in the field theory. The fermionic fields appear bilinear, hence they can be integrated out analytically with the generalized Gaussian integral over Grassmann numbers
	\begin{equation}
			\int \mathcal{D}[ f, f^*]\exp\left(-\sum_{\mu\nu} f^*_\mu \mathcal{A}_{\mu\nu} f_\nu
			\right)
			=\det{\underline{\mathcal{{A}}}}.
	\end{equation}
	The integration is performed in momentum space, where the matrix $\underline{\mathcal{A}}$ is diagonal. By applying the identity 
	\begin{equation}
		\det{\underline{\mathcal{A}}}=\exp\left(\Tr\log\underline{\mathcal{A}}\right),
	\end{equation}
	where the trace is to be understood as a sum over all momenta, Matsubara frequencies and spins
		\begin{align}
			\Tr(\underline{\mathcal{A}}) \equiv \sum_{q} \tr\left(\underline{\mathcal{A}}_{q}\right)\, .
		\end{align}
		The partition function in \eq{App:Fluctuations:Integration:PF} can be rewritten as a purely bosonic functional integral by means of an an effective action $\mathcal{S}_F$ which yields from the
		integration of the fermionic degrees of freedom
	\begin{subequations}
		\begin{gather}
			Z=\int\mathcal{D}[\psi^*,\psi] \e^{-\mathcal{S}_F[\psi]}\e^{-\mathcal{S}_B[\psi]},
		\end{gather}
		with
		\begin{equation}\label{App:Fluctuations:Exact:Pseudofermionic}
			\mathcal{S}_F=-\Tr\log\left(-i\varpi_n+\ul H[\psi]_{k_1,  k_2}\right)
		\end{equation}
		and
		\begin{equation}
			\begin{gathered}\label{App:Fluctuations:Exact:Bosonic}
				\mathcal{S}_B=\int_0^{1/T}d\tau\sum_{i} \Bigg[d^*_{i}(\partial_\tau + U)d^\nodag_{i}\\
				-\beta_{0,i}\left(1+|d_{i}|^2-e^2_{i}\right)
				-\bs\beta_{i}\cdot2\bs p^\nodag_{i} \sqrt{1-e^2_{i}-\bs p^2_{i}-|d_{i}|^2}\Bigg].
			\end{gathered}
		\end{equation}
The slave-boson dependent hopping matrix
$\ul H_{k_1,  k_2}[\psi]$ is the Fourier transformation of $\ul H_{\bk_1,  \bk_2}[\psi]$ defined in \eq{App:Hkk} with respect to time, labeled by the the multi-index
$k=(\bs k, \varpi_n)$, where $\varpi_n=2\pi T\left(n+\frac12\right)$ with $n\in\mathbb{Z}$ is a fermionic Matsubara frequency.
	\end{subequations}
	
	\subsection{Bosonic part of the fluctuation matrix}
	The bosonic part of the fluctuation matrix, corresponding to $\mathcal{S}_B$, is calculated by Fourier transformation of \eq{App:Fluctuations:Exact:Bosonic} and subsequently deriving by the respective bosonic fields as
	exemplary done for the $\mathcal{M}^B_{1,1}(\bs q,\omega_n)$ element:
	\begin{equation}
		\begin{gathered}
			\mathcal{M}^B_{1,1}(\bs q,\omega_n)
			=\frac12 \sqrt{\frac{T}{N}} \frac{\sum_{q_1,q_2} \partial^2 e_{q_1}e_{q_2}\beta_{0,-q_1-q_2}}{\partial e_{-q}\partial e_q}\\
			= \sqrt{\frac{T}{N}} \sum_{q_1,q_2}\delta_{q,q_1}\delta_{q,-q_2}\beta_{0,-q_1-q_2}= \sqrt{\frac{T}{N}}\beta_{0,0}=\mf{\beta}_0,
		\end{gathered}
	\end{equation}
	where $\mf \beta_0$ refers to the uniform mean field solution in real space whereas $\beta_{0,0}$ represents the Fourier transform of $\beta_0$ at momentum $q=0$. The results coincide with directly deriving the mean field Lagrangian by the respective fields,
	except there is one additional contribution resulting from the property, that $d$ is a complex field in opposite to the other slave-boson fields and therefore features a non vanishing derivative with respect to time (see \autoref{Chapter:App:Gauge}):
	\begin{subequations}
		\begin{align}
			\mathcal{M}^B_{2,3}(\bs q,\omega_n)
			&=\omega_n,\\
			\mathcal{M}^B_{3,2}(\bs q,\omega_n)&=-\omega_n.
		\end{align}
	\end{subequations}
	
	For completeness, we consider additional spin interaction terms, defined in \autoref{Chapter:App:Path:Spin} with the according Lagrangian
	\begin{subequations}
		\begin{equation}
		 \begin{gathered}
			\mc L_{S}\equiv \sum_{<ij>} \sum_\alpha J^\alpha \check{p}_{\alpha,i} \sqrt{1-e^2_{i}-\bs p^2_{i}-|d_{i}|^2} \\ \quad \quad \quad\ \quad \quad \quad \check{p}_{\alpha,j} \sqrt{1-e^2_{j}-\bs p^2_{j}-|d_{j}|^2} \\
			+\sum_i B^\alpha  \check{p}_{\alpha,i} \sqrt{1-e^2_{i}-\bs p^2_{i}-|d_{i}|^2}.
		 \end{gathered}
		 \end{equation}
		 
		The non vanishing contributions to the fluctuation matrix are given by
		\begin{align}
			\mathcal{M}^B_{\alpha,\beta}&=\frac12 \frac{\mf \psi_{\beta}}{\mf p_0}B^\alpha \quad \ \ \qquad \alpha \ \in \ \{5,6,7\}, \ \beta \in \{1,2,3\} \\
			\mathcal{M}^B_{\alpha,\alpha}&= \mf p_0^2 J^\alpha \sum_\Delta e^{i \bs q  \bs \Delta} \quad \alpha \ \in \ \{5,6,7\}
		\end{align}
		where $\bs \Delta \equiv \bs r_i-\bs r_j$ are the vectors connecting $i$ and $j$ and $\mf \psi$ represents the slave-bosons at the saddle point solution
		\end{subequations}

	\subsection{Pseudofermionic part of the fluctuation matrix}
	Now, we focus on the pseudoferminic part (\eq{App:Fluctuations:Exact:Pseudofermionic}) of the Fluctuation matrix
	and define the Green's function
	\begin{equation}
		\underline{G}_{k_1,k_2}[\psi]=\delta_{\varpi_{n},\varpi_{m}}\left(i\varpi_n-\underline{H}_{k_1,k_2}[\psi]\right)^{-1}
	\end{equation} 
	to expand the pseudofermionic part of the action around the saddle point solution
	\begin{subequations}
		\begin{align}
			\begin{split}
				\mc S_\mathrm{F}[\mf\psi]+\delta \mc S_\mathrm{F}[\delta\psi]
				&=-\Tr\log\left[-i\varpi_n+\ul H[\mf\psi]+\delta \ul H[\delta\psi] \right]
				\\
				= -\Tr\log\Big[(-i & \varpi_n+\ul H[\mf\psi])(1 - G[\mf\psi]\delta \ul H[\delta\psi]) \Big]
				\\
				=-\Tr\log\Big[(-i&\varpi_n+ \ul H[\mf\psi])\Big]+\Tr\sum_{l=1}^\infty \frac1l \left(\ul G[\mf\psi] \delta \ul H[\delta\psi]\right)^l
			\end{split}\label{App:Fluctuations:Pseudo:Expansion}
		\end{align}
		where the fluctuations $\delta \ul H[\psi]$ are defined as
		\begin{equation}
			\begin{gathered}
				\delta \ul H[\delta\psi]=\sum_{q}\sum_{\mu}\pddx{\ul H[\psi]}{\psi_{q,\mu}}{\mf\psi}\,\delta\psi_{q,\mu} \\
				+ \frac12 \sum_{qq'}\sum_{\mu\nu} \left.\frac{\partial^2 \ul H[\psi]}{\partial\psi_{q,\mu} \partial\psi_{q',\nu}}\right|_{\mf\psi}\,\delta\psi_{q,\mu}\delta\psi_{q',\nu}+ \mc{O}(\delta\psi^3).
			\end{gathered}
		\end{equation}
	\end{subequations}
	Now we expand \eq{App:Fluctuations:Pseudo:Expansion} up to the second order in $l$ and collect all terms which are of second order in $\delta\psi_\mu$
	\begin{equation}
		\begin{gathered}\label{App:Fluctuations:Pseudofermionic:S2}
			\delta\mc S_\mathrm{F}^{(2)}[\delta \psi]
			=\frac 12\sum_{qq'k}\sum_{\mu\nu}\delta\psi_{q,\mu}\delta\psi_{q',\nu}\,\tr\Bigg\{
			\ul G_{k}[\mf\psi]\left[\pdddx{\ul H[\psi]}{\psi_{q,\mu}}{\psi_{q',\nu}}{\mf\psi}\right]_{k,k}\\
			+\sum_{qq'k_1k_2}\sum_{\nu\mu}\ul G_{k}[\mf\psi]\left[\pddx{\ul H[\psi]}{\psi_{q,\mu}}{\mf \psi}\right]_{k_1,k_2}\,
			\ul G_{k_2}[\mf\psi]\left[\pddx{\ul H[\psi]}{\psi_{q',\nu}}{\mf \psi}\right]_{k_2,k_1}\Bigg\}.
		\end{gathered}
	\end{equation}
	The Green's function at the saddle point is given by
	\begin{equation}
		\begin{gathered}
			\ul G_{k}[\mf \psi]
			=\Big[\i\varpi_n-\ul H_\bk [\mf\psi]\Big]^{-1}
		\end{gathered}
	\end{equation}
	where $\ul H_\bk[\mf\psi]$ is the mean field Hamiltonian defined in \eq{App:Meanfield:Hamiltonian} at the saddle point.
	
	In order to evaluate Eq.~\eqref{App:Fluctuations:Pseudofermionic:S2}, we need to calculate derivatives of the $\underline{z}$-matrix in momentum space. It holds
	\begin{align}
		\begin{split}
			\delta \zz_q 
			&= \sqrt{\frac{T}{N}}\sum_{\br}\sum_{\mu}\int_0^{\frac{1}{T}}d \tau \ \e^{-\i \bq\br-i\omega_n \tau} \pdd{\zz_{r}}{\psi_{r,\mu}}\Bigg{|}_{\mf\psi} \,\delta\psi_{r,\mu} \\
			&= \sum_\mu \pdd{\zz}{\psi_\mu} \delta\psi_{q,\mu}
		\end{split}
	\end{align}
	and
	\begin{align}
		\delta^2 \zz_q&= \sqrt{ \frac{T}{N}}\sum_k\sum_{\mu\nu} \frac{\partial^2 \ul z}{\partial \psi_\mu \partial \psi_\nu}\Bigg{|}_{\mf\psi}\delta \psi_{k,\mu}\delta\psi_{q-k,\nu}
	\end{align} 
	since we evaluate the derivatives at the uniform static mean field solution, which does not depend on $r\equiv(\bs r$, $\tau)^T$. Consequently, we find 
	\begin{subequations}\label{App:Fluctuations:Pseudo:Derivatives}
		\begin{align}
			\pdd{\zz_q}{\psi_{q',\mu}}\Bigg{|}_{\mf \psi}&=\delta_{q,q'}\pdd{\zz}{\psi_\mu}\Bigg{|}_{\mf \psi}\\
			\quad \pdd{\zz^\dag_q}{\psi_{q',\mu}}\Bigg{|}_{\mf \psi}&=\delta_{q,-q'}\pdd{\zz^\dag}{\psi_\mu}\Bigg{|}_{\mf \psi}
			\\
			\pddd{\zz_q}{\psi_{q_1,\mu}}{\psi_{q_2,\nu}}\Bigg{|}_{\mf \psi}&= \sqrt{ \frac{T}{N}}\delta_{q,q_1+q_2}\pddd{\zz}{\psi_\mu}{\psi_\nu}\Bigg{|}_{\mf \psi}\\
			\quad \pddd{\zz^\dag_q}{\psi_{q_1,\mu}}{\psi_{q_2,\nu}}\Bigg{|}_{\mf \psi}&= \sqrt{ \frac{T}{N}}\delta_{q,-q_1-q_2}\pddd{\zz^\dag}{\psi_\mu}{\psi_\nu}\Bigg{|}_{\mf \psi}\,.
		\end{align}
	\end{subequations}
	Within path integral formalism, the $\ul z$-matrix defined in \eq{App:Meanfield:Noninteracting:z} is given by
	\begin{widetext}
	\begin{equation}\label{App:Fluctuations:Pseudo:z}
		\begin{gathered}
			\zz=\frac{\left[(e+d_1+id_2)\sqrt{1-e^2-\bs p^2-|d|^2}\ttau^0 + (e-d_1-id_2)(\bs p\cdot \bs\ttau)\right]}
			{\sqrt{2\left[\tfrac{1-|d|^2+e^2}{2}\ttau^0-\sqrt{1-e^2-\bs p^2-|d|^2}(\bs p\cdot \bs\ttau)\right]\left[\tfrac{1+|d|^2-e^2}{2}\ttau^0+\sqrt{1-e^2-\bs p^2-|d|^2}(\bs p\cdot \bs\ttau) \right]}}.
			\end{gathered}
	\end{equation}
	\end{widetext}
	To evaluate the derivatives, we make use of the fact, that every hermitian $2\times 2$ matrix can be diagonalized as
	\begin{equation}\label{App:Fluctuations:Pseudo:Pauli}
		U_{\hat{\bs a}}^\dag \left( a_0\ttau^0+\bs{a}\cdot\bs{\ttau} \right) U_{\hat{\bs a}}^\nodag = a_0 \ttau^0 + |a|\ttau^3\,
	\end{equation}
	where $U_{\hat{\bs a}}^\dag$ is a unitary matrix which only depends on $ \hat{\bs a}=\bs a/|\bs a|$. With \eq{App:Fluctuations:Pseudo:Pauli},
	one can diagonalize the three matrices in \eq{App:Fluctuations:Pseudo:z} and finds
	\begin{equation} 
		U_{\hat{\bs p}}^\dag \ \zz \ U^\nodag_{\hat{\bs p}}=
		\begin{pmatrix}
			z_+ & 0\\
			0 & z_- 
		\end{pmatrix}
		=\frac{z_+ + z_-}{2}\ttau^0+\frac{z_+ - z_-}{2}\ttau^3\,\\
	\end{equation}
	with
	\begin{widetext}
	\begin{align}
		z_\pm&=\frac{\left[\sqrt{1-e^2-\bs p^2-|d|^2}(e+d_1+\i d_2)\pm |\bs p|(e-d_1-\i d_2)\right]}{\sqrt{2\left[1-|d|^2-(\sqrt{1-e^2-\bs p^2-|d|^2}\pm|\bs p|)^2/2\right] \left[1-e^2-(\sqrt{1-e^2-\bs p^2-|d|^2}\mp |\bs p|)^2/2\right]}}
	\end{align}
	\end{widetext}
	where $|\bs p|\equiv\sqrt{p_1^2+p_2^2+p_3^2}$ and $|d|\equiv\sqrt{d_1^2+d_2^2}$.
	Rotating back with \eq{App:Fluctuations:Pseudo:Pauli} yields
	\begin{equation}\label{App:Fluctuations:Pseudo:zPauli}
		\zz=\frac{z_+ + z_-}{2}\ttau_0 +\frac{z_+ - z_-}{2} \sum_{\mu=1}^3 \frac{p_\mu}{|\bs p|}\,\ttau^\mu.
	\end{equation}
	We define
	\begin{subequations}
		\begin{align}
		\ul Z&=\ul z^T \\
		\ul z_0&=\ul Z \ \Big{|}_{\mf\psi} = z_0\Big{|}_{\mf\psi} \tau^0 \\
		\ul Z_\mu &=\frac{\partial \ul Z}{\partial \psi_\mu} \ \Bigg{|}_{\mf\psi}\\
		\ul Z_{\mu\nu} &=\frac{\partial^2 \ul Z}{\partial \psi_\mu\partial \psi_\nu} \ \Bigg{|}_{\mf\psi}\\
			\ul B_\mu&=\frac{\partial \ul \beta^T}{\partial \psi_\mu} \Bigg{|}_{\mf\psi}.
		\end{align}
	\end{subequations}
	where $\underline{\beta}$ has been defined in \eq{App:Pathintegral:Beta}.
	Note that 
	\begin{equation}
	 \ul Z \ \Big{|}_{\mf\psi} = z_0\Big{|}_{\mf\psi} \tau^0 
	\end{equation}
	is the unity matrix we have encountered already in equation \eqref{App:Meanfield:z0}. \\
	Now we can evaluate the derivatives of the slave-boson dependent hopping matrix at the saddle point. The first derivative yields
	
	\begin{equation}
		 \begin{gathered}
				\left[\pddx{\ul H[\psi]}{\psi_{q,\mu}}{\mf \psi}\right]_{k_1,k_2}
				= \sqrt{ \frac{T}{N}} \left(\delta_{q,0}\ \ul B_\mu
				+\delta_{q,k_1-k_2}z^\nodag_{0} \left[\ul Z^\dagger_\mu \underline{\mathcal{H}}_{\bk_2}
				+\underline{\mathcal{H}}_{\bk_1} \ul Z_\mu\right]\right).
		\end{gathered}
		\end{equation}
	For the second derivative, one finds
		\begin{align}
			\begin{gathered}
				\left[\left.\frac{\partial^2 \ul H[\psi]}{\partial\psi_{q,\mu} \partial\psi_{q',\nu}}\right|_{\mf\psi}\right]_{k_1,k_2}
				=\frac{T}{N}\delta_{k_2,k_1-q-q'}\Bigg[ \ul Z^\dagger_{\mu\nu} \underline{\mathcal{H}}_{\bk_2} z^\nodag_{0} \\
				+ z^\nodag_{0}  \underline{\mathcal{H}}_{\bk_1}  \ul Z_{\mu\nu}
				+\ul Z^\dagger_\mu \underline{\mathcal{H}}_{\bk_2+\bq'}  \ul Z_\nu
				+\ul Z^\dagger_\nu \underline{\mathcal{H}}_{\bk_1-\bq'}  \ul Z_\mu \Bigg]\,.
			\end{gathered}
		\end{align}
	Note, that the matrix $\underline{\mathcal{H}}_{\bk}$ as defined in equation \eqref{App:Def:Hamilton} contains only the $\bk$ dependent hopping elements of the bare system and is independent of the slave-bosons. This matrix has to be diagonal and spin degenerate in order to 
	get a paramagnetic solution for the mean field
	\begin{equation}
	 \left[\underline{\mathcal{H}}_{\bk}\right]_{s,s'} = \xi_{\bk}\delta_{s,s'}
	\end{equation}

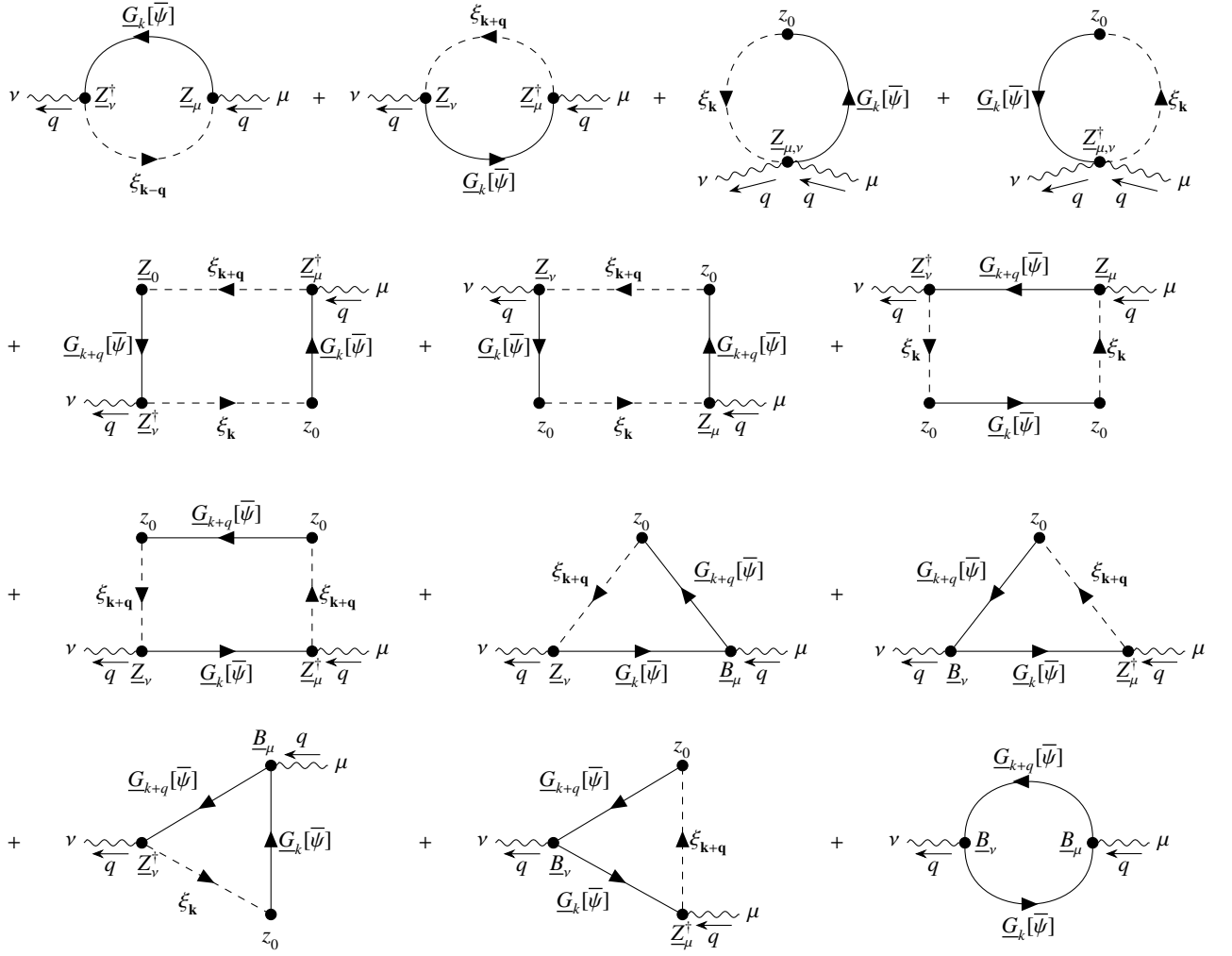
\begin{figure*}
		\tikzfeynmanset{every vertex={black,dot}}
		\tikzfeynmanset{/tikzfeynman/momentum/arrow distance=0.3}
		\begin{center}
			\begin{tikzpicture}
			\begin{feynman}
			\vertex (h1) ;
			\vertex at ($(h1) + (0.3cm, 0.0cm)$) (h12) {\(\ul Z^\dagger_{\nu\vphantom\mu}\)} ;
			\vertex at ($(h1) + (1.8cm, 0.0cm)$) (i1);
			\vertex at ($(i1) + (-0.3cm, 0cm)$) (h13) {\(\ul Z^\nodag_{\mu}\)} ;
			\vertex [right= 1 cm of i1] (m1){\(\mu\)};
			\vertex [left= 1 cm of h1] (m12){\(\nu\vphantom\mu\)};
			\diagram* {
				(h1) -- [anti fermion, half left, edge label=\(\ul G_k \lbrack \mf \psi \rbrack \)] (i1),(h1) -- [charged scalar, half right, edge label'=\( \xi_{\mathbf{k} - \mathbf{q}}^\nodag\)] (i1),
				(h1) -- [photon, momentum=\(q\)] (m12),
				(m1) -- [photon, momentum=\(q\)] (i1),   
			};
			
			\vertex [right= 0.5 cm of m1] (m1+){\(+\)};
			
			\vertex [right= 1.5 cm of m1+] (h2);
			\vertex at ($(h2) + (0.3cm, 0cm)$) (h22) {\(\ul Z^\nodag_{\nu\vphantom\mu}\)} ;
			\vertex at ($(h2) + (1.8cm, 0.0cm)$) (i2);
			\vertex at ($(i2) + (-0.3cm, 0cm)$) (h23) {\(\ul Z^\dagger_{\mu}\)} ;
			\vertex [right= 1 cm of i2] (m2){\(\mu\)};
			\vertex [left= 1 cm of h2] (m22){\(\nu\vphantom\mu\)};
			\diagram* {
				(i2) -- [charged scalar, half right, edge label'=\( \xi_{\mathbf{k} + \mathbf{q}}^\nodag\)] (h2),(i2) -- [anti fermion, half left, edge label=\(\ul G_k \lbrack \mf \psi \rbrack \)] (h2),
				(h2) -- [photon, momentum=\(q\)] (m22),
				(m2) -- [photon, momentum=\(q\)] (i2),   
			};
			
			\vertex [right= 0.5 cm of m2] (m2+){\(+\)};
			
			\vertex at ($(m2+) + (1.8cm, 0.9cm)$) (h3);
			\vertex at ($(h3) - (0cm, -0.3cm)$) (h32) {\( z^\nodag_{0}\)} ;
			\vertex at ($(h3) + (0cm, -1.8cm)$) (i3);
			\vertex at ($(i3) + (0.cm, 0.3cm)$) (h33) {\(\ul Z^\nodag_{\mu,\nu}\)} ;
			\vertex at ($(i3) + (1.2cm, -0.3cm)$) (m3) {\(\mu\)} ;
			\vertex at ($(i3) + (-1.2cm, -0.3cm)$) (m32) {\(\nu\vphantom\mu\)} ;
			\diagram* {
				(i3) -- [fermion, half right, looseness=1.5, edge label'=\(\ul G_k \lbrack \mf\psi \rbrack \)] (h3),(h3) -- [charged scalar, half right, looseness=1.5, edge label'=\( \xi_{\mathbf{k}}^\nodag\)] (i3),
				(i3) -- [photon, momentum=\(q\)] (m32),
				(m3) -- [photon, momentum=\(q\)] (i3),   
			};
			
			\vertex at ($(m2+) + (4cm, 0cm)$) (m3+) {\(+\)} ;
			
			\vertex at ($(m3+) + (2.2cm, 0.9cm)$) (h4);
			\vertex at ($(h4) - (0cm, -0.3cm)$) (h42) {\(z^\nodag_{0}\)} ;
			\vertex at ($(h4) + (0cm, -1.8cm)$) (i4);
			\vertex at ($(i4) + (0.cm, 0.3cm)$) (h43) {\(\ul Z^\dagger_{\mu,\nu}\)} ;
			\vertex at ($(i4) + (1.2cm, -0.3cm)$) (m4) {\(\mu\)} ;
			\vertex at ($(i4) + (-1.2cm, -0.3cm)$) (m42) {\(\nu\vphantom\mu\)} ;
			\diagram* {
				(h4) -- [fermion, half right, looseness=1.5, edge label'=\(\ul G_k \lbrack \mf\psi \rbrack \)] (i4),(i4) -- [charged scalar, half right, looseness=1.5, edge label'=\( \xi^\nodag_{\mathbf{k}}\)] (h4),
				(i4) -- [photon, momentum=\(q\)] (m42),
				(m4) -- [photon, momentum=\(q\)] (i4),   
			};
			
			\vertex at ($(h1) + (-1.0cm, -3.5cm)$) (m4+) {\(+\)} ;
			
			\vertex at ($(m4+) + (1.8cm, -0.8cm)$) (h5);
			\vertex at ($(h5) - (-0.1cm, 0.3cm)$) (h52) {\(\ul Z^\dagger_{\nu\vphantom\mu}\)} ;
			\vertex at ($(h5) + (2.4cm, 0.0cm)$) (i5);
			\vertex at ($(i5) - (0.cm, 0.3cm)$) (h53) {\(z^\nodag_{0}\)} ;
			\vertex at ($(i5) + (0cm, 1.6cm)$) (h54);
			\vertex at ($(h54) + (0.0cm, 0.3cm)$) (h55) {\(\ul Z^\dagger_{\mu}\)} ;
			\vertex at ($(h54) - (2.4cm, 0cm)$) (h56);
			\vertex at ($(h56) - (-0.1cm, -0.3cm)$) (h57) {\(\ul Z^\nodag_{0}\)} ;
			\vertex [right= 1 cm of h54] (m5){\(\mu\)};
			\vertex [left= 1 cm of h5] (m52){\(\nu\vphantom\mu\)};
			
			\diagram* {
				(h5) -- [charged scalar, edge label'=\( \xi^\nodag_{\mathbf{k}}\)] (i5),(i5) -- [fermion, edge label'=\(\ul G_k \lbrack \mf \psi \rbrack \)] (h54), (h54) -- [charged scalar, edge label'=\(\xi^\nodag_{\mathbf{k}+\mathbf{q}}\)] (h56), (h56) -- [fermion, edge label'=\(\ul G_{k+q} \lbrack \mf \psi \rbrack \)] (h5)
				(h5) -- [photon, momentum=\(q\)] (m52),
				(m5) -- [photon, momentum=\(q\)] (h54),   
			};
			
			\vertex at ($(m4+) + (5.8cm, 0cm)$) (m5+) {\(+\)} ;
			
			\vertex at ($(m5+) + (1.6cm, -0.8cm)$) (h6);
			\vertex at ($(h6) - (-0.1cm, 0.3cm)$) (h62) {\(z^\nodag_{0}\)} ;
			\vertex at ($(h6) + (2.4cm, 0.0cm)$) (i6);
			\vertex at ($(i6) - (0.cm, 0.3cm)$) (h63) {\(\ul Z^\nodag_{\mu}\)} ;
			\vertex at ($(i6) + (0cm, 1.6cm)$) (h64);
			\vertex at ($(h64) + (0.0cm, 0.3cm)$) (h65) {\(z^\nodag_{0}\)} ;
			\vertex at ($(h64) - (2.4cm, 0cm)$) (h66);
			\vertex at ($(h66) - (-0.1cm, -0.3cm)$) (h67) {\(\ul Z^\nodag_{\nu}\)} ;
			\vertex [right= 1 cm of i6] (m6){\(\mu\)};
			\vertex [left= 1 cm of h66] (m62){\(\nu\vphantom\mu\)};
			
			\diagram* {
				(h6) -- [charged scalar, edge label'=\( \xi^\nodag_{\mathbf{k}}\)] (i6),(i6) -- [fermion, edge label'=\(\ul G_{k+q} \lbrack \mf \psi \rbrack \)] (h64), (h64) -- [charged scalar, edge label'=\( \xi^\nodag_{\mathbf{k}+\mathbf{q}}\)] (h66), (h66) -- [fermion, edge label'=\(\ul G_{k} \lbrack \mf \psi \rbrack \)] (h6)
				(h66) -- [photon, momentum=\(q\)] (m62),
				(m6) -- [photon, momentum=\(q\)] (i6),   
			};
			
			\vertex at ($(m5+) + (5.8cm, 0cm)$) (m6+) {\(+\)} ;
			
			\vertex at ($(m6+) + (1.3cm, +0.8cm)$) (h7);
			\vertex at ($(h7) + (-0.1cm, 0.3cm)$) (h72) {\(\ul Z^\dagger_{\nu\vphantom \mu}\)} ;
			\vertex at ($(h7) + (0cm, -1.6cm)$) (i7);
			\vertex at ($(i7) - (0.cm, 0.3cm)$) (h73) {\(z^\nodag_{0}\)} ;
			\vertex at ($(i7) + (2.4cm, 0cm)$) (h74);
			\vertex at ($(h74) - (0.0cm, 0.3cm)$) (h75) {\(z^\nodag_{0}\)} ;
			\vertex at ($(h74) + (0cm, +1.6cm)$) (h76);
			\vertex at ($(h76) - (-0.1cm, -0.3cm)$) (h77) {\(\ul Z^\nodag_{\mu}\)} ;
			\vertex [right= 1 cm of h76] (m7){\(\mu\)};
			\vertex [left= 1 cm of h7] (m72){\(\nu\vphantom\mu\)};
			
			\diagram* {
				(h7) -- [charged scalar, edge label'=\( \xi^\nodag_{\mathbf{k}}\)] (i7),(i7) -- [fermion, edge label'=\(\ul G_{k} \lbrack \mf \psi \rbrack \)] (h74), (h74) -- [charged scalar, edge label'=\( \xi^\nodag_{\mathbf{k}}\)] (h76), (h76) -- [fermion, edge label'=\(\ul G_{k+q} \lbrack \mf \psi \rbrack \)] (h7)
				(h7) -- [photon, momentum=\(q\)] (m72),
				(m7) -- [photon, momentum=\(q\)] (h76),   
			};      
			
			\vertex at ($(m4+) + (0cm, -3.5cm)$) (m7+) {\(+\)} ;
			
			\vertex at ($(m7+) + (1.8cm, +0.8cm)$) (h8);
			\vertex at ($(h8) - (-0.1cm, -0.3cm)$) (h82) {\(z^\nodag_{0}\)} ;
			\vertex at ($(h8) + (0cm, -1.6cm)$) (i8);
			\vertex at ($(i8) - (0.cm, 0.3cm)$) (h83) {\(\ul Z^\nodag_{\nu}\)} ;
			\vertex at ($(i8) + (2.4cm, 0cm)$) (h84);
			\vertex at ($(h84) + (0.0cm, -0.3cm)$) (h85) {\(\ul Z^\dagger_{\mu}\)} ;
			\vertex at ($(h84) - (0cm, -1.6cm)$) (h86);
			\vertex at ($(h86) - (-0.1cm, -0.3cm)$) (h87) {\(z^\nodag_{0}\)} ;
			\vertex [right= 1 cm of h84] (m8){\(\mu\)};
			\vertex [left= 1 cm of i8] (m82){\(\nu\vphantom\mu\)};
			
			\diagram* {
				(h8) -- [charged scalar, edge label'=\( \xi^\nodag_{\mathbf{k}+\mathbf{q}}\)] (i8),(i8) -- [fermion, edge label'=\(\ul G_{k} \lbrack \mf \psi \rbrack \)] 
				(h84), (h84) -- [charged scalar, edge label'=\( \xi^\nodag_{\mathbf{k}+\mathbf{q}}\)] (h86), (h86) -- [fermion, edge label'=\(\ul G_{k+q} \lbrack \mf \psi \rbrack \)] (h8)
				(i8) -- [photon, momentum=\(q\)] (m82),
				(m8) -- [photon, momentum=\(q\)] (h84),   
			};   
			
			\vertex at ($(m5+) + (0cm, -3.5cm)$) (m8+) {\(+\)} ;
			
			\vertex at ($(m8+) + (1.8cm, -0.8cm)$) (h9);
			\vertex at ($(h9) - (-0.1cm, 0.3cm)$) (h92) {\(\ul Z^\nodag_{\nu}\)} ;
			\vertex at ($(h9) + (2.5cm, 0.0cm)$) (i9);
			\vertex at ($(i9) - (0.cm, 0.3cm)$) (h93) {\(\ul B^\nodag_{\mu}\)} ;
			\vertex at ($(i9) + (-1.25cm, 1.6cm)$) (h94);
			\vertex at ($(h94) + (0.0cm, 0.3cm)$) (h95) {\(z^\nodag_{0}\)} ;
			\vertex [left= 1 cm of h9] (m9){\(\nu\vphantom\mu\)};
			\vertex [right= 3.5 cm of h9] (m92){\(\mu\)};
			
			\diagram* {
				(h9) -- [fermion, edge label'=\(\ul G_{k} \lbrack \mf \psi \rbrack \)] (i9),(i9) -- [fermion, edge label'=\(\ul G_{k+q} \lbrack \mf \psi \rbrack \)] (h94), (h94) --[charged scalar, edge label'=\(\xi^\nodag_{\mathbf{k}+\mathbf{q}} \)] (h9), 
				(m92) -- [photon, momentum=\(q\)] (i9),
				(h9) -- [photon, momentum=\(q\)] (m9),   
			};  
			
			\vertex at ($(m6+) + (0cm, -3.5cm)$) (m9+) {\(+\)} ;
			
			\vertex at ($(m9+) + (1.6cm, -0.8cm)$) (h10);
			\vertex at ($(h10) - (-0.1cm, 0.3cm)$) (h102) {\(\ul B^\nodag_{\nu}\)} ;
			\vertex at ($(h10) + (2.5cm, 0.0cm)$) (i10);
			\vertex at ($(i10) - (0.cm, 0.3cm)$) (h103) {\(\ul Z^\dagger_{\mu}\)} ;
			\vertex at ($(i10) + (-1.25cm, 1.6cm)$) (h104);
			\vertex at ($(h104) + (0.0cm, 0.3cm)$) (h105) {\(z^\nodag_{0}\)} ;
			\vertex [left= 1 cm of h10] (m10){\(\nu\)};
			\vertex [right= 3.5 cm of h10] (m102){\(\mu\)};
			
			\diagram* {
				(h10) -- [fermion, edge label'=\(\ul G_{k} \lbrack \mf \psi \rbrack \)] (i10),(i10) -- [charged scalar, edge label'=\(\xi^\nodag_{\mathbf{k}+\mathbf{q}} \)] (h104), (h104) -- [fermion, edge label'=\(\ul G_{k+q} \lbrack \mf \psi \rbrack \)] (h10), 
				(m102) -- [photon, momentum=\(q\)] (i10),
				(h10) -- [photon, momentum=\(q\)] (m10),   
			};  
			
			\vertex at ($(m7+) - (0cm, 3.5cm)$) (m10+) {\(+\)} ;
			
			\vertex at ($(m10+) + (1.8cm, -0.0cm)$) (h11);
			\vertex at ($(h11) - (-0.1cm, 0.3cm)$) (h112) {\(\ul Z^\dagger_{\nu}\)} ;
			\vertex at ($(h11) + (1.82cm, -1.01cm)$) (i11);
			\vertex at ($(i11) - (0.cm, 0.3cm)$) (h113) {\(z^\nodag_{0}\)} ;
			\vertex at ($(i11) + (0cm, 2.1cm)$) (h114);
			\vertex at ($(h114) + (-0.1cm, 0.35cm)$) (h115) {\(\ul B^\nodag_{\mu}\)} ;
			\vertex [left= 1 cm of h11] (m11){\(\nu\vphantom\mu\)};
			\vertex [right= 1 cm of h114] (m112){\(\mu\)};
			
			\diagram* {
				(h11) -- [charged scalar, edge label'=\( \xi^\nodag_{\mathbf{k}}\)] (i11),(i11) -- [fermion, edge label'=\(\ul G_{k} \lbrack \mf \psi \rbrack \)] (h114), (h114) -- [fermion, edge label'=\(\ul G_{k+q} \lbrack \mf \psi \rbrack \)] (h11), 
				(m112) -- [photon, momentum'=\(q\)] (h114),
				(h11) -- [photon, momentum=\(q\)] (m11),   
			}; 
			
			\vertex at ($(m8+) + (0cm, -3.5cm)$) (m11+) {\(+\)} ;
			
			\vertex at ($(m11+) + (1.8cm, -0.0cm)$) (h12);
			\vertex at ($(h12) - (-0.1cm, 0.3cm)$) (h122) {\(\ul B^\nodag_{\nu}\)} ;
			\vertex at ($(h12) + (1.82cm, -1.01cm)$) (i12);
			\vertex at ($(i12) - (0.cm, 0.3cm)$) (h123) {\(\ul Z^\dagger_{\mu}\)} ;
			\vertex at ($(i12) + (0cm, 2.1cm)$) (h124);
			\vertex at ($(h124) + (0.0cm, 0.3cm)$) (h125) {\(z^\nodag_{0}\)} ;
			\vertex [left= 1 cm of h12] (m12){\(\nu\vphantom\mu\)};
			\vertex [right= 1 cm of i12] (m122){\(\mu\)};
			
			\diagram* {
				(h12) -- [fermion, edge label'=\( \ul G_{k} \lbrack \mf \psi \rbrack\)] (i12),(i12) -- [charged scalar, edge label'=\(\xi^\nodag_{\mathbf{k}+\mathbf{q}} \)] (h124), (h124) -- [fermion, edge label'=\(\ul G_{k+q} \lbrack \mf \psi \rbrack \)] (h12), 
				(m122) -- [photon, momentum=\(q\)] (i12),
				(h12) -- [photon, momentum=\(q\)] (m12),   
			};  
			
			\vertex at ($(m9+) - (0cm, 3.5cm)$) (m12+) {\(+\)} ;
			
			\vertex at ($(m12+) + (1.8cm, -0.0cm)$) (h13);
			\vertex at ($(h13) - (-0.3cm, -0.0cm)$) (h132) {\(\ul B^\nodag_{\nu\vphantom\mu}\)} ;
			\vertex at ($(h13) + (1.8cm, 0.0cm)$) (i13);
			\vertex at ($(i13) - (0.3cm, 0cm)$) (h133) {\(\ul B^\nodag_{\mu}\)} ;
			\vertex [left= 1 cm of h13] (m13){\(\nu\vphantom\mu\)};
			\vertex [right= 1 cm of i13] (m132){\(\mu\)};
			
			\diagram* {
				(h13) -- [fermion, half right, looseness=1.5, edge label'=\(\ul G_k \lbrack \mf \psi \rbrack \)] (i13),(h13) -- [anti fermion, half left, looseness=1.5, edge label=\( \ul G_{k+q} \lbrack \mf \psi \rbrack\)] (i13),
				(m132) -- [photon, momentum=\(q\)] (i13),
				(h13) -- [photon, momentum=\(q\)] (m13),   
			};                             
			\end{feynman}
			\end{tikzpicture}
		\end{center} 
		\caption{\eq{App:Fluctuations:Pseudofermionic:KernelF} recast in Feynman diagrams. Each loop contains a trace over the spin indices of the matrices and a 
		sum over $ k =(\bk, \omega_n)$. The diagrams are read in the arrow direction of the propagators, with an arbitrary starting point due to the invariance of the trace under cyclic permutations. 
		The first four terms originate from the first four terms of Eq. \eqref{App:Fluctuations:Pseudofermionic:KernelF}, and the the following nine from the rest.}
		\label{Fig:App:Feynman}
	\end{figure*}

	Combining all these results, we can write the fermionic part of the second order expansion of the action as
	\begin{widetext}
		\begin{align}
			\begin{gathered}
				\delta\mc S_F^{(2)}[\delta \psi]=
				\frac{T}{2N}\sum_{qq'}\sum_{\mu\nu}\delta\psi_{q',\mu}\delta\psi_{q,\nu}\,\delta_{q,-q'}\sum_k \tr\Bigg[
				 \ul G_k[\mf\psi]\left[\ul Z^\dagger_{\mu\nu} \xi_{\bk} z^\nodag_{0}
				+z^\nodag_{0} \xi_{\bk} \ul Z_{\mu\nu}
				+ \ul Z^\dagger_\mu \xi_{\bk+\bq} \ul Z_\nu
				+\ul  Z^\dagger_\nu \xi_{\bk-\bq} \ul  Z_\mu \right]
				\\
				+
				\ul G_k[\mf\psi]
				\left[\ul  Z^\dagger_\mu \xi_{\bk+\bq} z^\nodag_{0} +z^\nodag_{0}  \xi_{\bk} \ul  Z_\mu+\ul B_\mu \delta_{q,0} \right]
				\ul G_{k+q}[\mf\psi]
				\left[\ul Z^\dagger_\nu \xi_{\bk} z^\nodag_{0} +z^\nodag_{0}  \xi_{\bk+\bq} \ul  Z_\mu+ \ul B_\mu \delta_{\bq,0} \right]\Bigg].
			\end{gathered}
		\end{align}
		Then, the pseudofermionic part of the fluctuation matrix $\mc M_{\mu\nu}$ defined in Eq.~\eqref{App:Fluctuations:Kernel} in is given by
		\begin{align}\label{App:Fluctuations:Pseudofermionic:KernelF}
			\begin{gathered}
				\mc M^F_{\mu\nu}(q)
				=\frac{T}{2N} \sum_k \tr\Bigg[
				 \ul G_k[\mf\psi]\left[\ul Z^\dagger_{\mu\nu} \xi_{\bk} z^\nodag_{0} 
				+z^\nodag_{0} \xi_{\bk} \ul Z_{\mu\nu}
				+ \ul Z^\dagger_\mu \xi_{\bk+\bq} \ul Z_\nu
				+\ul  Z^\dagger_\nu \xi_{\bk-\bq} \ul  Z_\mu \right]
				\\
				+
				\ul G_k[\mf\psi]
				\left[\ul  Z^\dagger_\mu \xi_{\bk+\bq} z^\nodag_{0} +z^\nodag_{0}  \xi_{\bk} \ul  Z_\mu+\ul B_\mu \delta_{q,0} \right]
				\ul G_{k+q}[\mf\psi]
				\left[\ul Z^\dagger_\nu \xi_{\bk} z^\nodag_{0} +z^\nodag_{0}  \xi_{\bk+\bq} \ul  Z_\mu+ \ul B_\mu \delta_{\bq,0} \right]\Bigg].
			\end{gathered}
		\end{align}
	\end{widetext}
	Equation \eqref{App:Fluctuations:Pseudofermionic:KernelF} can be recast as Feynman diagrams, which are shown in Figure \ref{Fig:App:Feynman}.
	
	\begin{subequations}
		The occurring fermionic Matsubara summations can be carried out analytically. The Green's Matrix on mean field level is diagonal and degenerate
		\begin{equation}
	  \left[\underline G_{k}[\mf \psi] \right]_{ss'}=\left(\i\varpi_n-\epsilon_{\bk}\right)^{-1}\delta_{ss'}
		\end{equation}\label{App:Fluctuations:eigensystem}
		where $s$ is the spin index. Note, that in contrast to $\xi_\bk$ the eigenvalues  $\epsilon_{\bk,s}=\epsilon_{\bk,s}[\mf\psi
	]$ depend on the mean field values of the slave-bosons. The summations yield
		\begin{align}
			\begin{gathered}
				\left[T\sum_n \ul G_{(\omega_n,\bk)}[\mf\psi]\right]_{ss'}=
				\nF(\epsilon_{\bk})\delta_{s,s'}.
			\end{gathered}
		\end{align}
		and
		\begin{align}
			\begin{gathered}
				\left[T\sum_n \ul G_{(\varpi_n,\bk)}[\mf\psi] \ul M_{\bk,\bq} \ul G_{(\varpi_n+\omega_m,\bk+\bq)}[\mf\psi]\right]_{ss'}\\
				=\sum_{ss'} \frac{\nF(\epsilon_{\bk})-\nF(\epsilon_{\bk+\bq})}{i\omega_m+\epsilon_{\bk}-\epsilon_{\bk+\bq}}
				\left( \ul M_{\bk,\bq}\right)_{ss'}
			\end{gathered}
		\end{align}
		where 
		\begin{align}
			\ul M_{\bk,\bq}\equiv\left[\ul Z^\dagger_\mu \xi^\nodag_{\bk+\bq}z_0 +z_0  \xi^\nodag_{\bk}\ul Z_\mu+\ul B_\mu \right].
		\end{align}
	\end{subequations}
	\\
	\subsection{Result for the fluctuation matrix}
	Using all previous results, we obtain the final result for the fluctuation matrix.
	\begin{widetext}
	\begin{subequations}
	\begin{equation}
	\mc M_{\mu\nu}(\bs q, \omega_n)= \mc M^B_{\mu\nu}(\omega_n)+\mc M^F_{\mu\nu}(\bs q, \omega_n)
	\end{equation}
	\begin{equation}
	 \begin{gathered}
	 \mc M^B_{\mu\nu}(\omega_n)=\frac{\partial^2}{\partial\psi_\mu \partial \psi_\nu}
	 \frac12\Big[U \left(d_1^2+d_2^2\right) 
	-\beta_{0}\left(1 + d_1^2+d_2^2 - e^2\right)
	-2\bs \beta \bs p \sqrt{1- d_1^2- d_2^2 - e^2 -\bs p^2}   \Big]\Bigg{|}_{\mf\psi}\\
	+ \omega_n \left(\delta_{\mu,2}\delta_{\nu,3}-\delta_{\mu,3}\delta_{\nu,2} \right)		
	\end{gathered}
	\end{equation}
		\begin{align}\label{App:Fluctuations:Fluctuationmatrix}
			\begin{gathered}
				\mc M^F_{\mu\nu}(\bq, \omega_n)
				=\frac{1}{2N}\sum_\bk \Bigg\{
				\sum_s\nF(\epsilon_{\bk})\left[\ul Z^\dagger_{\mu\nu} \xi_{\bk} z_0
				+z_0 \xi_{\bk} \ul Z_{\mu\nu} 
				+\ul Z^\dagger_\mu \xi_{\bk+\bq} \ul Z_\nu
				+\ul Z^\dagger_\nu \xi_{\bk-\bq} \ul Z_\mu\right]_{s,s}
				\\
				{} +\sum_{ss'} \frac{\nF(\epsilon_{\bk})-\nF(\epsilon_{\bk+\bq})}{i\omega_n+\epsilon_{\bk}-\epsilon_{\bk+\bq}}
				\left[\ul Z^\dagger_\mu \xi_{\bk+\bq} z_0+ z_0 \xi_\bk \ul Z_\mu+\ul B_\mu\right]_{ss'}  
				\left[\ul Z^\dagger_\nu \xi_{\bk} z_0+ z_0 \xi_{\bk+\bq} \ul Z_\nu+\ul B_\nu\right]_{s's}
				\Bigg\} \,.
			\end{gathered}
		\end{align}
	\end{subequations}
	\end{widetext}
	
	In order to numerically evaluate the fluctuation matrix, we have to Wick rotate the fluctuation matrix to the real axis by analytic continuation and introduce a finite broadening $\eta$ such that we replace
	\begin{equation}
		i\omega_n \rightarrow \omega+i\eta
	\end{equation}
	with $\eta \rightarrow 0$. In practice, $\eta$ should be smaller than any physically relevant energy scale.
	
	Remember, that all slave-boson fields except for the $d=d_1+id_2$ field are real valued since their phase has been gauged away in \autoref{Chapter:App:Gauge}. Due to that, there is an $\omega_n$-dependent term in the bosonic 
	part of the fluctuation matrix which couples $d_1$ and $d_2$. Moreover, in the pseudo fermionic part, it is $\ul Z^\dagger \neq \ul Z^\nodag$, but the only difference is, that $id_2\rightarrow -id_2$ in the 
	$\ul Z$-matrix, which is only relevant if $\mu=3$ and/or $\nu=3$ which corresponds to the $d_2$ field in our basis. The matrix employs the symmetry
	\begin{align}
	\begin{split}
	 \mathcal{M}_{\nu\mu}(q)&=-\mathcal{M}_{\mu\nu}(q) \qquad (\mu=3, \nu \neq 3)\  \cup (\mu\neq3, \nu = 3) \\
	 \mathcal{M}_{\nu\mu}(q)&=\mathcal{M}_{\mu\nu}(q) \qquad \qquad \quad \qquad \mathrm{otherwise}.
	\end{split}
	\end{align}

	\section{Correlation functions}\label{Chapter:App:Correlation}
	In this section, it will be shown how to obtain correlation functions from the Fluctuation matrix $\mathcal{M}_{\mu\nu}$. Correlation functions can be written as a functional integral
	\begin{align}
		\begin{split}
			\langle \delta \psi^\nodag_\mu(-q) \delta\psi^\nodag_\nu(q)\rangle &=\frac{1}{Z^{(2)}}\int D[\delta \psi^*,\delta \psi] \delta \psi^\nodag_\mu(-q) \delta\psi^\nodag_\nu(q) e^{-\delta \mathcal{S}^{(2)}}\\
			\text{with} \quad Z^{(2)} &=\int D[\delta \psi^*,\delta \psi]  e^{-\delta \mathcal{S}^{(2)}},
		\end{split}
	\end{align}
	which can be integrated with the genereralized Gaussian integral
	\begin{equation}
		\begin{gathered}
			\int D\left[\psi^*,\psi^\nodag \right] \exp\left(-\sum_{\alpha \beta} \psi_{\alpha}^* \mathcal{A}_{\alpha \beta} \psi^\nodag_{\beta}+J^*_\alpha \psi^\nodag_\alpha + \psi^*_\alpha J^\nodag_\alpha\right)\\
			=\frac{\exp(J^*_\alpha \mathcal{A}^{-1}_{\alpha \beta}J^\nodag_\beta)}{\det\left(\underline{\mathcal{A}}\right)}
		\end{gathered}
	\end{equation}
	and the fluctuation matrix $\mathcal{M}_{\mu\nu}$:
	\begin{equation}\label{App:Fluctuations:Calculate:Correlation}
		\begin{gathered}
			\langle \delta \psi_\mu^*(q) \delta\psi^\nodag_\nu(q)\rangle=
			\lim_{J\rightarrow 0} \frac{1}{Z^{(2)}}\int D[\delta \psi^*,\delta \psi] \ \partial J^*_\nu(q) \partial J^\nodag_\mu(q) \ \times \\
			\exp \left(\sum_{\tilde{q}}[-\delta\psi^*_\mu(\tilde{q}) \mathcal{M}_{\mu\nu}(\tilde{q}) \delta \psi_\nu(\tilde{q})+J^*_\mu(\tilde{q}) \delta \psi^\nodag_\nu(\tilde{q})+\delta \psi^*_\mu(\tilde{q}) J^\nodag_\nu(\tilde{q})]\right)\\
			=\mathcal{M}^{-1}_{\mu \nu}(q). \\
		\end{gathered}
	\end{equation}
	
	\subsection{Bare susceptibility}
	The bare susceptibility is defined as
	\begin{equation}
		\begin{gathered}
			\chi_0(q):= \frac{1}{Z^{(0)}}\int D[f^*,f] \hat{n}_{-q}\hat{n}_q  e^{-\mathcal{S}^{(0)}}\\
		\end{gathered}
	\end{equation}
	with $Z^{(0)} =\int D[f^*,f]  e^{-\mathcal{S}^{(0)}}$
	and the mean field action
	\begin{equation}
		\mathcal{S}^{(0)}=\sum_{\bk, \omega_n} \bs  f^\dagger_k \left[-i\omega_n+\ul H[\mf \psi]_{\bk}\right]\bs f^\nodag_k=-\sum_{\bk, \omega_n} \bs f^\dagger_k \ul G_k^{-1}[\mf \psi]\bs f^\nodag_k
	\end{equation}
	where $\ul H_{\bk}[\mf \psi]$ is the mean field renormalized Hamiltonian defined in \eq{App:Meanfield:Hamiltonian}, $\bs f$ represents the collection of (pseudo-) fermionic fields and $n_q$ is the 
	pseudofermion density defined in \eq{App:Operators:Numberoperator} in Fourier space. Consequently, it is
	\begin{equation}
		\begin{gathered}
			\chi_0(q)= \frac{1}{Z^{(0)}}\int D[f^*,f] f^*_{k_1 +q}f^\nodag_{k_1}f^*_{k_2-q}f^\nodag_{k_2} \times \\ 
			\exp\left(\sum_{k} \bs f^\dagger_k \ul G_k^{-1}[\mf \psi] \bs f^\nodag_k\right)  =-\frac{T}{N} \sum_k \tr \left(\ul G_{k+q}[\mf \psi]\ul G_{k}[\mf \psi]\right).
		\end{gathered}
	\end{equation}
	Comparing this result with, \eq{App:Fluctuations:Pseudofermionic:KernelF}, one finds that the bare susceptibility can be expressed with the fluctuation matrix
	\begin{equation}
		-\frac{1}{2}\chi_0(q)=\mathcal{M}_{4,4}(q).
	\end{equation}
	If the system is spin rotational invariant we moreover find
	\begin{equation}
		-\frac{1}{2}\chi_0(q)=\mathcal{M}_{8,8}(q)=\mathcal{M}_{9,9}(q)=\mathcal{M}_{10,10}(q).
	\end{equation}
	The bare susceptibility carries a ``hidden'' dependence of the interaction via the mean field Greens function $G[\mf \psi]$.
	\subsection{Charge susceptibility}
	The charge susceptibility is defined by
	\begin{equation}
		\chi_c(q)\equiv \langle \delta n_{-q} \delta n_q\rangle
	\end{equation}
	where $n_q$ is the charge density given in \eq{App:Operators:NumberoperatorBoson} in Fourier space
	\begin{equation}
		\begin{gathered}
			n_q=\sqrt{\frac{N}{T}}\delta_{q,0}+\sqrt{\frac{T}{N}}\sum_k\left( d_{1,q+k} d_{1,-k} 
			+d_{2,q+k}d_{2,-k}-e_{q+k}e_{-k}\right).
		\end{gathered}
	\end{equation}
	Note, that terms like $\langle d_{1,q}^2 \rangle$ and $\langle d_{1,q}e_q \rangle$ vanish which can be seen with Eq.~\eqref{App:Fluctuations:Calculate:Correlation}. Thus, we find
	\begin{equation}
		\chi_c(q)=2\bar{d}^2_{1}\mathcal{M}^{-1}_{2,2}(q)+2\bar{e}^2
		\mathcal{M}^{-1}_{1,1}(q)-2\bar{d}_{1}\bar{e}\left[\mathcal{M}^{-1}_{1,2}(q)+\mathcal{M}^{-1}_{2,1}(q)\right].
	\end{equation}

	\subsection{Spin susceptibility}\label{App:Correlationfunctions:Spinsusceptibility}
	The general spin susceptibility is defined by
	\begin{equation}
		\chi^{\alpha\beta}_s(q):=\Big{\langle}  \delta S^\alpha_{-q} \delta S^\beta_q \Big{\rangle}.
	\end{equation}
	where $S^\alpha_q$ is the $\alpha$-th component of the spin density in three dimensions and $\delta S^\alpha$ is the respective fluctuation around the mean field solution.
	With the slave-boson spin density vector in real space given in \eq{App:Gauge:Spinoperator}, one finds
	\begin{equation}
		S^\alpha_q=\sqrt{\frac{T}{N}}\sum_k \check{p}_{\alpha,k+q}p_{0,-k}
	\end{equation}
	which yields
	\begin{equation}
		\chi^{\alpha\beta}_s(q) =\bar{p}_0^2\langle \delta \check{p}_{\alpha,-q}  \delta \check{p}_{\beta,q} \rangle
	\end{equation}
	where $\bar{p}_0^2$ denotes the mean field value.
	With \eq{App:Fluctuations:Calculate:Correlation} we find
	\begin{equation}
		\chi^{\mu\nu}_s(q)=  \bar{p}_0^2 \mathcal{M}^{-1}_{\mu \nu}(q) \quad \text{where }\mu,\nu \in [5,6,7].
	\end{equation}
	For spin rotation invariant models, the off-diagonal elements of the susceptibility vanish, while the diagonal elements are identical. 
 Moreover, due to the paramagnetic mean field, one finds for the derivatives for models without spin orbit coupling
	\begin{subequations}
		\begin{align}
			\dfrac{\partial\zz}{\partial \psi_c} \Bigg|_{\mf\psi} &\propto \underline{\mathbb{1}}_2, \\
			\dfrac{\partial\zz}{\partial p_\mu} \Bigg|_{\mf\psi} &\propto \ttau^\mu, \\
			\dfrac{\partial^2\zz}{\partial p_\mu \partial p_\nu} \Bigg|_{\mf\psi} &= 0 \quad \text{for~} \mu \neq \nu \\
			\dfrac{\partial^2\zz}{\partial p_\mu \partial \psi_c} \Bigg|_{\mf\psi} &\propto \ttau^\mu.
		\end{align}
	\end{subequations}
	Inserting these results into \eq{App:Fluctuations:Pseudofermionic:KernelF} and calculating the trace, one finds, that only matrix elements which couple charge fields $\psi_c = (e,d_1,d_2,\beta_0)$ to 
	charge fields or $ p $-fields to their respective $ \beta $-fields (e.g $ p_1 $ and $ \beta_1 $) are non zero. Consequently,	
		fluctuations between spin 
	fields $\psi_s = (\bs p, \bs \beta)$ and 
	charge fields $\psi_c = (e,d_1,d_2,\beta_0)$ vanish, therefore $\mathcal{M}_{\mu \nu}$ is block diagonal.
	
	The resulting scalar susceptibility is then found by the simple formula
	\begin{equation} \label{eq:Spinsusceptibility}
		\chi_s(q)=\bar{p}_0^2\frac{ \mathcal{M}_{10,10}(q)}{\mathcal{M}_{7,7}(q)\mathcal{M}_{10,10}(q)-\mathcal{M}_{7,10}(q)\mathcal{M}_{10,7}(q)}.
	\end{equation}

\section{Spiral magnetic mean field in slave-boson formalism} \label{Chapter:App:MaME}
On the bases of the paramagnetic mean field discussed in \app{Chapter:App:MF}, we want to expand the ansatz to incorporate a spin spiral with ordering vector $\bq$.
Following reference\cite{PhysRevB.48.10320}, we define a new static mean field for the bosonic fields:	
\begin{subequations} \label{App:MagMeanfield:Ansatz}
	\begin{align}
			e_i&\rightarrow \langle e \rangle \in \mathbb{R}_0^+\\
			p_{0,i}&\rightarrow \langle p_0\rangle \in \mathbb{R}_0^+\\
			d_i&\rightarrow \langle d\rangle \in \mathbb{R}^+_0,\ \partial_\tau \langle d \rangle := 0 \\
			i\beta_{0,i}&\rightarrow \langle \beta_0\rangle \in \mathbb{R}\\
			i\alpha_{i}&\rightarrow \langle\alpha\rangle \in  \mathbb{R}, \\
			\bs p_i&\rightarrow\langle p\rangle 
			\begin{pmatrix}
				\cos(\phi_i)\\
				 \sin(\phi_i)\\
				  0
			\end{pmatrix},\ \langle p \rangle \in \mathbb{R}_0^+ \\
			i\beta_i&\rightarrow\langle \beta \rangle
			\begin{pmatrix}
				\cos(\phi_i)\\
				\sin(\phi_i)\\
				0
			\end{pmatrix},\ \langle \beta \rangle \in \mathbb{R}\\
			\phi_i&\equiv \bs q \bs r_i.
	\end{align}
\end{subequations}	
Again, we drop the brackets $\langle \rangle$ in the following equations to keep the notation short. 
Further we reemploy \eq{App:Fluctuations:Pseudo:zPauli} and find using \eq{App:MagMeanfield:Ansatz}

\begin{subequations}
		\begin{align}\label{App:MagMeanfield:Z_i}
				\zz_i=
		\begin{pmatrix}
		\mathcal{Z}_+ & \mathcal{Z}_- e^{i\phi_i}\\
		\mathcal{Z}_- e^{-i\phi_i} & \mathcal{Z}_+
		\end{pmatrix}
	\quad \text{with} \quad \mathcal{Z}_\pm= \frac{z_+ \pm z_-}{2}
	\end{align}
and
\begin{equation}
	z_\pm=\frac{p_0(e+d)\pm p(e-d)}{\sqrt{2\left[1-d^2-(p_0\pm p)^2/2\right] \left[1-e^2-(p_0\mp p)^2/2\right]}}.
\end{equation}
\end{subequations}

Note, that in contrast to the paramagnetic mean field $\zz_i$ is now not proportional to the unity matrix and not uniform. As stated before in \app{Chapter:App:MF}, the normalization fixes the non interacting limit and does not change $\zz_i$ on operator level before inserting the mean field ansatz.

\subsection{Free Energy}

Analogously to \autoref{Chapter:App:MF}, the Free energy $F$ given by
	\begin{equation}\label{App:MagMeanfield:Free:def}
		F=-T\ln Z+\mu_0 \mathcal{N},
	\end{equation}
	where $\mathcal{N}$ is the total number of electrons in the system.
	\begin{subequations}
		The Lagrangian in the magnetic mean field reads
		\begin{equation}
			\begin{gathered}
				\mc L_{\bq}= \sum_\bk \bs f^\dagger_\bk \left(\underline{H}_\bk[\bq,\psi] + \partial_\tau \right) \bs f^\nodag_\bk\\
				+ J\sum_{\left\langle ij \right\rangle}\bs S_i\bs S_j +N \big[ U d^2
				-\beta_{0}(p_{0}^2+p^2+2d^2)\\-2\beta p_0 p+\alpha(e^2+p_{0}^2+d^2-1+p^2)\big],
			\end{gathered}
		\end{equation}
with 
	\begin{equation} \label{eq:MagMeanfield:basis}
		\bs f_\bk:=
		\begin{pmatrix}
			f_{\uparrow,\bk} \\
			f_{\downarrow,\bk-\bq} 
		\end{pmatrix}.
	\end{equation}
The mean field renormalized hopping matrix $ \underline{H}_\bk[\bq,\psi] $ is found by transforming the real space hopping elements $\zz_i^{\dagger} t_{ij}^{\nodag} \zz_j^{\nodag}$ introduced in \eq{App:Def:Hamilton} to Fourier space
using the basis defined in \eq{eq:MagMeanfield:basis} and reads
	\begin{widetext}
		\begin{equation}\label{App:Meanfield:Hamiltonian}
			\underline{H}_{\bk}[\bq,\psi]\equiv 
			\begin{pmatrix}
				 \mathcal{Z}_+^2 \xi_\bk + \mathcal{Z}_-^2  \xi_{\bk-\bq} +\beta_0-\mu_0 & \mathcal{Z}_+\mathcal{Z}_- (\xi_{\bk-\bq}+\xi_\bk)+\beta \\
				 \mathcal{Z}_+\mathcal{Z}_- ( \xi_{\bk-\bq}+ \xi_\bk)+\beta &  \mathcal{Z}_+^2 \xi_{\bk -\bq} + \mathcal{Z}_-^2 \xi_{\bk }+\beta_0-\mu_0
			\end{pmatrix}
		\quad \text{with} \quad \left[\underline{\mathcal{H}}_{\bk}\right]_{s,s'} = \xi_{\bk}\delta_{s,s'},
		\end{equation}
	\end{widetext}
	\end{subequations}
where $\underline{\mathcal{H}}_{\bk}$ is the bare hopping Hamiltonian with the spin degenerate eigenvalues $\xi_{\bk}$.
In contrast to the paramagnetic mean field, we can involve a uniform spin-spin-interaction term proportional to $J$, which takes a purely bosonic form. 
The pseudofermions in the mean field Lagrangian can be integrated out with \eq{App:Atomic:Z0}. The slave-boson dependent eigenvalues of the matrix 
$\underline{H}_{\bk}[\psi]$ are labeled by $\epsilon_{\bs k,\pm}$.\\
The mean field free energy per lattice site is then found to be
	\begin{equation}
		\begin{gathered}
			f_\bq \equiv
			\frac{F_\bq}{N}=-T\frac{1}{N}\sum_{\bk,\pm} \log\left(1+\e^{-\epsilon_{\bk,\pm}/T}\right) \\
		    + Jp_0^2 p^2\sum_{\bs \Delta} \cos(\bs q \bs \Delta) + U d^2	-\beta_{0}(p_{0}^2+p^2+2d^2) \\
			-2\beta p_0 p +\alpha(e^2+p_{0}^2+d^2-1+p^2)-\mu_0 n
		\end{gathered}
	\end{equation}
where $n=\mathcal{N}/N$ is the total electron filling, $N$ is the number of lattice sites and $\bs\Delta = \bs r_i-\bs r_j$ denotes all vectors connecting the sites $i$ and $j$.
Note, that for $p = \beta =0$ the mean field ansatz is reduces to the paramagnetic mean field discussed in \app{Chapter:App:MF}.
	
\subsection{Saddle point equations}
In order to find the mean field solution for the ground state, we need to minimize the free energy with respect to the fields $e,p_0,p$ and $d$, while enforcing the constraints, which can be recovered by deriving the Free energy by the respective Lagrange parameter. The resulting saddle point equations are given by

\begin{widetext}
\begin{subequations}\label{App:Meanfield:Saddlepoint}
\begingroup
\allowdisplaybreaks
	\begin{align}
		\frac{\partial f_\bq}{\partial e}&=\frac{1}{N}\sum_{\bs k, \pm}n_F(\epsilon_{\bs k, \pm})\frac{\partial \epsilon_{\bs  k, \pm}}{\partial e} +2\alpha e=0, \label{App:MagMeanfield:Saddlepoint:e}\\
		\frac{\partial f_\bq}{\partial p_0}&=\frac{1}{N}\sum_{\bs k, \pm}n_F(\epsilon_{\bs k, \pm})\frac{\partial \epsilon_{\bs  k, \pm}}{\partial p_0} +2p_0(\alpha -\beta_0) -2\beta p +2J p_0 p^2\sum_{\bs \Delta} \cos(\bs q \bs \Delta)  =0 , \label{App:MagMeanfield:Saddlepoint:p_0} \\
		\frac{\partial f_\bq}{\partial p}&=\frac{1}{N}\sum_{\bs k, \pm}n_F(\epsilon_{\bs k, \pm})\frac{\partial \epsilon_{\bs  k, \pm}}{\partial p} +2p(\alpha -\beta_0) -2\beta p_0 +2J p_0^2 p^{\phantom{2}}\sum_{\bs \Delta} \cos(\bs q \bs \Delta)  =0 , \label{App:MagMeanfield:Saddlepoint:p_0} \\
		\frac{\partial f_\bq}{\partial d}&=\frac{1}{N}\sum_{\bs k, \pm}n_F(\epsilon_{\bs ,\pm})\frac{\partial \epsilon_{\bs  k,\pm}}{\partial d} +2d(U+\alpha-2\beta_0)=0 , \\
		\frac{\partial f_\bq}{\partial \alpha}&=e^2+p_0^2+p^2+d^2-1=0,\label{App:MagMeanfield:Saddlepoint:alpha} \\
		\frac{\partial f_\bq}{\partial \beta_0}&=\frac{1}{N}\sum_{\bs k,\pm}n_F(\epsilon_{\bs k,\pm})\frac{\partial \epsilon_{\bs  k,\pm}}{\partial \beta_0} -2d^2-p_0^2-p^2=0, \label{App:MagMeanfield:Saddlepoint:beta0}  \\
		\frac{\partial f_\bq}{\partial \beta}&=\frac{1}{N}\sum_{\bs k,\pm}n_F(\epsilon_{\bs k,\pm})\frac{\partial \epsilon_{\bs  k,\pm}}{\partial \beta} -2p_0p=0, \label{App:MagMeanfield:Saddlepoint:beta}  \\
		\frac{\partial f_\bq}{\partial \mu_0}&=-\frac{1}{N}\sum_{\bs k,\pm}n_F(\epsilon_{\bs k,\pm})+n=0,  \label{App:MagMeanfield:Saddlepoint:mu}  \\
		\frac{\partial f_\bq}{\partial \bq}&=0,  \label{App:MagMeanfield:Saddlepoint:q} 
	\end{align}
	\endgroup
\end{subequations}
\end{widetext}
where $n_F(\epsilon_{\bs k})=(1+\exp(\epsilon_{\bs k}/T))^{-1}$ is the Fermi-Dirac distribution. The second to last equation has to be enforced additionally to ensure the correct electron filling, instead of fixing the chemical potential and the last equation fixes the ordering vector $\bq$.
\subsection{Reduction of mean field equations}\label{App:Meanfield:Numerical}
Analogously to the paramagnetic mean field, one can reduce the degrees of freedom, yielding only five independent mean field variables by substituting $\beta_0=-\mu_{\text{eff}}+\mu_0$.
We then exploit the the two constraints which only couple to bosonic degrees of freedom, i.e. the constraint which ensures, that there is only one boson per site associated with $\alpha$ and the constraint which fixes the total number of particles associated with $\beta_0$ by setting
\begin{subequations}\label{App:MagneticMF:reduction}
\begin{align}
1&=e^2+d^2+p_0^2+ p ^2 \label{App:MagneticMF:reduction:alpha} \\
n&=2d^2+p_0^2+p^2 \\
\mu_{\text{eff}}&=\mu_0-\beta_0.
\end{align}
\end{subequations}
This way, the redundant degrees of freedom $\alpha$, $\beta_0$ and two arbitrary slave-boson fields (we choose $d$, and $e$ without loss of generality) are removed from the mean field equations.
The mean field solution is given by the saddle of the free energy given by
\begin{widetext}
\begin{equation}\label{key}
f_\bq~\Big{|}_{\eqref{App:MagneticMF:reduction}}=-\frac{T}{N}\sum_{\bk,\pm} \ln\left[1+e^{-\epsilon_{\bk,\pm}/T}\right]-\frac{U}{2}(p_0^2+p^2-n)+\mu_{\text{eff}}n-2\beta p_0 p .
\end{equation}
with the energy eigenvalues
\begin{subequations}
 \begin{equation}
\epsilon _{k,\pm }=\frac{1}{4}\left[\zeta _{+}\xi _{k,+}-4\mu _{\text{eff}}\pm 
\sqrt{(\zeta _{+}^{2}-\zeta _{-}^{2})\xi _{k,-}^{2}+(\zeta _{-}\xi
_{k,+}+4\beta )^{2}}\right]
 -\mu_{\text{eff}}
 \end{equation}
where $\zeta _{\pm }=z_{+}^{2}\pm z_{-}^{2}$, $\xi _{k,\pm }=\xi _{\mathbf{k}}\pm \xi _{\mathbf{k-Q}}$ and
\begin{gather} 
z_\pm~\Big{|}_{\eqref{App:MagneticMF:reduction}}=\frac{\left(p_0\pm p\right)\sqrt{2-n-p^2-p_0^2}+(p_0\mp p)\sqrt{n-p^2-p_0^2}}{\sqrt{\left(2-\left(p_0\mp p\right)^2-(2-n-p^2-p_0^2)\right)\left((p_0\mp p)^2+(2-n-p^2-p_0^2)\right)}}.
\end{gather}
\end{subequations}
\end{widetext}
Notice, that the spin degenerate paramagnetic energy eingenvalues are recovered if $p=\beta=0$ since in that case $\mathcal{Z}_+=z_0$ and $\mathcal{Z}_-=0$. 
We are left to determine
\begin{equation}
\frac{\partial f_\bq}{\partial p} ~\Bigg{|}_{\eqref{App:MagneticMF:reduction}}
=\frac{\partial f_\bq}{\partial p_0} ~\Bigg{|}_{\eqref{App:MagneticMF:reduction}}
=\frac{\partial f_\bq}{\partial \beta} ~\Bigg{|}_{\eqref{App:MagneticMF:reduction}}
=\frac{\partial f_\bq}{\partial \mu_{\text{eff}}} ~\Bigg{|}_{\eqref{App:MagneticMF:reduction}}
=\frac{\partial f_\bq}{\partial \bq} ~\Bigg{|}_{\eqref{App:MagneticMF:reduction}}
=0
\end{equation}
which we do by minimizing $f_\bq$ with respect to $p$ and $p_0$ and maximizing with respect to $\beta$ and $\mu_{\text{eff}}$ between each minimization step.

The original chemical potential is recovered by evaluating
\begin{equation}
\mu_0=\frac{1}{2 \bar{d}}\frac{1}{N}\sum_{\bs k,\pm}n_F(\epsilon_{\bs k,\pm})\frac{\partial \epsilon_{\bs  k,\pm}}{\partial d}\Bigg{|}_{\mf\psi,\eqref{App:MagneticMF:reduction:alpha}}+\mu_{\text{eff}} + \frac{U}{2}.
\end{equation}
Note, that the previous equation is to be understood such, that only \eq{App:MagneticMF:reduction:alpha} is applied to reduce the degrees of freedom to eliminate the $e$ field. Consequently to assign a unique value to $\mu_0$, one has to leave the constrained subspace of the mean field, which is only relevant for a subsequent fluctuation calculation.

\end{document}